\pdfoutput=1
\documentclass[12pt]{article}
\usepackage[utf8]{inputenc}      
\usepackage[colorlinks=true,citecolor=blue,urlcolor=blue,linktocpage=true,linkcolor=blue]{hyperref}
\usepackage{amsmath,amssymb,mathtools}
\usepackage[T1]{fontenc}          
\usepackage{booktabs,tabularx}
\usepackage{graphicx}
\usepackage{xspace}
\usepackage[usenames]{xcolor}\definecolor{fscolor}{RGB}{44,118,255}
\usepackage{geometry}
\usepackage{todonotes}
\usepackage{listings}
\usepackage[absolute]{textpos}
\usepackage[many]{tcolorbox}
\usepackage{xparse}
\usepackage[font=small,labelfont=bf,format=plain,margin=0.05\textwidth]{caption}
\usepackage{bbm}
\usepackage{tabularx}
\usepackage{soul}
\usepackage[capitalize]{cleveref}
\usepackage{slashed}
\usepackage{pifont}
\usepackage[shortlabels]{enumitem}
\usepackage[sorting=none,%
  citestyle=numeric-comp,%
  bibstyle=mynumeric,%
  giveninits=true]{biblatex}

\usepackage{subfigure}
\usepackage[super]{nth}

\usepackage{lineno}

\allowdisplaybreaks

\evensidemargin \oddsidemargin
\newcommand{\cp}{\ensuremath{{\cal CP}}\xspace}
\newcommand{\SM}{{\text{SM}}}

\newcommand{\NP}{{\text{NP}}}
\newcommand{\BR}{{\rm{BR}}}

\newcommand{\gev}{\,\, \mathrm{GeV}}

\newcommand{\hife}{Higgs--fermion\xspace}

\newcommand{\fV}{fermion+V\xspace}

\newcommand{\lv}{lepton-vector\xspace}

\definecolor{Darkgreen}{rgb}{0.,.7,0.2}
\definecolor{Darkblue}{rgb}{0.,.2,0.7}
\definecolor{Magenta}{rgb}{0.7,0.,0.7}

\newcommand{\kgamma}{\ensuremath{\kappa_\gamma}\xspace}
\newcommand{\kg}{\ensuremath{\kappa_g}\xspace}

\newcommand{\YBobs}{\ensuremath{Y_B^{\text{obs}}}\xspace}
\newcommand{\YBVIA}{\ensuremath{Y_B^{\text{VIA}}}\xspace}
\newcommand{\YBratio}{\ensuremath{Y_B^{\text{VIA}}/Y_B^{\text{obs}}}\xspace}

\newcommand{\cf}{\ensuremath{c_f}\xspace}
\newcommand{\ct}{\ensuremath{c_t}\xspace}
\newcommand{\cb}{\ensuremath{c_b}\xspace}
\newcommand{\cc}{\ensuremath{c_c}\xspace}
\newcommand{\cs}{\ensuremath{c_s}\xspace}
\newcommand{\cd}{\ensuremath{c_d}\xspace}
\newcommand{\cu}{\ensuremath{c_u}\xspace}
\newcommand{\ce}{\ensuremath{c_e}\xspace}
\newcommand{\cmu}{\ensuremath{c_\mu}\xspace}
\newcommand{\ctau}{\ensuremath{c_\tau}\xspace}
\newcommand{\cq}{\ensuremath{c_q}\xspace}
\newcommand{\cl}{\ensuremath{c_l}\xspace}
\newcommand{\cqu}{\ensuremath{c_{q_u}}\xspace}
\newcommand{\cqd}{\ensuremath{c_{q_d}}\xspace}
\newcommand{\cfff}{\ensuremath{c_{f_3}}\xspace}
\newcommand{\cff}{\ensuremath{c_{f_2}}\xspace}

\newcommand{\cft}{\ensuremath{\tilde c_{f}}\xspace}
\newcommand{\ctt}{\ensuremath{\tilde c_{t}}\xspace}
\newcommand{\cbt}{\ensuremath{\tilde c_b}\xspace}
\newcommand{\cct}{\ensuremath{\tilde c_c}\xspace}
\newcommand{\cst}{\ensuremath{\tilde c_s}\xspace}
\newcommand{\cdt}{\ensuremath{\tilde c_d}\xspace}
\newcommand{\cut}{\ensuremath{\tilde c_u}\xspace}
\newcommand{\cet}{\ensuremath{\tilde c_e}\xspace}
\newcommand{\cmut}{\ensuremath{\tilde c_\mu}\xspace}
\newcommand{\ctaut}{\ensuremath{\tilde c_\tau}\xspace}
\newcommand{\cqt}{\ensuremath{\tilde c_{q}}\xspace}
\newcommand{\clt}{\ensuremath{\tilde c_{l}}\xspace}
\newcommand{\cqut}{\ensuremath{\tilde{c}_{q_u}}\xspace}
\newcommand{\cqdt}{\ensuremath{\tilde{c}_{q_d}}\xspace}
\newcommand{\cffft}{\ensuremath{\tilde{c}_{f_3}}\xspace}
\newcommand{\cfft}{\ensuremath{\tilde{c}_{f_2}}\xspace}

\newcommand{\cv}{\ensuremath{c_V}\xspace}
\newcommand{\cvt}{\ensuremath{\tilde{c}_V}\xspace}
\newcommand{\cgamma}{\ensuremath{c_\gamma}\xspace}
\newcommand{\cgammatilde}{\ensuremath{\tilde c_{\gamma}}\xspace}
\newcommand{\cg}{\ensuremath{c_g}\xspace}
\newcommand{\cgtilde}{\ensuremath{\tilde c_{g}}\xspace}

\newcommand{\chimin}{\ensuremath{\chi^2_{\rm{min}}}\xspace}

\newcommand{\udl}{\textbf{up-down-lepton}}

\newcommand{\order}[1]{\ensuremath{{\cal O}(#1)}}

\newcommand{\address}{\sl\footnotesize}

\NewBibliographyString{refname}
\NewBibliographyString{refsname}
\DefineBibliographyStrings{english}{%
  refname = {Ref\adddot},
  refsname = {Refs\adddot}
}
\DeclareCiteCommand{\ccite}
  {%
  \ifnum\thecitetotal=1
    \bibstring{refname}%
  \else%
    \bibstring{refsname}%
  \fi%
  \addnbspace\bibopenbracket%
  \usebibmacro{cite:init}%
   \usebibmacro{prenote}}
  {\usebibmacro{citeindex}%
   \usebibmacro{cite:comp}}
  {}
  {\usebibmacro{cite:dump}%
   \usebibmacro{postnote}%
   \bibclosebracket}
\newrobustcmd*{\Ccite}{\bibsentence\ccite}

\begin{document}

\thispagestyle{empty}
\def\thefootnote{\fnsymbol{footnote}}

\begin{flushright}
CERN-TH-2021-231, DESY-22-033, EFI-22-1, IFT--UAM/CSIC--21-148
\end{flushright}
\vspace{3em}
\begin{center}
{\large{\bf
Constraining the \cp structure of Higgs-fermion couplings\\[.5em]
with a global LHC fit, the electron EDM and baryogenesis
}}
\\
\vspace{3em}
 {

Henning Bahl$^{1}$\footnotetext[0]{\footnotesize emails: hbahl@uchicago.edu, elina.fuchs@cern.ch, sven.heinemeyer@cern.ch, judith.katzy@desy.de,\\
\mbox{}\hspace{19mm}marco.menen@ptb.de, krisztian.peters@desy.de, matthias.saimpert@cea.fr,\\ \mbox{}\hspace{19mm}georg.weiglein@desy.de},
Elina Fuchs$^{2,3,4}$,
Sven Heinemeyer$^{5}$,
Judith Katzy$^{6}$,
Marco Menen$^{4,7,8}$,\\[.3em]
Krisztian Peters$^{6}$,
Matthias Saimpert$^{9}$,
Georg Weiglein$^{6,10}$
 }\\[2em]
 {\address $^1$    University of Chicago, Department of Physics, 5720 South Ellis Avenue, Chicago, IL 60637 USA}\\[0.2em]
 {\address $^2$    CERN, Department of Theoretical Physics, 1211 Geneve 23, Switzerland}\\[0.2em]
 {\address $^3$    Institut für Theoretische Physik, Leibniz Universität Hannover, Appelstraße 2, 30167 Hannover, Germany}\\[0.2em]
 {\address $^4$    Physikalisch-Technische Bundesanstalt, Bundesallee 100, 38116 Braunschweig, Germany}\\[0.2em]
 {\address $^5$    Instituto de F\'isica Te\'orica, (UAM/CSIC), Universidad Aut\'onoma de Madrid,\\[0.2em] Cantoblanco, E-28049 Madrid, Spain}\\[0.2em]
{\address $^6$    Deutsches Elektronen-Synchrotron DESY, Notkestr.\ 85, 22607 Hamburg, Germany}\\[0.2em]
{\address $^7$    Institut f\"ur Kernphysik, Universit\"at zu K\"oln, Z\"ulpicher Straße 77, 50937 K\"oln, Germany}\\[0.2em]
{\address $^8$    Physikalisches Institut, Universit\"at Bonn, Nu{\ss}allee 12, 53115 Bonn, Germany}\\[0.2em]
  {\address $^9$    IRFU, CEA, Université Paris-Saclay, Gif-sur-Yvette; France.}\\[0.2em]
 {\address $^{10}$ II.\  Institut f\"ur  Theoretische  Physik, Universit\"at  Hamburg, Luruper Chaussee 149,\\[0.2em] 22761 Hamburg, Germany}
\def\thefootnote{\arabic{footnote}}
\setcounter{page}{0}
\setcounter{footnote}{0}
\end{center}
\vspace{1.5ex}
\begin{abstract}
{}

\cp violation in the Higgs couplings to fermions is an intriguing, but not yet extensively explored possibility. We use inclusive and differential LHC Higgs boson measurements to fit the \cp structure of the Higgs Yukawa couplings. Starting with simple effective models featuring \cp violation in a single \hife coupling, we probe well-motivated models with up to nine free parameters. We also investigate the complementarity of LHC constraints with the electron electric dipole moment 
bound, 
taking into account the possibility of a modified electron Yukawa coupling, and assess to which extent \cp violation in the \hife couplings can contribute to the observed baryon asymmetry of the universe.
Even after including the recent analysis of angular correlations in $H\to\tau^+\tau^-$ decays, we find that a complex tau Yukawa coupling alone may be able to account for the observed baryon asymmetry, but with large uncertainties in the baryogenesis calculation.
A combination of complex top and bottom quark Yukawa couplings yields a result four times larger than the sum of their separate contributions, but remains insufficient to account for the observed baryon asymmetry.

\end{abstract}

\newpage
\tableofcontents
\newpage
\def\thefootnote{\arabic{footnote}}


\section{Introduction}
\label{sec:intro}

In 2012 the ATLAS and CMS collaborations discovered a new particle whose production and decay rates are consistent with the predictions for the Higgs boson of the Standard Model~(SM) with a mass of about $125 \gev$~\cite{Aad:2012tfa,Chatrchyan:2012xdj,Khachatryan:2016vau} within the present theoretical and experimental uncertainties. The latter amount to roughly $20\%$ for the most important production/decay rates of the detected state \cite{ATLAS:2019nkf,CMS:2018uag}. While so far no conclusive signs of beyond the SM (BSM) physics have been found at the LHC, the measured properties of the signal as well as the existing limits from the searches for additional particles are also compatible with the predictions of a wide range of BSM scenarios. Consequently, one of the main tasks of the LHC Run~3 as well as the High-Luminosity LHC (HL-LHC) is to probe the Higgs-boson couplings and quantum numbers more thoroughly and with higher accuracy.

An important target in this context is to determine the \cp structure of the Higgs-boson couplings. The possibility that the observed Higgs boson is a pure \cp-odd state could be ruled out already based on Higgs boson decays to gauge bosons recorded during Run~1~\cite{Khachatryan:2014kca,Aad:2015mxa}. These direct searches employed \cp-odd observables for probing \cp-violating effects. However, experimental access to a possible \cp admixture of the observed Higgs boson requires a much higher precision. Various experimental and theoretical analyses have been 
carried out in which the possibility of a \cp-mixed state has been taken into account. Experimental results became available mainly for the Higgs couplings to massive gauge bosons and gluons, and to a lesser extent also for the \hife couplings to top quarks and tau leptons.

\cp-violating effects in Higgs boson decays to gauge bosons were also investigated using the ``optimal observable method''~\cite{Aad:2016nal}. This analysis has been updated with the first year of the Run~2 data~\cite{Aad:2020mnm}. Anomalous Higgs couplings to massive gauge bosons were analyzed with the first year of the Run~2 data in the Higgs decay to four leptons ($H \to 4\ell$)~\cite{Sirunyan:2017tqd} and in weak boson fusion production followed by the decay $H \to \tau^+\tau^-$~\cite{Sirunyan:2019nbs}, as well as via a comparison of on- and off-shell production with $H \to 4\ell$~\cite{Sirunyan:2019twz}. Using the full Run~2 data, CMS studied \cp violation and anomalous couplings in Higgs production and the decay $H \to 4\ell$~\cite{CMS:2020dkv}. The \cp structure of the effective coupling of the Higgs boson to gluons was investigated by ATLAS with the first year of the Run~2 data using the decay $H \to WW^*$~\cite{ATLAS:2020uti}.

The experimental analyses described above were mainly based on observables involving either the $HZZ$ or the $HW^+W^-$ coupling. However, even for the case where $H$ would have a relatively large \cp-odd component, in many BSM models its effect in these couplings would be heavily suppressed in comparison to the \cp-even part. This is caused by the absence of a tree-level coupling between a \cp-odd Higgs boson and two vector bosons, yielding a large loop suppression. On the other hand, for the \hife couplings no such loop suppression is expected. As a consequence, the effects of a \cp-mixed state could manifest themselves in the \hife couplings in a much more pronounced way than in the Higgs couplings to massive gauge bosons. A thorough experimental investigation of the \hife couplings is therefore crucial for assessing the Higgs \cp properties. 

A very significant progress in this direction was recently 
made with a CMS analysis based on the full Run~2 data set~\cite{CMS:2021sdq}, where the \cp properties of the Higgs coupling to tau leptons were investigated using angular correlations between the decay planes of the tau leptons~\cite{Grzadkowski:1995rx,Berge:2008dr,Harnik:2013aja}. Furthermore, the \cp structure of the Higgs coupling to top quarks was analyzed by effectively mixing \cp-even and \cp-odd observables based on the full set of Run~2 data in the top quark associated Higgs production mode with decays to two photons~\cite{Sirunyan:2020sum,Aad:2020ivc} and to four leptons~\cite{CMS:2020dkv,CMS:2021nnc}. 

The searches for a possible \cp admixture of the observed Higgs boson described above did not give rise to any evidence for \cp-violating couplings, albeit with large experimental uncertainties.
In order to assess how well the \cp properties of the observed Higgs boson could be constrained on the basis of the experimental information available at that time, global fits have been carried out. These fits involve indirect searches where  \cp-even observables, e.g.\ Higgs-boson production and decay rates, are measured 
and their sensitivity to \cp violation in the Higgs boson couplings is explored. Although these are powerful tests, deviations from the SM predictions may also be caused by other BSM effects and so cannot uniquely be associated with the presence of \cp-violating effects. 
Early fits to a possible \cp admixture of the observed Higgs have been performed using Run~1 data (and partial Run~2 data), either investigating all Higgs-boson couplings~\cite{Freitas:2012kw,Djouadi:2013qya}, or focusing on the Higgs-top quark sector~\cite{Ellis:2013yxa,Boudjema:2015nda,Kobakhidze:2016mfx}. These analyses could set only very weak bounds on possible \cp violation in the Higgs sector. In addition to early fits, many studies have been performed investigating new observables and analysis techniques to constrain \cp violation in the Higgs sector at the LHC~\cite{Agrawal:2012ga,Harnik:2013aja,Bishara:2013vya,Demartin:2014fia,Demartin:2015uha,Demartin:2016axk,Cao:2019ygh,Buckley:2015vsa,Casolino:2015cza,Hagiwara:2016zqz,Gritsan:2016hjl,Azevedo:2017qiz,Brehmer:2017lrt,Barger:2018tqn,Goncalves:2018agy,Hou:2018uvr,Ren:2019xhp,Kraus:2019myc,Grojean:2020ech,Martini:2021uey,Bahl:2021dnc,Bhardwaj:2021ujv}.
A global fit to the \cp structure of the Higgs--top quark coupling was performed in \ccite{Bahl:2020wee}, taking into account inclusive and differential Higgs boson measurements from the ATLAS and CMS experiments. Future prospects for constraining the \cp-properties of the Higgs--top quark coupling at the HL-LHC were also investigated in this context.

Information that can be obtained about the \cp properties of the observed Higgs boson has important implications for cosmology. Indeed, since the Cabbibo-Kobayashi-Maskawa (CKM) matrix, which is the only source of \cp violation in the SM (unless one allows for a QCD phase, which is severely constrained experimentally), can only account for an amount of the baryon asymmetry of the universe (BAU) that is many orders of magnitude smaller than the observed one~\cite{Gavela:1993ts,Huet:1994jb}, additional sources of \cp violation must exist in nature. In the attempt to account for this observational gap, several possibilities to generate a sufficiently large BAU were analysed, for a recent review see \ccite{Bodeker:2020ghk}. 
One attractive possibility is electroweak baryogenesis (EWBG)~\cite{Trodden:1998ym,Morrissey:2012db} where the baryon asymmetry is produced during the electroweak phase transition, which relates this mechanism to Higgs physics and makes it potentially testable at colliders. While successful baryogenesis requires a modification of the scalar potential in order to render the electroweak phase transition first order, in this work we focus on the other necessary modification compared to the SM, namely providing additional sources of \cp violation via a \cp-admixed Higgs boson.

On the other hand, electric dipole moment (EDM) measurements put severe experimental constraints on the possible \cp structure of the couplings of the Higgs boson to fermions. The most stringent bounds on the electron EDM (eEDM) are currently set by the ACME collaboration, $|d_e| \le 1.1 \times 10^{-29} \, e$\,cm~\cite{Andreev:2018ayy}. Concerning the neutron EDM, the most stringent bound has been obtained by the nEDM collaboration, finding that $|d_n| \le 1.8 \times 10^{-26} \, e$\,cm~\cite{Abel:2020gbr}. The experimental progress was accompanied by theoretical investigations for applying these bounds to constrain the \cp structure of the fermion Yukawa couplings. An early review emphasizing the importance of EDMs in the analysis of \cp-violating Higgs couplings can be found in \ccite{Pospelov:2005pr}. Bounds on \cp-odd Higgs--top-quark couplings of \order{0.5} were set in \ccite{Brod:2013cka}, see also \ccite{Cirigliano:2016nyn}. First stringent bounds on the imaginary part of the Higgs-electron coupling were derived in \ccite{Altmannshofer:2015qra} based on the assumption of a SM-like top quark Yukawa coupling. Including two-loop QCD corrections, bounds on \cp violation in the Higgs coupling to bottom and charm quarks were analyzed in \ccite{Brod:2018pli}.
One way to evade the above-mentioned bounds are cancellations between various contributions to the EDMs. This was explored, e.g., in the context of BAU for the electron EDM in the Two Higgs Doublet Model (2HDM) in~\ccite{Fuyuto:2019svr}. The interplay between LHC, EDM and BAU constraints on complex Yukawa couplings has been investigated in \ccite{Fuchs:2019ore,Fuchs:2020uoc,Fuchs:2020pun} in the SM Effective Field Theory (SMEFT) including operators of dimension six for the top quark, bottom quark, tau and muon Yukawa couplings, and in \ccite{Shapira:2021mmy} in the $\kappa$-framework for all fermions. The emphasis of these papers was on predictions for the BAU, while the treatment of the LHC bounds was performed in an approximate manner. 
Furthermore, \ccite{deVries:2017ncy} investigated \cp violation for EWBG in the SMEFT of up to dimension eight, and \ccite{Bonnefoy:2021tbt} provided \cp-odd flavor invariants in the general SMEFT at dimension six and their relation to collective \cp violation.

Making use of up-to-date experimental information, we investigate in the present paper the question to which extent \cp violation may be present in the couplings of the observed Higgs boson to fermions, in particular to the top, bottom and charm quarks, as well as to the charged leptons of all three generations. This extends the scope of the previous analysis of~\ccite{Bahl:2020wee}, which focused specifically on the top quark Yukawa coupling and found that on the basis of the available LHC data a significant \cp-odd component in the top quark Yukawa coupling was compatible with the experimental results.

In our present analysis,  we also extend the scope of \ccite{Fuchs:2019ore,Fuchs:2020uoc,Shapira:2021mmy} by a more precise and complete evaluation of the LHC constraints and by a variation of the electron Yukawa coupling. Inclusive and differential Higgs-boson measurements as well as the recent CMS $H\to\tau^+\tau^-$ \cp analysis~\cite{CMS:2021sdq} are used to derive bounds on possible \cp-violating couplings to third and second generation fermions as well as to gauge bosons. These bounds are complemented with limits from the most recent EDM measurements. Based on this analysis, we investigate how much BAU can have been generated in the early universe, including the possible interplay of \cp violation in several \hife couplings.

This paper is organized as follows. The framework for the evaluation of the constraints in this work is the Higgs characterization model, summarized in \cref{sec:model_couplings}. In \cref{sec:model_pheno}, we present the phenomenological effective models with an increasing number of free coupling modifiers as well as their physics motivation from various concrete new physics (NP) models. In \cref{sec:LHC_constraints}, we 
discuss the LHC constraints and describe our sampling algorithm used to perform the fits with \texttt{HiggsSignals} in the high-dimensional parameter space. We then discuss the eEDM constraint in \cref{sec:EDM_constraints} and the prediction of the BAU in \cref{sec:BAU_constraints}, where in both cases we provide simple numerical formulas summarizing the different contributions and highlight the approximations they are based on. Our results are presented in \cref{sec:results}: in \cref{sec:results_LHC}, we analyze the LHC constraints on the different effective models in detail before confronting them with the complementary eEDM bound and the corresponding prediction for the BAU in \cref{sec:results_complementarity}. 
We conclude in \cref{sec:conclusions}. Details on the fit results of models with a high number of free parameters are provided in Appendix \ref{sec:appendix_additional_fits}.


\section{Effective model description}
\label{sec:model}

\subsection{Higgs characterization model}
\label{sec:model_couplings}

The basis of our investigation is the ``Higgs characterization model'', a framework based on an effective field theory (EFT) approach. This framework allows one to introduce \cp-violating couplings and to perform studies in a consistent, systematic and accurate way, see e.g.~\ccite{Artoisenet:2013puc}. The Yukawa part of the Lagrangian is modified with respect to the SM and reads as
\begin{align}\label{eq:Yuk_lagrangian}
\mathcal{L}_\text{yuk} = - \sum_{f=u,d,c,s,t,b,e,\mu,\tau} \frac{y_f^\SM}{\sqrt{2}} \bar f \left(\cf + i \gamma_5 \cft\right) f H,
\end{align}
where $H$ denotes the Higgs boson field and $f$ the fermion fields. The sum runs over all SM fermions. The coupling $y_f^\SM$ is the SM Yukawa coupling of the fermion $f$; the parameter \cf parameterizes deviations of the \cp-even $H f \bar f$ coupling from the SM, for which $\cf = 1$; the parameter \cft is used to introduce a \cp-odd $Hf\bar f$ coupling, with $\cft=0$ in the SM.

In the literature, the modified Yukawa couplings are also often parameterized in terms of an absolute value $|g_f|$ and a \cp-violating phase $\alpha_f$,
\begin{align}
|g_f| \equiv \sqrt{\cf^2 + \cft^2}, \qquad \qquad \tan\alpha_f = \frac{\cft}{\cf}.
\label{eq:phase}
\end{align}
In addition to modifications of the Yukawa Lagrangian, we also allow for an $SU(2)_L$ conserving modification of the Higgs interaction with massive vector bosons,
\begin{align}\label{eq:HVV}
\mathcal{L}_V = \cv H \left(\frac{M_Z^2}{v} Z_\mu Z^\mu + 2\frac{M_W^2}{v} W_\mu^+ W^{-\mu}\right).
\end{align}
Here, $Z$ and $W$ are the massive vector boson fields with the masses $M_Z$ and $M_W$, respectively, and $v \simeq 246\gev$ is the Higgs vacuum expectation value. The SM interaction is rescaled by the parameter \cv, which is equal to one in the SM.\footnote{The Higgs interaction with massive vector bosons can also be modified by introducing additional non-SM-like operators (e.g.\ $Z_{\mu\nu}Z^{\mu\nu}H$, where $Z_{\mu\nu}$ is the field strength of the $Z$ boson). However, we decided not to include these operators to keep the focus on the Higgs interaction with fermions.}

We further include the following operators to parameterize the effect of additional BSM particles affecting the Higgs production via gluon fusion and the Higgs decay into two photons,
\begin{align}\label{eq:dim5}
\mathcal{L}_{Hgg,H\gamma\gamma} ={}& - \frac{1}{4 v} H \left(- \frac{\alpha_s}{3\pi}\cg G_{\mu\nu}^a G^{a,\mu\nu} + \frac{\alpha_s}{2\pi}\cgtilde G_{\mu\nu}^a \widetilde{G}^{a,\mu\nu} \right)  \nonumber\\
& - \frac{1}{4 v} H\left( \frac{47\alpha}{18\pi}\cgamma A_{\mu\nu}A^{\mu\nu} + \frac{4\alpha}{3\pi}\cgammatilde A_{\mu\nu}\widetilde{A}^{\mu\nu}\right),
\end{align}
where $G_{\mu\nu}^a$ and $A_{\mu\nu}$ are the gluon and photon field strengths, respectively. Here $\alpha_s = g_3^2/4\pi$, where $g_3$ is the strong gauge coupling, and $\alpha = e^2/4\pi$, where $e$ is the elementary electric charge.\footnote{The form of the operators and their prefactors (e.g.\ $\alpha_s/3\pi$) are identical to those induced at the loop level in the SM when the top quark and the $W$ boson are decoupled from the theory.} The parameters $\cg, \cgtilde, \cgamma$ and $\cgammatilde$ parameterize these BSM effects in the Higgs couplings to gluons and photons. The SM is recovered for $\cg = \cgtilde = \cgamma = \cgammatilde = 0$.

The modifications of the Higgs production cross section via gluon fusion --- parameterized by $\kg^2 \equiv \sigma_{gg\to H}/\sigma^\SM_{gg\to H}$ --- and of the Higgs decay width into two photons --- parameterized by $\kgamma^2 \equiv \Gamma_{H\to\gamma\gamma}/\Gamma^\SM_{H\to\gamma\gamma}$---, are calculated in terms of the other model parameters. While we use analytical equations in our fit, which can be found e.g.\ in \ccite{Freitas:2012kw}, we provide here numerical formulas allowing to quickly assess the size of the different contributions,
\begin{align}
  \kg^2 ={}& 1.11 \ct^2 + 2.56 \ctt^2 - 0.12 \ct \cb - 0.20 \ctt \cbt + 0.01 \cb^2 + 0.01 \cbt^2 \nonumber\\
  & + 1.04\cg^2 + 2.34\cgtilde^2 + 2.15 \ct\cg + 4.90 \ctt \cgtilde  - 0.11\cb\cg - 0.19\cbt\cgtilde, \label{eq:kg}\\
  \kgamma^2 ={}& 0.08 \ct^2 + 0.18 \ctt^2 - 0.002 \ct \cb - 0.004 \ctt \cbt \nonumber\\
  & + 4\cdot 10^{-5}\cb^2 + 4\cdot 10^{-5}\cbt^2 - 0.002 \ct \ctau - 0.003 \ctt \ctaut \nonumber\\
  & + 6\cdot 10^{-5} \cb \ctau + 6\cdot 10^{-5} \cbt \ctaut + 2\cdot 10^{-5}\ctau^2 + 2\cdot 10^{-5} \ctaut^2 \nonumber\\
  & + 1.62 \cv^2 - 0.71 \cv \ct + 0.009 \cv \cb + 0.009 \cv \ctau \nonumber\\
  & + 0.64\cgamma^2 + 0.17\cgammatilde^2 - 0.45 \ct \cgamma + 0.35 \ctt\cgammatilde \nonumber\\
  & + 6\cdot 10^{-3}\cb\cgamma - 3\cdot 10^{-3}\cbt\cgammatilde + 6\cdot 10^{-3}\ctau\cgamma - 3\cdot 10^{-3}\ctaut\cgammatilde. \label{eq:kgamma}
\end{align}
Here we neglect the small dependencies on the first and second generation couplings. 
The modifiers of the \cp-even and \cp-odd Higgs couplings to the fermions are parameterized as $\ct, \ctt, \cb, \cbt, \ctau, \ctaut$ for top quark, bottom quark and tau lepton, respectively.
In scenarios where we do not assume $\cg = \cgtilde = 0$ ($\cgamma = \cgammatilde = 0$), we directly float for convenience \kg (\kgamma) instead of \cg, \cgtilde (\cgamma, \cgammatilde).

\begin{figure}
    \centering
    \includegraphics[width=.48\linewidth]{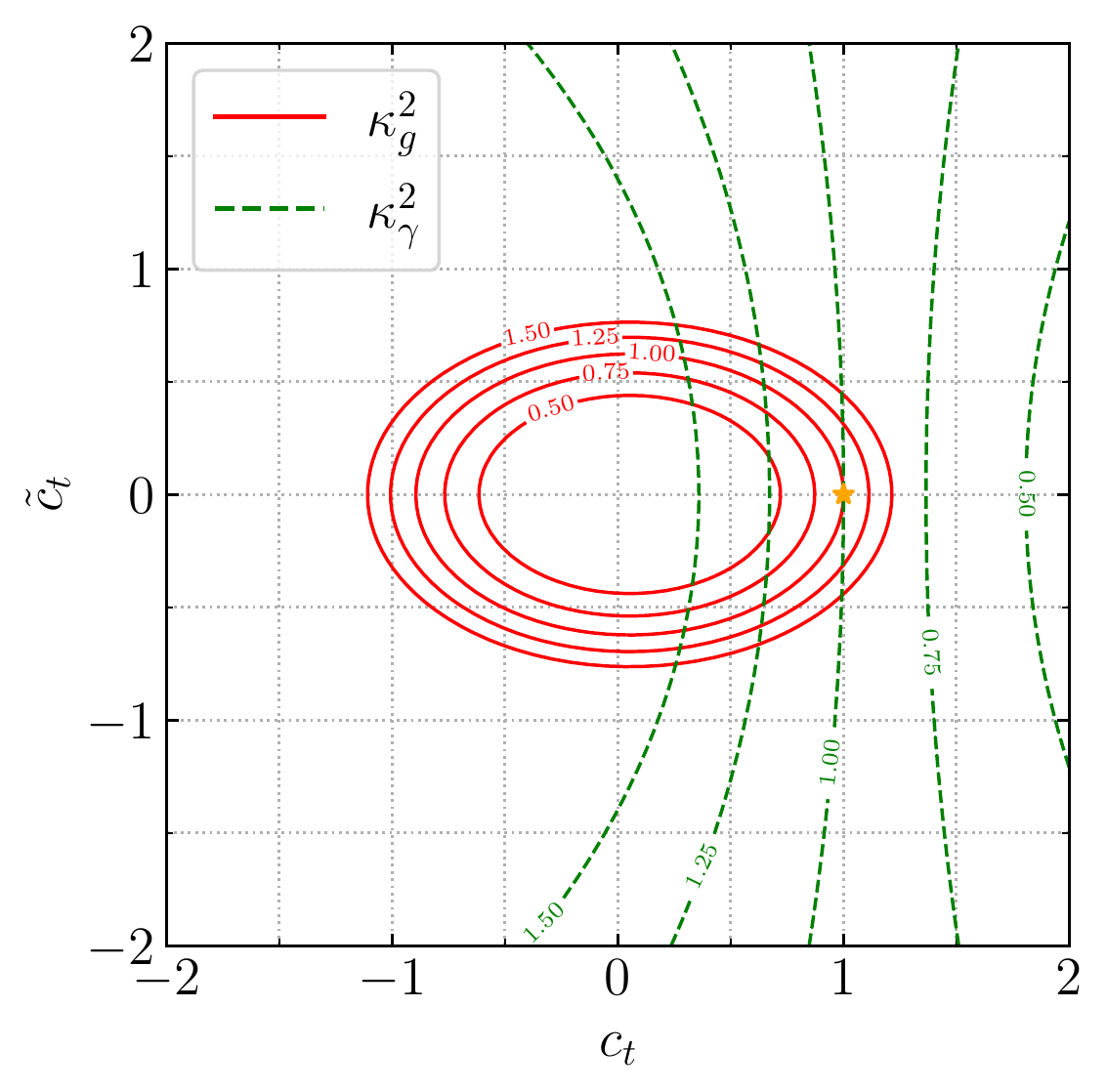}
    \includegraphics[width=.48\linewidth]{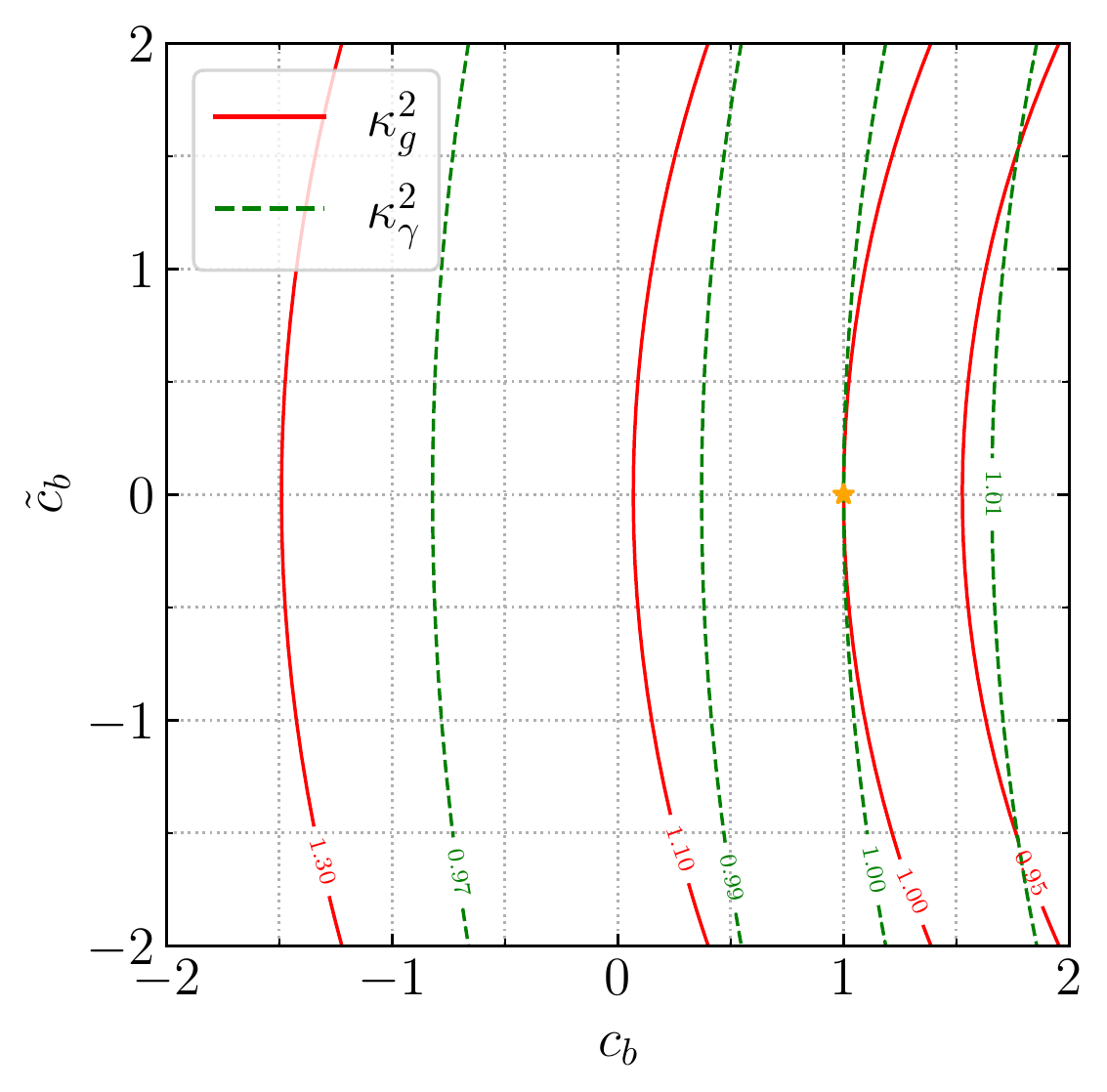}
    \caption{$\kg^2$ and $\kgamma^2$ as a function of (left) \ct and \ctt and (right) \cb and \cbt, with all other parameters fixed to their SM values. The orange star denotes the SM point.}
    \label{fig:ggH_Hgaga}
\end{figure}

We illustrate the numerical size of these dependencies in \cref{fig:ggH_Hgaga}. In the left panel, we vary \ct and \ctt keeping all other parameters fixed to their SM values. The gluon fusion cross section is equal to its SM value alongside an ellipsis stretching to the left of the SM point which can approximately be described by the equation $\ct^2 + 9/4 \, \ctt^2 = 1$. The gluon fusion cross section is subject to relatively large deviations from its SM prediction even in case of small deviations from the SM ellipsis. The dependence of \kgamma on \ct and \ctt is less pronounced. While \kg only has a weak sensitivity to the sign of \ct, \kgamma is very sensitive to it  as a consequence of the additional dependence on \cv.

In the right panel of \cref{fig:ggH_Hgaga}, the dependence of \kg and \kgamma on \cb and \cbt is displayed. In comparison to the left panel, the variations of \kg and \kgamma are much smaller for similar variations of the coupling modifiers. The gluon fusion cross section can be enhanced by about $\sim 26\%$ for a negative \cb within the $2\,\sigma$ allowed region $-1.23\le |\cb| \le -1.08$ (see \cref{sec:results}), while the di-photon decay width is reduced in this region by $\sim 3\%$. The dependence of the same quantities on \cbt is only at the sub-percent level. The dependence of the gluon fusion cross section on \ctau and \ctaut is zero to very good approximation. The dependence of the di-photon decay width on \ctau and \ctaut is even smaller than the dependence on \cb and \cbt, reaching maximally $\sim 1\%$ in the parameter space $0.78 \le |\ctau| \le 1.08$, $|\ctaut| \le 0.74$  still allowed experimentally at the $2\,\sigma$ level (see \cref{sec:results}).

\medskip
\subsection{Phenomenological models}
\label{sec:model_pheno}

In order to disentangle the impact of the various experimental constraints and to adjust the level of complexity in our analysis, we explore several simplified effective models with a restricted number of free parameters. We summarize them in \cref{tab:models}, starting with more specific but simpler models with fewer free fit parameters before moving to more general models with up to nine free parameters. Our models are intended as phenomenological characterizations of the Higgs coupling structure and can be mapped onto concrete model realizations, as discussed below. Besides the coupling modifications, exotic Higgs decays into BSM states can occur if they are kinematically allowed. We will leave an investigation of this possibility to future work, but note that preliminary results are available in \ccite{Menen:2021MScThesis}. We will use the following notation for common coupling modifiers: the common quark modifier $c_q$ implies that all quark Yukawa couplings are modified equally, $c_{q_i}\equiv c_q$ with $q_i=\{u, d, c, s, t, b\}$, likewise for \cqt; analogously for the global lepton modifiers with $c_{l_i}\equiv \cl$ of the charged leptons $l_i=\{e, \mu, \tau\}$ . Furthermore, we denote $f_3 \equiv \{t, b, \tau\}$, $f_2 \equiv \{c, s, \mu\}$, as well as $q_u=\{u, c, t\}$ and $q_d = \{d, s, b\}$.

\begin{table}[ht!]
\renewcommand{\arraystretch}{1.5}
    \centering
    \begin{tabular}{c|c|c|c|c}
    \hline
    Model &Free modifiers& $N$  &Motivation &Fig(s).\\ \hline\hline
          \textbf{1-flavor} & $c_{f_a}, \tilde c_{f_a}$ (1 fermion) &2 & flavor-specific, simplest &\ref{fig:cms_analysis}, \ref{fig:1flavour}, \ref{fig:edm_1flavour_a}, \ref{fig:edm_1flavour_b}\\ \hline
          \textbf{2-flavor} &$c_{f_a}, \tilde c_{f_a}, c_{f_b}, \tilde c_{f_b}$&4  &interplay of 2 flavors &\ref{fig:2flavour_1}, \ref{fig:2flavour_2}, \ref{fig:edm_2flavour_a}, \ref{fig:edm_2flavour_b}
          \\ \hline
          \textbf{quark}    &\cq, \cqt (all quarks)&2 &NP in quark sector &\ref{fig:shared_phases_q}\\ \hline
          \textbf{lepton}   &\cl, \clt (all leptons)&2  &NP in lepton sector &\ref{fig:shared_phases_l}\\ \hline
          \textbf{quark--lepton} &\cq, \cqt, \cl, \clt &4 &NP specific to $q$s, $l$s &\ref{fig:quark_lepton}\\ \hline
         \begin{tabular}{c}
              \textbf{up--down--} \\
               \textbf{lepton}
         \end{tabular} & 
            \begin{tabular}{c}
                 \cqu, \cqut, \cqd, \cqdt, \\
                 \cl, \clt
            \end{tabular} &6
           & \begin{tabular}{c}
               2HDM-like distinguishing\\
               up/down-type $q$s and $l$s
          \end{tabular} & \ref{fig:quark_lepton}             \\ \hline
          \textbf{\nth{2}/\nth{3} gen.} &\cfff, \cffft, \cff, \cfft & 4
             &\begin{tabular}{c}
               NP per generation\\
               recent sensitivity to \\
               \nth{2} gen. muon anomalies
          \end{tabular} &\ref{fig:generations}  \\ \hline
          \textbf{fermion} &\cf, \cft (all fermions)&2
             &\begin{tabular}{c}
                 NP universal\\
                 in fermion sector
             \end{tabular} & \ref{fig:shared_phases_f}, \ref{fig:edm_global}\\ \hline
          \textbf{universal}   &
          \begin{tabular}{c}
               $\cf \equiv \cv, \cft$\\
                $\tilde c_V \equiv 0$
          \end{tabular} &2
           &\begin{tabular}{c}
              mixing with a pseudoscalar\\
              (apart from $\cvt=0$)
           \end{tabular}&\ref{fig:shared_phases_f=V} \\ \hline
           \textbf{\fV}   &
           \begin{tabular}{c}
                $\cf, \cft, \cv$\\
                 $\tilde c_V \equiv 0$
           \end{tabular} &3
            &\begin{tabular}{c}
                NP in fermion sector \\
                + gauge sector
            \end{tabular}&\ref{fig:shared_phases_f_V} \\ \hline
           \begin{tabular}{c}
                \textbf{fermion-}    \\
                 \textbf{vector} 
           \end{tabular}
           &$\cf, \cft, \cv, \kg, \kgamma$  &5
            &\begin{tabular}{c}
                separate NP contr.\\
                to fermions, to vectors,\\
                and to loops
            \end{tabular} &\ref{fig:cv_constraints}              \\ \hline
            \begin{tabular}{c}
                \textbf{up-down-}\\
                \textbf{\lv}
            \end{tabular} &
          \begin{tabular}{c}
               $\cqu, \cqut, \cqd, \cqdt,$\\
               $\cl, \clt, \cv, \kg, \kgamma$
          \end{tabular}&9
            &\begin{tabular}{c}
               most general\\
               considered here
          \end{tabular} &\ref{fig:up_down_lepton_full} \\ \hline
    \end{tabular}
    \caption{Sets of coupling modifiers that are floated simultaneously. Coupling modifiers that are not listed in the second column are fixed to the SM values, $c_X=1$, $\tilde c_X=0$. The notation of the fermion coupling modifiers is described in the text. The vector coupling modifier is denoted as $\cv$ with $V=W, Z$, whereas $\kg$ and $\kgamma$ are given by \cref{eq:kg,eq:kgamma} unless they are listed as free parameters in the second column. The third column summarizes the number $N$ of free parameters. In the fourth column, we provide a brief theory motivation behind each phenomenological model (see text for more details), while  the last column references the relevant figures in which the considered model is analyzed.}
    \label{tab:models}
\end{table}

The \textbf{1-flavor} model is the simplest case where only the Yukawa coupling of one fermion is modified. This can be realized in models where the new physics (NP) is flavor-specific and only affects one \hife interaction, or the modifications of other couplings are suppressed. Several models mainly affect the top quark, such as composite Higgs models~\cite{Bruggisser:2018mus,Bruggisser:2018mrt}. Here, the Higgs-dilaton potential enables a sufficiently strong first-order electroweak phase transition and the \cp violation enters via the top quark Yukawa coupling, thus leading to successful baryogenesis. \ccite{Bruggisser:2018mrt} also considers \cp violation in the charm quark Yukawa coupling. 
A complex tau Yukawa coupling may play an important role for lepton-flavored electroweak baryogenesis (see \cref{sec:BAU_constraints}), and the recent CMS analysis~\cite{CMS:2021sdq} directly probes its \cp structure. Furthermore, in the Minimal Supersymmetric Standard Model large (possibly complex) corrections to the bottom quark Yukawa coupling, parameterized by a quantity called $\Delta_b$, can arise especially for large values of $\mu$ and $\tan\beta$~\cite{Hall:1993gn,Carena:1994bv,Carena:1999py,}.
Studying the muon Yukawa coupling is motivated by the observation of anomalies in its magnetic moment, $(g-2)_\mu$, and for bottom and strange quarks by the $B$-anomalies of $b\to s$ transitions, see e.g. \ccite{Alguero:2019ptt,Aebischer:2019mlg,Arbey:2019duh}. More generally, it can be motivated for all fermions to explore various scenarios where lepton universality and lepton flavor are violated, see e.g.~\ccite{Greljo:2021npi,Greljo:2021xmg,Crivellin:2021rbq,Darme:2021qzw}. Therefore, we consider the \nth{3}-generation fermions, top quark, bottom quark, tau, and the \nth{2}-generation fermions, muon and charm quark, as well-motivated candidates for modified, complex Yukawa couplings. We treat them separately in the 1-flavor, or jointly in the \nth{2}/\nth{3} generation model (see below). We do not consider the strange quark separately due to its currently weak collider bounds~\cite{ATLAS:2017gko,Duarte-Campderros:2018ouv,Erdmann:2020ovh} and its small contribution to baryogenesis~\cite{Shapira:2021mmy}.

The \textbf{2-flavor} model takes the interplay of complex Yukawa couplings of two specific fermions into account. As a theory realization, one can consider models where coupling modifications are enhanced for two fermions compared to the remaining fermions while the couplings of the latter are not required to be exactly those of the SM, but only to have negligible modifications. The combination of two coupling variations allows for the (partial) cancellation of the contributions of two different fermions to, e.g., an electric dipole moment. Combinations of two fermions from the third generation lead to an interesting interplay in the LHC constraints. Another relevant combination is to include the electron as one of the selected fermions. Large variations of the electron Yukawa coupling including an imaginary part can be realized e.g.~in the 2HDM~\cite{Dery:2017axi}.

The \textbf{quark} model assumes that all quark couplings are modified universally. Such a scenario can arise when new bosonic particles $X$ couple universally to any quark in a loop as a correction to the SM Yukawa coupling, e.g.\ in a triangle diagram to each $hq\bar q$ vertex containing one $X$ and two quarks or two $X$ and one quark. As the experimental constraints on the Yukawa couplings of the top and bottom quarks are the tightest, in this model the lighter quarks will be simultaneously constrained by the strong bounds on the third generation, more strongly than if they were independent.

Likewise, the \textbf{lepton} model assumes a universal modification of all lepton Yukawa couplings by New Physics that couples exclusively or dominantly to charged leptons while its effects on the quarks (and neutrinos) is negligible. One possibility is a $U(1)_L$ extension of gauged lepton number~\cite{Jeong:2015bbi}.

In contrast, in the \textbf{quark-lepton} model, we consider NP that affects quarks and leptons, but separately. Here, models with several new particles coupling either to quarks or to leptons, or with one new particle coupling differently to quarks and leptons are a possible realization. An example of the latter case is the $B-L$ extension of the SM by gauging the difference of baryon ($B$) and lepton ($L$) number. As all quarks have the same charge of $z_q = 1/3$ under the new gauge group $U(1)_{B-L}$ and all leptons have $z_l=-1$, the predicted $Z'$ boson couples more strongly to leptons than to quarks by a factor of $-3$. As the $Z'$ can modify the $l\bar l h$ and $q\bar q h$ vertices, this model motivates the separate fit of a quark and a lepton Yukawa coupling modifier.

In the \udl~ model, the NP couples to the fermions according to their category: up-type or down-type quarks, and charged leptons. This is inspired by the Two-Higgs Doublet Model (2HDM) where --- according to its type --- all or two of these three categories are summarized by shared Yukawa coupling modifiers. In the 2HDM, however, also the vector couplings are modified (albeit weaker than the couplings to fermions). Furthermore, two instead of three independent fermion coupling modifiers would give tighter constraints. Nevertheless, this model can be considered as a prototype for more specific choices.

The  \textbf{\nth{2}/\nth{3} generation} model distinguishes changes of the Yukawa couplings in the third generation from the second. This is motivated by scenarios where the NP couples differently to each generation, possibly also to the first generation, but there is not sufficient experimental sensitivity to the latter yet. Therefore we restrict the analysis to the heavier two generations. In view of the muon and $B$-meson anomalies this setup is not only motivated by theory, but also by recent data~\cite{Greljo:2021npi,Greljo:2021xmg,Crivellin:2021rbq,Darme:2021qzw}. For example, a $U(1)'$ extension of the SM with generation-dependent couplings~\cite{Bause:2021prv} can account for several $B$-anomalies. 

In the \textbf{fermion} model, all fermions are modified by the same coupling modifiers, see e.g.\ \ccite{Li:2012ku} about universal Yukawa modifiers.

In the \textbf{universal} model, the real parts of the Higgs couplings to all fermions and the vector bosons $W, Z$ are varied universally. In addition, a universal imaginary Yukawa coupling modifier is included as in the fermion model. This can be realized in the relaxion framework~\cite{Graham:2015cka} where the relaxion as a light pseudoscalar scans the Higgs mass and stops its evolution at a local minimum of its potential that breaks \cp. As a consequence, the relaxion mixes with the Higgs boson~\cite{Flacke:2016szy,Frugiuele:2018coc,Fuchs:2020cmm}, and the Higgs couplings to all fermions and the massive vector bosons are reduced by the universal mixing angle $\cf=\cv=\cos\theta$. 
Furthermore, Minimal Composite Higgs models (MCHM, see e.g.\ Refs.~\cite{Kaplan:1983sm,Agashe:2004rs,Contino:2006qr,Contino:2013kra,Goertz:2018dyw}) predict the Higgs couplings to vector bosons to be reduced by a factor of $\cv = \sqrt{1-\xi}$, where $\xi = v^2/f^2$ is the squared ratio of the electroweak vacuum expectation value $v$ and the scale $f$ of global symmetry breaking. The Yukawa coupling modifiers depend on the chosen symmetry breaking pattern; in the minimal composite Higgs model of $SO(5)/SO(4)$ with the top partner in the spinorial representation, denoted as MCHM$_4$, also the \hife couplings are reduced by $\cf= \sqrt{1-\xi}$~\cite{Agashe:2004rs}. Furthermore, in the Twin Higgs framework~\cite{Chacko:2005pe}, the coupling of the observed Higgs boson to SM particles is suppressed by a universal factor $c_f=c_V= \cos\left(\frac{v}{\sqrt{2}f}- \theta\right)$~\cite{Chacko:2017xpd}, where $\theta$ is the mixing angle between the observed and a heavier Higgs boson and $f$ is the energy scale that breaks the global $SU(4)\times U(1)$ symmetry. Hence, there are several examples of a universal modifier model.

The \textbf{fermion+V} model, in contrast, distinguishes between universal modifiers for the fermions $f$ and vector bosons $V$. In addition, we allow for BSM contributions to the loop-induced couplings of the Higgs bosons to gluons and photons.
In the Minimal Composite Higgs Model of $SO(5)/SO(4)$ with the top partner in the fundamental representation, denoted as MCHM$_5$, the Yukawa coupling modifiers are predicted to differ from $\cv$ (see above) and are given by $\cf = \frac{1-2\xi}{\sqrt{1-\xi}}$~\cite{Contino:2006qr,Contino:2013kra}. The fermion-vector model can also originate from the Two-Higgs Doublet Model (2HDM) of Type~I where all fermions couple to the second Higgs doublet $H_2$ such that the Yukawa couplings of the lighter neutral Higgs boson are modified by $\cf = \cos\alpha/\sin\beta$ with respect to the SM Higgs boson. The couplings to vector bosons are modified by $\cv=\sin(\alpha-\beta)$ where $\alpha$ is the mixing angle of the neutral Higgs states, and $\tan\beta=v_2/v_1$ is the ratio of the vacuum expectation values of the Higgs doublets.

The \textbf{fermion-vector} model extends the \fV model by additional modifiers for Higgs production via gluon fusion and the Higgs decay into two photons. These can be used to parameterize the effect of additional colored or electrically charged BSM states like the top partner in the Minimal Composite Higgs Model.

The \textbf{up-down-lepton-vector} model is the most general model considered in this work. It allows one to vary the three types of fermion couplings as well as the real parts of the tree-level coupling to the massive gauge bosons, and independently the loop-induced couplings to gluons and photons.


\section{Constraints}
\label{sec:constraints}

In this Section, we describe the different types of measurements used to constrain the models introduced in \cref{sec:model}.


\subsection{LHC measurements and constraints}
\label{sec:LHC_constraints}

\paragraph{Included data}
In order to derive LHC constraints on \cp-odd admixtures in the Higgs couplings we fit Higgs boson rates as measured by ATLAS and CMS with a focus on those involving fermions. We use the same set of observables as in~\ccite{Bahl:2020wee} which includes in particular the inclusive measurements of $t\overline{t}H$ in the multi-lepton~\cite{ATLAS:2019nvo,Aaboud:2018jqu,Aad:2020mkp,Aad:2019lpq,CMS:2019kqw,CMS:2019chr}, di-photon~\cite{Sirunyan:2020sum,Aad:2020ivc,ATLAS:2019jst,CMS:1900lgv} and $b\overline{b}$ decay channels~\cite{Aaboud:2018gay,Aaboud:2017rss,Sirunyan:2017elk,Sirunyan:2017dgc,CMS:2019lcn,Sirunyan:2018mvw} and the  measurements for $ZH$ production in bins of Higgs transverse momentum ($p_T$)~\cite{Aaboud:2019nan,Aad:2020jym,ATLAS:2019jst}, which are sensitive to the \cp nature of the top quark Yukawa interaction. 

We furthermore include the latest experimental results on $H\to \mu^+\mu^-$ \cite{ATLAS:2020fzp, CMS:2020xwi} and $H\to c \bar c$~\cite{ATLAS:2021zwx,CMS:2019hve,ATLAS:2022ers} measurements from ATLAS and CMS. The charm quark Yukawa coupling has not been observed yet at the LHC, but the searches for $H\to c\bar c$ decays~\cite{CMS:2019hve,ATLAS:2021zwx,ATLAS:2022ers} in the $VH(c\bar c)$ channel with charm tagging constrain $|g_c|<8.5$ at the 95\% CL. This direct limit includes the effect of a modified charm quark Yukawa coupling on the partial width $\Gamma(H\to c\bar c)$ as well as on the total Higgs width $\Gamma_h^{\rm{tot}}$, see Eq.~(\ref{eq:BRf}). Other measurements involving the charm quark Yukawa coupling are not as sensitive to $|g_c|$ as the indirect bound via $\BR(H\to\gamma\gamma)$\footnote{The upper bound on the exclusive decay of $\BR(H\to J/\psi \gamma)<3.5\cdot 10^{-4}$~\cite{ATLAS:2018xfc,CMS:2018gcm} yields a limit of $|g_c|<\mathcal{O}(100)$~\cite{Bodwin:2013gca,Perez:2015aoa,Coyle:2019hvs}, i.e.\ an order of magnitude weaker than the limit from the inclusive Higgs decay into charm quarks. A further possibility to determine the charm quark Yukawa coupling directly is the charm-induced Higgs + jet production via the cross section measurement and the shape of the Higgs-$p_T$ distributions~\cite{Bishara:2016jga}. The very recent ATLAS analysis~\cite{ATLAS:2022vpz} reports $-10.1<g_c<18.3$. Although it has the advantage of relying on different assumptions and uncertainties, it is not included in our fit because it is not competitive with the indirect bound from the precise measurement of $\BR(H\to\gamma\gamma)$ which constrains $|g_c|$ more strongly than the inclusive search due to the sensitivity to the total width modification, see Eq.~(\ref{eq:BRgamma}).}. Finally, we add the CMS analysis of the \cp structure of the tau Yukawa coupling~\cite{CMS:2021sdq}. The \cp analysis in the $\tau^+\tau^-$ final state is the only measurement targeting a \hife coupling based on a dedicated \cp-odd observable, where the \cp-violating phase $\alpha_\tau$ is directly inferred by measuring the angle between the decay planes of the two tau leptons. It is included in the list of measurements by adding the $\chi^2$ contribution corresponding to the data of~\ccite{hepdata.104978.v1/t4}. We refer to these measurements as the ``LHC data set'' in the following.

Several \cp sensitive measurements could not be included such as the $p_T$-binned STXS measurement of $t\overline{t}H$~\cite{ATLAS:2020pvn} as it relies on a separation of $tH$ from $t\bar tH$ production based on the assumption of SM-like kinematics. As shown in \ccite{Bahl:2020wee}, this assumption does not hold in the presence of a \cp-odd Yukawa coupling. We also did not include analyses using full Run 2 data fitting the rates of the $t\bar t H$, $tH$ and $tWH$ processes in the Higgs to di-photon decay channel~\cite{ATLAS:2020ior,CMS:2020cga}, as the data is not available in a sufficiently model-independent format, see the discussion in \ccite{Bahl:2020wee}.

In order to include the LHC constraints, we use \texttt{HiggsSignals} (version 2.5.0)~\cite{Bechtle:2013xfa,Bechtle:2013wla,Bechtle:2014ewa,Bechtle:2020uwn} which incorporates the available inclusive and differential Higgs boson rate measurements from ATLAS and CMS from Run 2~\cite{Aaboud:2018gay,Aaboud:2017rss,ATLAS:2019nvo,Aaboud:2018jqu,Aad:2020mkp,ATLAS:2019jst,Aaboud:2018pen,Aad:2019lpq,Aad:2020jym,Sirunyan:2020sum,Sirunyan:2018hbu,Sirunyan:2017elk,Sirunyan:2017dgc,CMS:2019lcn,Sirunyan:2018mvw,Sirunyan:2018shy,CMS:2018dmv,CMS:2019kqw,CMS:2019chr,CMS:1900lgv,CMS:2019pyn}, as well as the combined measurements from Run 1~\cite{Khachatryan:2016vau}.
On top of the channels currently implemented in \texttt{HiggsSignals} we included the most recent results on $H \to c \bar c, \mu^+\mu^-, \tau^+\tau^-$~\cite{CMS:2019hve,ATLAS:2021zwx,ATLAS:2022ers,ATLAS:2020fzp,CMS:2020xwi,CMS:2021sdq}\footnote{These additional measurements will be part of the upcoming \texttt{HiggsSignals~3} release, which will be part of the new \texttt{HiggsTools} framework.} as described above.

The experimental measurements are implemented with correlation matrices and detailed information about the composition of the signal in terms of the various relevant Higgs boson production and decay processes. If this information is not available (e.g.\ in the recent CMS $H\to\tau^+\tau^-$ \cp study), the signal is assumed to be composed of the relevant Higgs processes with equal acceptances, and only correlations of the luminosity uncertainty (within one experiment) and theoretical rate uncertainties are taken into account. Further details on the cross section calculations and the fit implementation can be found in~\cite{Bahl:2020wee}.

Based on this large collection of measurements and the predicted cross sections for different parameter choices, \texttt{HiggsSignals} is used to determine favored regions in parameter space by calculating the $\chi^2$ difference, $\Delta \chi^2 \equiv \chi^2 - \chi^2_\text{min}$, with $\chi^2_\text{min}$ being the minimal $\chi^2$ value found at the best-fit point. The parameter spaces are sampled using a combination of two different types of Markov Chain Monte Carlo samplers. Alongside a common Metropolis Hastings approach~\cite{metropolis, 10.2307/2334940}, a realization of the Stretch Move Algorithm in \texttt{Emcee}~\cite{10.2140/camcos.2010.5.65, Foreman_Mackey_2013} was used. We found this method to provide the best convergence behaviour. Furthermore, it ensures that parameter spaces in which regions of high likelihood are either narrow or separated by large potential barriers are sampled efficiently~\cite{Menen:2021MScThesis}.

While for gluon fusion and the Higgs decay into two photons fit formulas are available at leading-order (LO) in analytic form, this is not the case for other Higgs boson production processes. For calculating cross-sections in the ``Higgs characterization model''~\cite{Artoisenet:2013puc,Maltoni:2013sma,Demartin:2014fia}, we use \texttt{MadGraph5\_aMC@NLO~2.7.0}~\cite{Alwall:2014hca} with \texttt{Pythia~8.244}~\cite{Sjostrand:2007gs} as parton shower employing the A14 set of tuned parameters~\cite{ATL-PHYS-PUB-2014-021}. The cross-sections are computed at LO in the five-flavor scheme and rescaled to the state-of-the-art SM predictions reported in Ref.~\cite{deFlorian:2016spz}.

\paragraph{Approximate dependences of branching ratios on coupling modifiers}

For an arbitrary combination of Yukawa coupling modifications, but $\cv = 1$, the branching ratios into a pair of fermions $f\bar f$ are given by\footnote{If the fermion mass is negligible in comparison to the Higgs boson mass, the corresponding Higgs boson decay width is rescaled by $g_f^2=\cf^2+\cft^2$ in comparison to the SM.}
\begin{align}
     \BR_f &= \frac{g_f^2\Gamma_f^\SM}{\sum_{f'}g_{f'}^2\Gamma_{f'}^\SM + \Gamma^\SM(H\to VV) + \Gamma(H\to \NP)}\,,
\end{align}
where $g_f$ is defined in \cref{eq:phase}, and $\Gamma_f^\SM = \Gamma^{\SM}(h\to f\bar f)$ is the partial width in the SM, and the sum over $f'$ includes all fermions.
Therefore, in the case of the modification of only one Yukawa coupling of fermion $f$ and no decay into new particles, the branching ratio into  $f\bar f$ can be expressed as
\begin{align}
 \BR_f &= \frac{g_f^2\Gamma_f^\SM}{g_f^2\Gamma_f^\SM + \Gamma_{\rm{tot}}^\SM - \Gamma_f^\SM}
 = \frac{g_f^2 \BR_f^\SM}{(g_f^2-1) \BR_f^\SM +1}\,,
 \label{eq:BRf}
\end{align}
where $\BR_f^\SM$ denotes the branching ratio in the SM and $\Gamma_{\rm{tot}}^\SM$ denotes the total Higgs width in the SM.
Hence, a particular measured value of the
branching ratio,  $\BR_f^{\rm{exp}}$,
yields a circle in the $\cf,\,\cft$ plane of squared radius
\begin{align}
 g_f^2 \equiv \cf^2 + \cft^2 = \frac{\BR_f^{\rm{exp}}}{\BR_f^\SM}\,\cdot\, \frac{\BR_f^\SM-1}{\BR_f^{\rm{exp}}-1}\,. \label{eq:circle}
\end{align}
This results in rings corresponding to the upper and lower bound on $\BR_f^{\rm{exp}}$ if the constraints are dominated by the decay rate information\footnote{In the case of $\cv \neq 1$, \cref{eq:circle} generalizes to 
\begin{equation}
g_f^2 \equiv \cf^2 + \cft^2 = \frac{\BR_f^{\rm{exp}}}{\BR_f^\SM}\,\cdot\, \frac{\BR_f^\SM-1 + \BR_V^\SM(1-c_V^2)}{\BR_f^{\rm{exp}}-1}\,.
\nonumber
\end{equation}
}, see \cref{fig:cms_analysis,fig:1flavour}.
 
Likewise, we consider the modification of one Yukawa coupling and its impact on the $H\to\gamma\gamma$ decay via the modification of the total Higgs width, i.e.\,the free coupling modifiers $\cf$, $\cft$ and $\kgamma$,
\begin{align}
     \BR_\gamma = \frac{\kgamma^2 \Gamma_\gamma^\SM}{\Gamma_f^\SM (g_f^2-1) + \Gamma_\gamma^\SM (\kgamma^2-1) + \Gamma_{\rm{tot}}^\SM 
     }\, .\label{eq:BRgamma}
\end{align}
Here. $\Gamma_\gamma^\SM\equiv \Gamma^\SM(H\to\gamma\gamma)$ is the SM value of the partial width into a pair of photons. This implies that $g_f^2$ can be expressed as
\begin{align}
     g_f^2 \equiv \cf^2 + \cft^2 = 1 + \frac{1}{\BR_f^\SM} \cdot \left[
     \BR_\gamma^\SM\cdot\,\left(\frac{\kgamma^2}{\BR_\gamma^{\rm{exp}}} - (\kgamma^2-1) \right)-1 \right]\, ,\label{eq:gfBRgamma}
\end{align}
where $\BR_\gamma^{\rm{exp}}\equiv\BR^{\rm{exp}}(H\to\gamma\gamma)$ denotes the measured branching ratio of the Higgs boson into photons, and $\BR_\gamma^\SM$ is its prediction in the SM.
As discussed in Sec.~\ref{sec:model_couplings}, $\kgamma$ can be calculated as a result of $\cf$, $\cft$, see \cref{eq:kgamma}, it can be floated independently, or it can be set to its SM value of 1. 
Unless \kgamma is treated as depending on $\cf$, $\cft$, Eq.~(\ref{eq:gfBRgamma}) shows that also the constraint from measuring $\BR_\gamma=\BR_\gamma^{\rm{exp}}$ leads to a circular ring in the $\cf$, $\cft$ plane. 
If one further assumes $\kgamma=1$ (in case of a negligible contribution of the considered fermion $f$ to \kgamma), Eq.~(\ref{eq:gfBRgamma}) can be simplified to
\begin{align}
     g_f^2\Big|_{\kgamma=1} = 1 + \frac{1}{\BR_f^\SM} \cdot \left[ \frac{\BR_\gamma^\SM}{\BR_\gamma^{\rm{exp}}} -1\right]  \,.\label{eq:gfBRgammaSimple}
\end{align}


\subsection{EDM constraint}
\label{sec:EDM_constraints}

Several EDMs are sensitive to the \cp nature of the Higgs boson. The most sensitive ones are the electron EDM (eEDM) and the neutron EDM (nEDM).\footnote{In \ccite{Brod:2018pli} it has been shown in particular that the constraints on the bottom and charm quark coupling modifiers $\cb$, $\cbt$, $\cc$, $\cct$ from the electron EDM are always significantly stronger than those from the neutron and mercury EDMs (if the electron-Yukawa coupling is SM-like).} Besides the experimental results~\cite{Andreev:2018ayy,Abel:2020gbr}, much work has been done to provide precise theory predictions~\cite{Pospelov:2005pr,Brod:2013cka,Altmannshofer:2015qra,Cirigliano:2016nyn,Brod:2018pli,Panico:2018hal,Altmannshofer:2020shb,Brod:2022inPreparation}.

The main focus of the present work are the LHC constraints. Therefore, we take into account only constraints from the eEDM, which is theoretically the cleanest EDM. Since the various EDMs are independent measurements, taking into account additional EDM measurements (e.g.\ the nEDM) could potentially tighten the constraints on the Higgs \cp nature. Correspondingly, our EDM constraint, based only on the eEDM, can be regarded as conservative.

The dominant contribution from \cp-violating \hife couplings to the eEDM appear at the two-loop level in the form of Barr-Zee diagrams. For their evaluation, we make use of the analytical results given in \ccite{Brod:2013cka,Altmannshofer:2015qra,Panico:2018hal,Altmannshofer:2020shb}.\footnote{In comparison to \ccite{Panico:2018hal} (and also \ccite{Fuchs:2020pun} which applied \ccite{Panico:2018hal}), we corrected a factor of $1/\sqrt{2}$. It should also be noted that in comparison to \ccite{Shapira:2021mmy}, a relative sign between the contributions proportional to \ce and \cet has been corrected.} While we use the full analytical expressions for our numerical analysis, these expressions can also be translated into a simple numerical formula allowing to easily assess the relative importance of the various \cp-violating Higgs couplings,
\begin{align}
\frac{d_e}{d_e^\text{ACME}} ={}& \ce\left(870.0\ctt + 3.9\cbt + 2.8\cct + 0.01\cst + 8\cdot 10^{-5} \cut+ 7\cdot 10^{-5} \cdt + 3.4 \ctaut + 0.03\cmut \right) \nonumber\\
& + \cet\left(610.1\ct + 3.1\cb + 2.3\cc + 0.01\cs + 7\cdot 10^{-5} \cu + 6\cdot 10^{-5} \cd + 2.8 \ctau + 0.02\cmu \right.\nonumber\\
& \left.\hspace{1cm} - 1082.6\cv\right) \nonumber\\
& + 2\cdot 10^{-6}\ce\cet. \label{eq:EDM_fit_formula}
\end{align}
Here, $d_e^\text{ACME} = 1.1\cdot 10^{-29} e\,\rm{cm}$ is the 90\% CL upper bound on the eEDM obtained by the ACME collaboration~\cite{Andreev:2018ayy}. Correspondingly, a parameter point is regarded as excluded at the 90\% CL if $|d_e/d_e^\text{ACME}| > 1$.

In general, the contributions of two or more particles to $d_e$ can cancel each other, partially or fully. Especially, \cref{eq:EDM_fit_formula} shows that a \cp-violating top-Yukawa coupling can induce a large contribution to the eEDM, depending on the value of $|c_e|$. For a non-zero value of $\cet$, large additional contributions proportional to $\ct$ and $\cv$ can occur. As we will investigate below (see also \ccite{Fuyuto:2019svr}), these two types of potentially large contributions can cancel each other. Currently, the most constraining experimental bound on the electron-Yukawa coupling is from the ATLAS Run 2 measurement \cite{ATLAS:2019old} and yields $g_e \leq 268$ at 95\% CL.
Since the current bound is too loose for meaningful fit results, we will restrict to the SM values $\ce = 1$, $\cet = 0$ in the analyses presented in this work unless otherwise stated. On the other hand, there exists no experimental lower bound on the electron-Yukawa coupling, and also in the future it will remain very difficult to establish evidence of a non-zero electron Yukawa coupling. Thus, in case $\ce$ and $\cet$ are very small, i.e.\ in particular if $\ce$ is much below the SM value, the BSM contributions to $d_e$ would be heavily suppressed and therefore the impact of the limit on the eEDM would be drastically reduced.\footnote{It should be noted that in our effective model approach the limit $\ce = \cet = 0$ would not imply that the electron mass is zero. While this is true e.g.\ in the SM effective field theory of dimension six, it is not true if additionally dimension-eight operators are taken into account.}


\subsection{BAU constraint}
\label{sec:BAU_constraints}

The baryon asymmetry in the universe, $Y_B$, was measured by PLANCK to be~\cite{Zyla:2020zbs}
\begin{equation}
    \YBobs = (8.59 \pm 0.08)\times 10^{-11}\,. \label{eq:YBobs}
\end{equation}
The description of the BAU requires additional sources for \cp violation beyond the CKM phase that is present in the SM. An attractive framework for explaining the BAU is electroweak baryogenesis (EWBG), for reviews see e.g.\ Refs.~\cite{Cline:2006ts,Morrissey:2012db,White:2016nbo,Bodeker:2020ghk}. For EWBG a non-vanishing baryon number density is achieved during the electroweak phase transition, implying that the mechanism can potentially be tested with Higgs processes at the LHC. In the phase transition, bubbles of the broken phase with $v\neq 0$ form, whereas the electroweak symmetry is unbroken outside of the bubbles; the bubbles expand until the universe is filled by the broken phase. Across the bubble wall, \cp-violating interactions create a chiral asymmetry that is partially washed out by the \cp-even interactions and the sphalerons. A part of the generated chiral asymmetry diffuses through the bubble wall into the symmetric phase where it is converted into a baryon asymmetry by the weak sphaleron process. Then the expanding bubble wall reaches the region where the baryon asymmetry was created, which is then maintained in the broken phase.

While the experimental precision of the PLANCK measurement in Eq.~(\ref{eq:YBobs}) is around 1\%, the theoretical uncertainties of predicting the BAU in different models of electroweak baryogenesis are up to now much larger. The largest uncertainty can be associated with the deviations between different approaches that are employed for calculating the source term for the baryon asymmetry, namely the perturbative so-called vev-insertion-approximation (VIA)~\cite{Huet:1994jb,Huet:1995mm,Huet:1995sh,Carena:1996wj,Riotto:1997vy,Lee:2004we}, and the semi-classical Wentzel-Kramer-Brillouin (WKB) approach~\cite{Joyce:1994zt,Kainulainen:2001cn,Kainulainen:2002th,Prokopec:2003pj,Prokopec:2004ic,Konstandin:2013caa,Konstandin:2014zta}. 
While both formalisms yield similar outcomes of the baryon asymmetry for an equivalent source term, they largely differ in the calculation of the source term~\cite{Cline:2020jre,Cline:2021dkf}. For a systematic comparison of both approaches see in particular \ccite{Cline:2021dkf}.
The perturbative approach starts with Green's functions in a Closed Time Path formalism. The interaction rates and the \cp-violating source term are computed from the self-energies, and the vev-dependent contributions to the particle masses are included as a perturbation. In contrast, the WKB approach starts with the Boltzmann equations, and the interactions are described as semi-classical forces in the plasma. 
There is a long-standing controversy in the literature about which approach to apply. Recent studies have shown that the VIA leads to systematically higher predictions of the amount of the baryon asymmetry due to an additional derivative in the WKB source term~\cite{Cline:2020jre,Cline:2021dkf,Kainulainen:2021oqs}.
This has been evaluated for a source term from a complex tau Yukawa coupling with an additional singlet scalar term of dimension 5 and a maximal relative \cp-violating phase between the terms of dimension 4 and 5. The evaluation furthermore assumed the benchmark value of the wall velocity of $v_w=0.05$ (as used in our work), and a wall thickness of $L_w=5/T$ with $T=88\gev$ (i.e.\ about half of the value adopted in our work, $L_w=0.11\,\mathrm{GeV}^{-1}$). Using these parameters \ccite{Cline:2021dkf} reports a discrepancy of about five orders of magnitude between the VIA and WKB approach. For a charm quark source, the discrepancy is found to be around one order of magnitude. The discrepancy for a top quark source was found to reach a factor of 10 -- 50 depending on $v_w$ and $L_w$~\cite{Cline:2020jre}. Furthermore, the applicability of the VIA depends on the width of the bubble wall~\cite{Postma:2021zux}. 
Another recent study~\cite{Kainulainen:2021oqs} stresses the impact of thermal corrections at the one-loop level, raising concerns about the validity of the VIA because of an apparently vanishing source term, but also pointing to potential open issues in the WKB approach. 

Besides the large differences between the VIA and WKB approaches, which we treat as a theoretical uncertainty, there are also smaller theoretical uncertainties that are inherent to the individual approaches. We comment here on the case of the VIA approach.
Theoretical uncertainties of this kind arise within the VIA approach from the perturbative expansion and from the not precisely known bubble parameters. Especially the NLO terms in the VIA can be large as was shown in Ref.~\cite{Postma:2019scv}, around $\mathcal{O}(1)$ for the top quark, but at the sub-per-mille level for the bottom quark and the tau lepton because of their smaller Yukawa couplings. In addition, the prediction of $Y_B$ depends on the bubble wall properties, in particular on the bubble wall width, $L_w$, and velocity, $v_w$,  as well as on the bubble wall profile (i.e.\ the variation of the vev from the inner to the outer bubble wall). In contrast to the long-standing expectation that small velocities should lead to higher $Y_B$, EWBG can also be successful with supersonic bubble walls, as shown recently in Refs.~\cite{Dorsch:2021ubz,Cline:2021dkf}.

In view of the described uncertainties affecting the prediction of the BAU, we adopt the following strategy for assessing the impact of the BAU constraint. As the predictions based on the VIA approach tend to yield significantly higher values for $Y_B$ than the WKB approach, we employ the VIA approach for obtaining an ``optimistic'' reference value for the BAU. Specifically, we apply the bubble wall parameters $v_w$ and $L_w$ as in the benchmark used in \ccite{deVries:2018tgs,Fuchs:2020uoc,Fuchs:2020pun,Shapira:2021mmy} such that they yield values of $\YBVIA$ for the given couplings that are near the maximally possible values. Accordingly, the obtained value for $\YBVIA$ corresponds to an approximate upper bound on $Y_B$ .For this reason we do not attempt to show confidence levels, but restrict ourselves to displaying the nominal value of the BAU. We regard a parameter point as disfavored by the observed BAU if the value predicted for $Y_B$ in the (optimistic) VIA approach is such that $\YBratio < 1$. On the other hand, values with $\YBratio > 1$ may well be phenomenologically viable if the VIA approach turns out to overestimate the predicted value of $Y_B$. We therefore indicate the parameter regions fulfilling $\YBratio \geq 1$ as those that are favored by the observed BAU. For illustration, contour lines for fixed values of \YBratio are shown in our plots.

The prediction for \YBratio consists of the contributions from the different fermions that are proportional to the respective parameters $\cft$. We use the evaluations from Refs.~\cite{Fuchs:2020uoc,Shapira:2021mmy} of all fermions as given by the simple formula
\begin{align}
\YBratio ={}& 28 \ctt- 0.2 \cbt - 0.03 \cct - 2\cdot 10^{-4} \cst - 9\cdot 10^{-8} \cut - 4\cdot 10^{-7} \cdt \nonumber\\
& - 11 \ctaut - 0.1 \cmut - 3\cdot 10^{-6} \cet, \label{eq:BAU_fit_formula}
\end{align}
where the coefficients have been evaluated by employing the benchmark parameters as defined in \ccite{Fuchs:2020uoc,Shapira:2021mmy}.


\section{Results}
\label{sec:results}

In this Section, we present the results of our numerical fits for specific realizations of the scenarios defined in \cref{sec:model}. First, we focus on the constraints set by LHC measurements (supplementary results are provided in \cref{sec:appendix_additional_fits}). In a second step, we investigate the interplay with the eEDM constraint and the obtainable BAU in the VIA.

\subsection{LHC results}

\label{sec:results_LHC}

In the following, all presented results are based on the LHC data set, defined in \cref{sec:LHC_constraints}, except for \cref{fig:cms_analysis_a}, where the CMS $H\to\tau^+\tau^-$ \cp  measurement is excluded. Accordingly, the $\chi^2$ value of the SM point in the plots below is always $\chi^2_{\mathrm{SM}} = 89.36$ (except for \cref{fig:cms_analysis_a}).

\subsubsection{1-flavor models}
\label{subsubsec:1-flavor_LHC}

\begin{figure}[!htp]
    \centering
    \subfigure[]{\includegraphics[width=.48\linewidth]{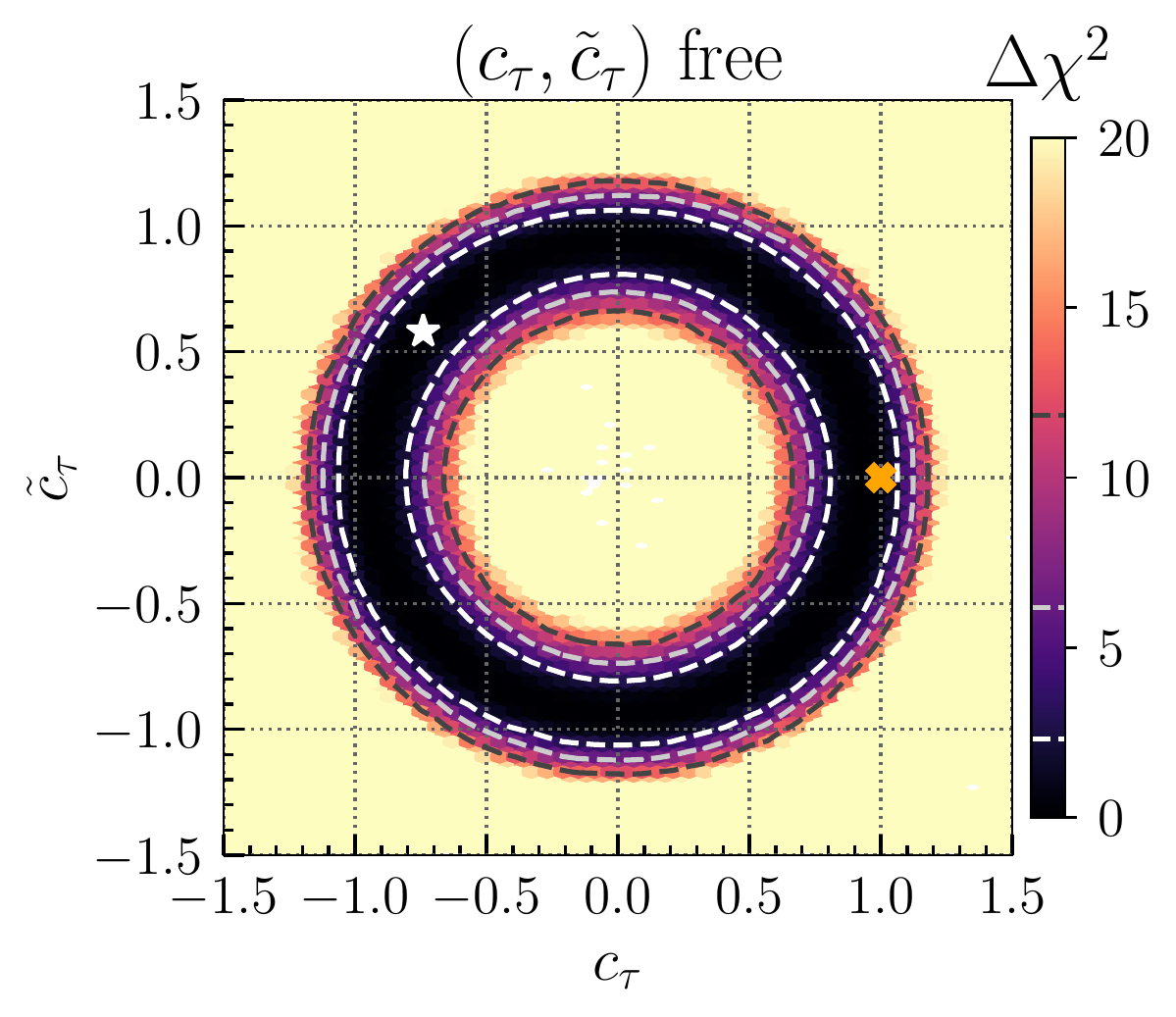}\label{fig:cms_analysis_a}}
    \subfigure[]{\includegraphics[width=.48\linewidth]{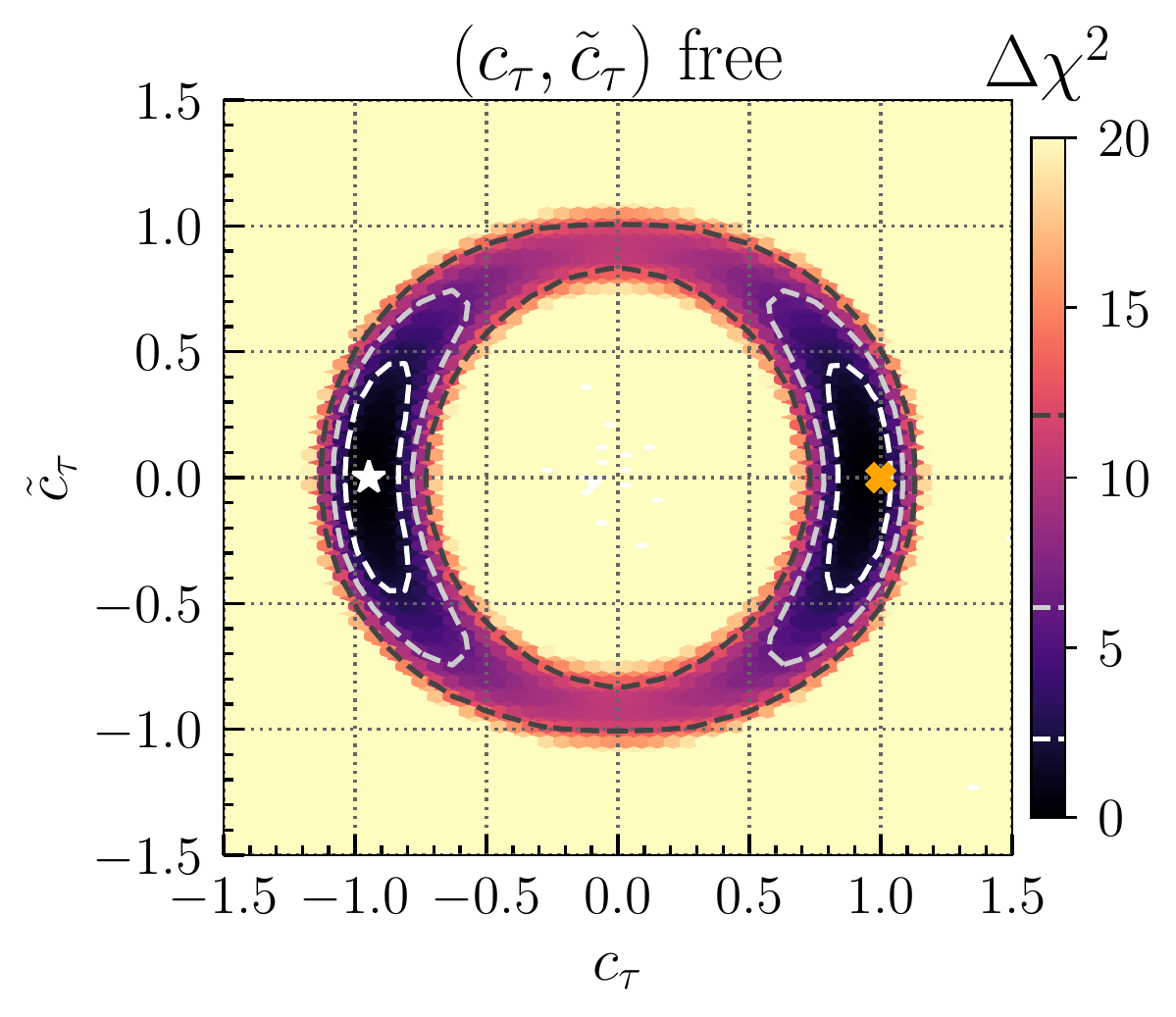}\label{fig:cms_analysis_b}}
    \caption{Results of fits to the LHC measurements in the (\ctau, \ctaut) parameter plane where in the set of input measurements
    the CMS $H\to\tau^+\tau^-$ \cp result~\cite{CMS:2021sdq}
    is (a) omitted and (b) included. The coupling modifiers \ctau\ and \ctaut\ are treated as free parameters while all other parameters are fixed to their SM values. The color corresponds to the profile $\Delta\chi^2$ of the global fit, and the $1\sigma$, $2\sigma$ and $3\sigma$ confidence regions are shown as white, light-gray and dark-gray dashed contours, respectively. The best-fit point and the SM case are marked by a white star and an orange cross, respectively.}
    \label{fig:cms_analysis}
\end{figure}

\paragraph{\boldmath{$\tau$} Yukawa coupling}
We first investigate the two-dimensional plane of the \cp-even and \cp-odd tau Yukawa coupling modifiers, \ctau\ and \ctaut, respectively, treating only these two parameters as free-floating in the fit.
The tau Yukawa coupling is constrained by measurements of $H\to\tau^+\tau^-$ decays, and by measurements of $H\to\gamma\gamma$ decay rates in which tau leptons enter at the loop-level. In practice, the former dominates the current constraint due to the predominance of the $W$ boson, top quark and bottom quark loop diagrams in $H\to\gamma\gamma$ decays. In the absence of \cp-sensitivity, one expects $H\to\tau^+\tau^-$ decay rate measurements to show a dependence on $\ctau^2 + \ctaut^2$ forming a ring-shaped constraint (see~\cref{eq:gfBRgamma}). This pattern is observed in \cref{fig:cms_analysis_a}, where only the inclusive $H\to\tau^+\tau^-$ decay rate but not the recent
CMS $H \to \tau \tau$ \cp measurement~\cite{CMS:2021sdq} is included as input to the fit. The current precision of the $H\to\gamma\gamma$ measurement has no visible effect on the ring structure. Furthermore, there is no statistically significant sensitivity to the precise location of the best-fit point (shown as a white star in the plots) within the ring.
We find $\chimin = 87.48$ for the best-fit point, while the SM point ($\ctau=1,~\ctaut=0$, shown as an orange cross in the plots) has $\chi^2_{\mathrm{SM}} = 88.33$.
\cref{fig:cms_analysis_b} shows the result based on the full set 
of input measurements, i.e.\ including the CMS $H\to\tau^+\tau^-$ \cp measurement~\cite{CMS:2021sdq}. This experimental result
excludes large $|\ctaut|$ values (i.e.\ $|\ctaut| < 0.75$ at the 95\% CL) in the fit, as expected from the unique sensitivity brought by this analysis. The best-fit point has $\chimin = 87.63$. The fact that the best-fit point is located at a negative rather than a positive $\ctau$ value is again not statistically significant and only corresponds to a small difference of $\Delta\chi^2=0.23$ with respect to the best-fit point at $\ctau>0$. The SM point is located within the 1$\sigma$ area.

\begin{figure}
    \centering
    \subfigure[]{\includegraphics[width=.487\linewidth]{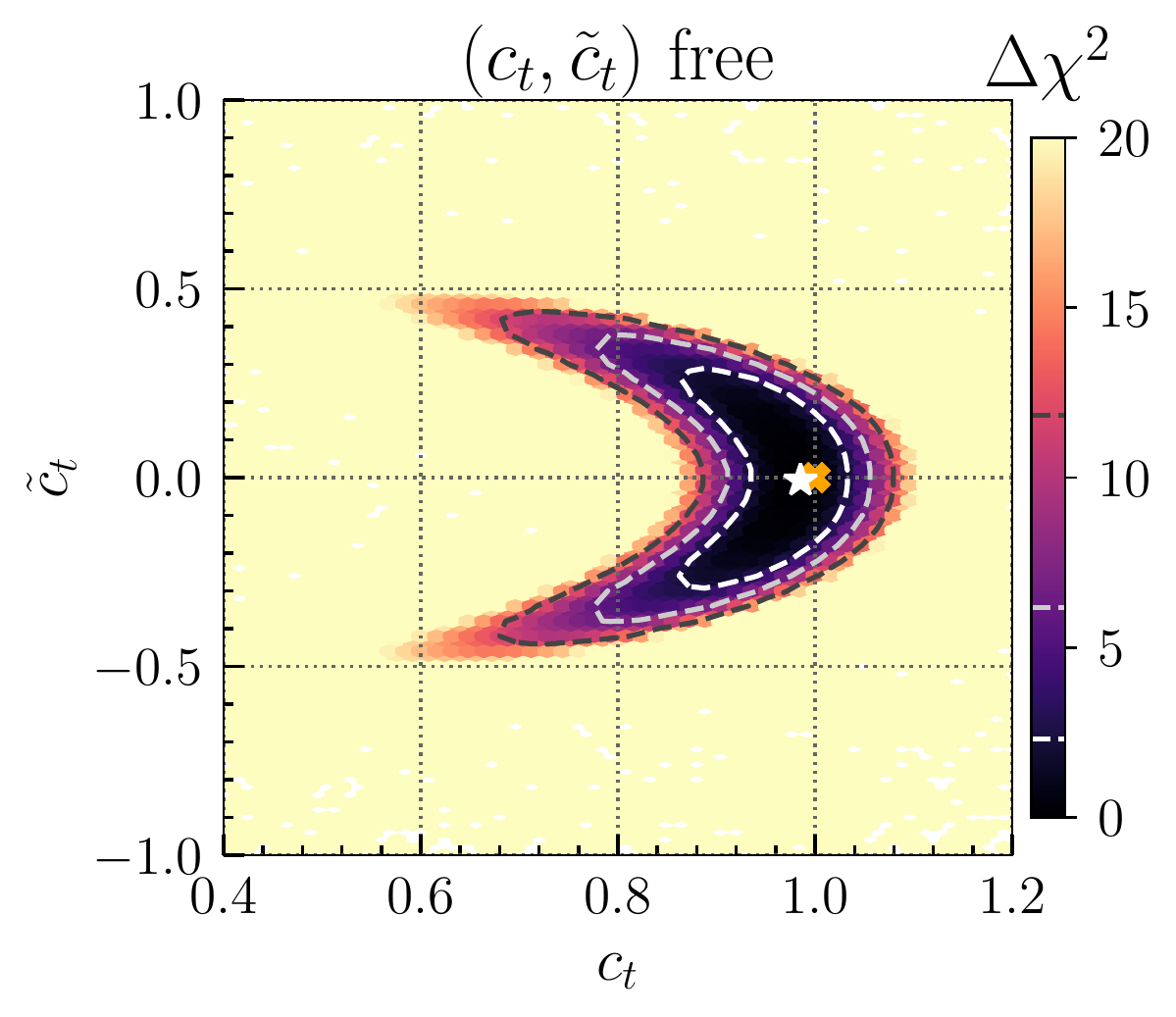}\label{fig:1flavour_a}}
    \subfigure[]{\includegraphics[width=.473\linewidth]{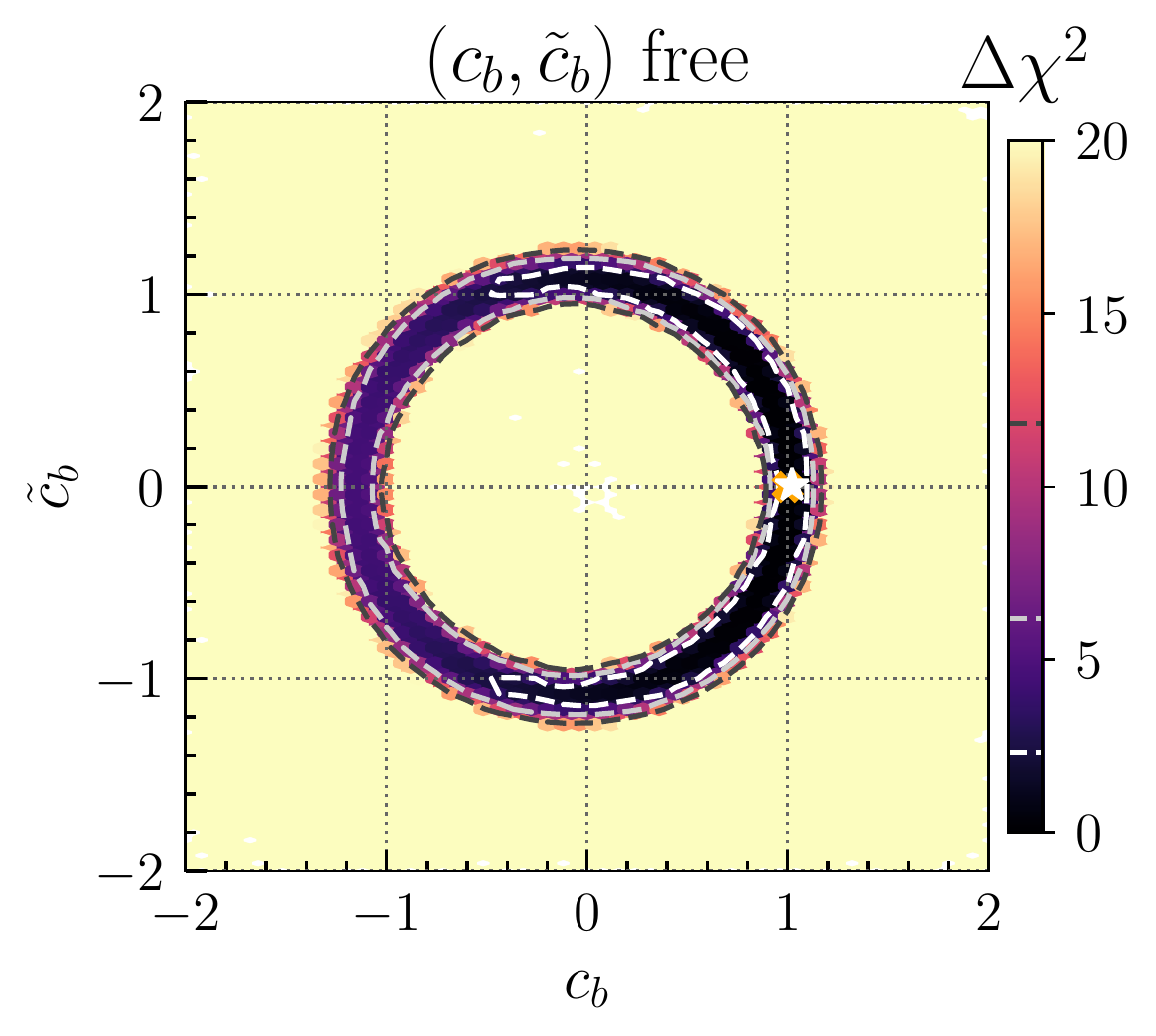}\label{fig:1flavour_b}}
    \subfigure[]{\includegraphics[width=.477\linewidth]{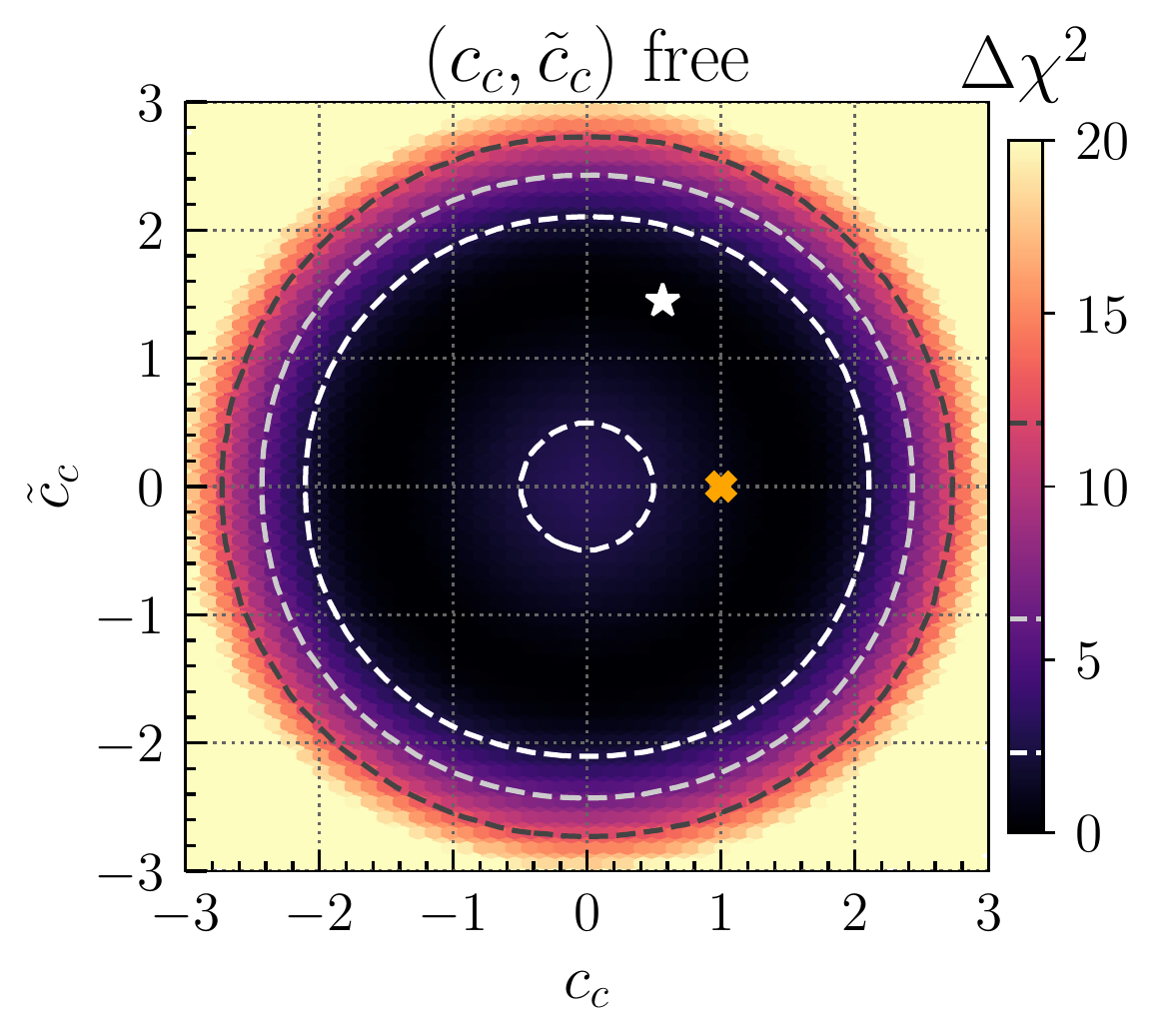}\label{fig:1flavour_c}}
    \subfigure[]{\includegraphics[width=.483\linewidth]{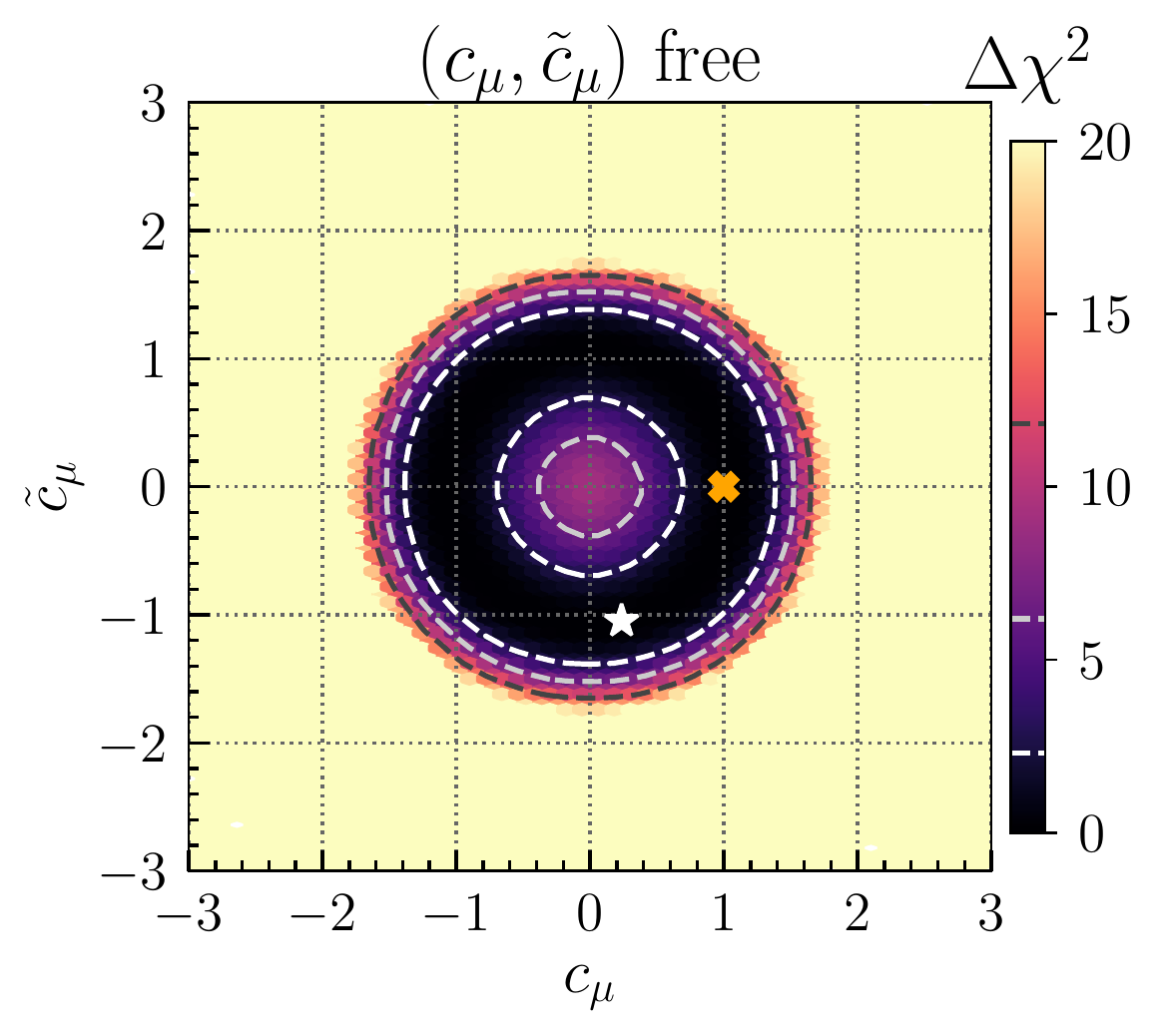}\label{fig:1flavour_d}}
    \caption{Results of fits to the LHC measurements in the (a) (\ct, \ctt), (b) (\cb, \cbt), (c) (\cc, \cct) and (d) (\cmu, \cmut) parameter planes. In each case, the two parameters shown in the plot are free-floating while all the other parameters are set to their SM values. The legend corresponds to the one  in \cref{fig:cms_analysis}.}
    \label{fig:1flavour}
\end{figure}

\paragraph{Quark and \boldmath{$\mu$} Yukawa couplings}

We now consider similar two-dimensional planes for the top quark, the bottom quark, the charm quark and the muon, see \cref{fig:1flavour}. The LHC data set is included for these fits, and in each case only the two plotted parameters are free-floating, while all the others are set to their SM values. For a discussion of the correlation between the top quark coupling modifiers that is displayed in \cref{fig:1flavour_a} we  refer to \ccite{Bahl:2020wee}. In that fit, the best-fit point, with $\chimin = 89.28$, is close to the SM point.

The bottom quark Yukawa coupling is constrained predominantly by measurements of $H\to bb$ decays. The rate measurement, similarly to $H\to\tau^+\tau^-$, depends on $\cb^2 + \cbt^2$, but in this case no additional \cp measurement is available, and consequently the ring-shape constraint is preserved. The fit result is shown in \cref{fig:1flavour_b}. In contrast to \cref{fig:cms_analysis_a} we observe that the region corresponding to $\cb < 0$ is disfavored. This is mostly because of the significant increase of the $ggH$ production cross section in this region, caused by the positive interference with the top quark contribution ($ggH$ production is enhanced by $\sim 23\%$ for $\cb = -1$). Note also that due to the same effect, the ring structure of the one- and two-sigma regions are slightly asymmetric around $\cb=0$. The best-fit point has $\chimin = 89.35$.

The results for the charm quark Yukawa coupling are shown in 
\cref{fig:1flavour_c}. While the direct search for $VH,~H\to c\overline{c}$ yields a limit on $|g_c|$ (see \cref{eq:phase}) of $|g_c|<8.5$ at the 95\% CL, the precise measurement of $\BR(H\to\gamma\gamma)$ sets tighter constraints on $|g_c|$ due to the modification of the total width. Therefore, our global fit is dominated by this indirect constraint. The result shown in the plot corresponds to $|g_c|<2.45$ at the 95\% CL, in agreement with the estimate in Eq.~(\ref{eq:BRgamma}). The best-fit point has $\chimin = 89.00$. Similar to \cref{fig:cms_analysis_a}, its precise location inside the ring is not statistically significant.

\cref{fig:1flavour_d} shows the results for the muon Yukawa coupling. Since the contribution of the muon loop to $\BR(H\to\gamma\gamma)$ is negligible, the constraints on $\cmu$ and $\cmut$ stem from the $H\to\mu^+\mu^-$ decay. The searches for this decay at ATLAS and CMS~\cite{ATLAS:2020fzp, CMS:2020xwi} reach a higher sensitivity compared to the case of the charm quark, reducing the ring width such that the point with $\cmu=0,~\cmut=0$ is outside the $2\sigma$ region, but still inside the $3\sigma$ region. The best-fit point has $\chimin = 89.21$, and again there is no significant sensitivity to its precise location.

\subsubsection{2-flavor models}

\begin{figure}
    \centering
    \subfigure[]{\includegraphics[width=.478\linewidth]{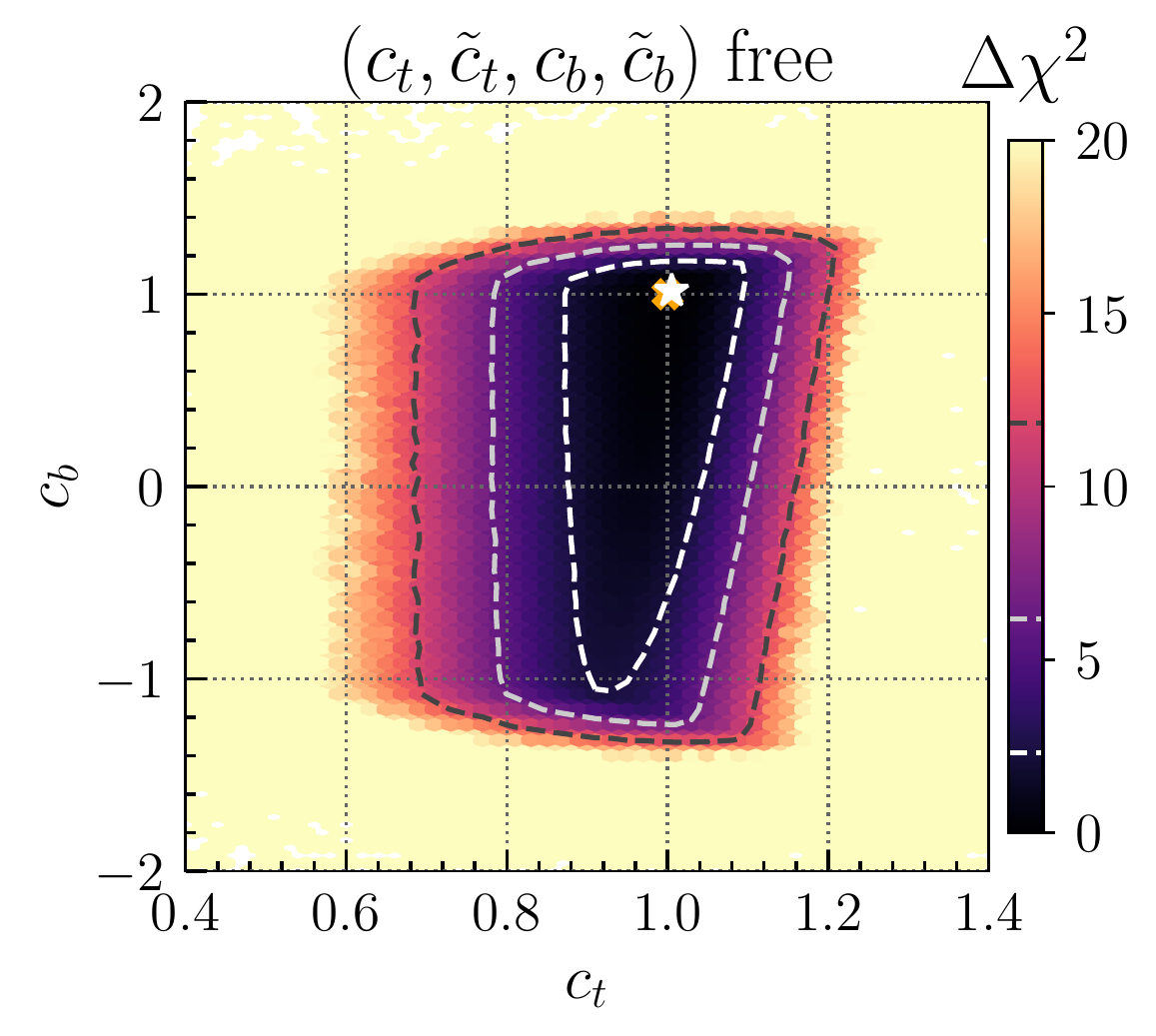}\label{fig:2flavour_1_a}}
    \subfigure[]{\includegraphics[width=.482\linewidth]{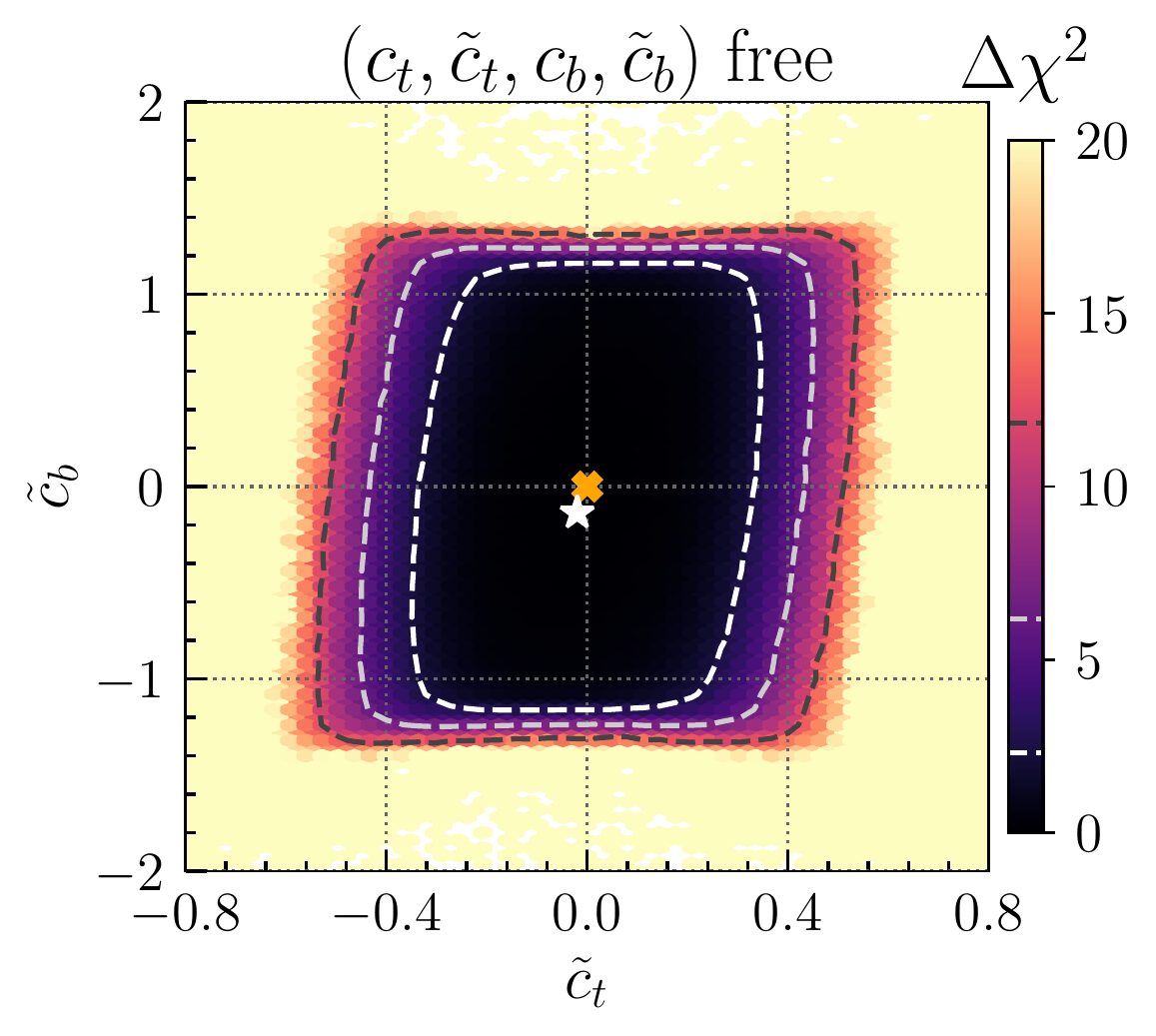}\label{fig:2flavour_1_b}}
    \caption{Fit results in the (a) (\ct, \cb) and (b) (\ctt, \cbt) parameter planes. In each case, the \cp-even and \cp-odd parts of the couplings shown in the plot are free-floating while all the other parameters are set to their SM values. The legend corresponds to the one in \cref{fig:cms_analysis}.}
    \label{fig:2flavour_1}
\end{figure}

\begin{figure}
    \centering
    \subfigure[]{\includegraphics[width=.479\linewidth]{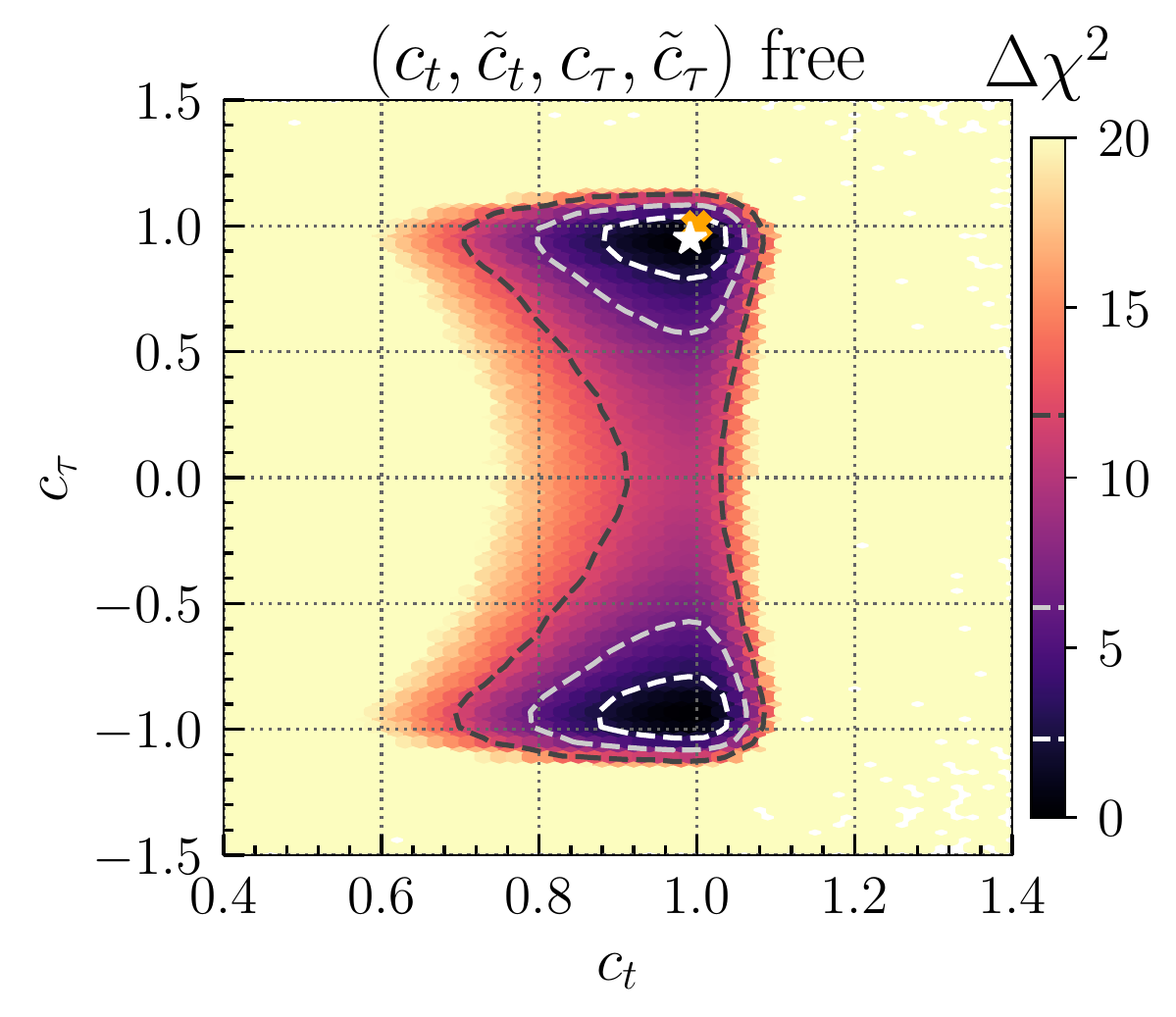}\label{fig:2flavour_2_a}}
    \subfigure[]{\includegraphics[width=.481\linewidth]{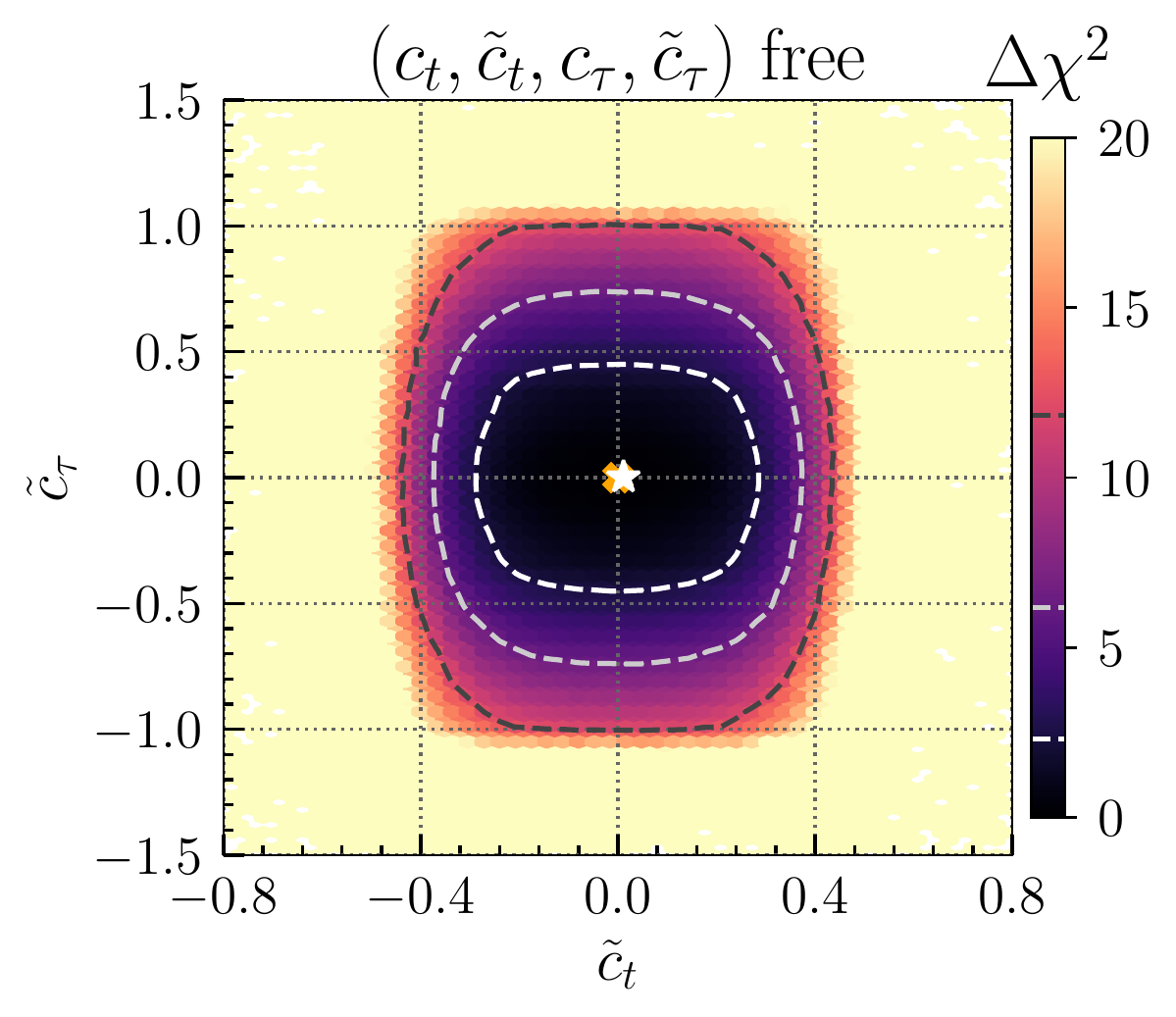}\label{fig:2flavour_2_b}}
    \subfigure[]{\includegraphics[width=.479\linewidth]{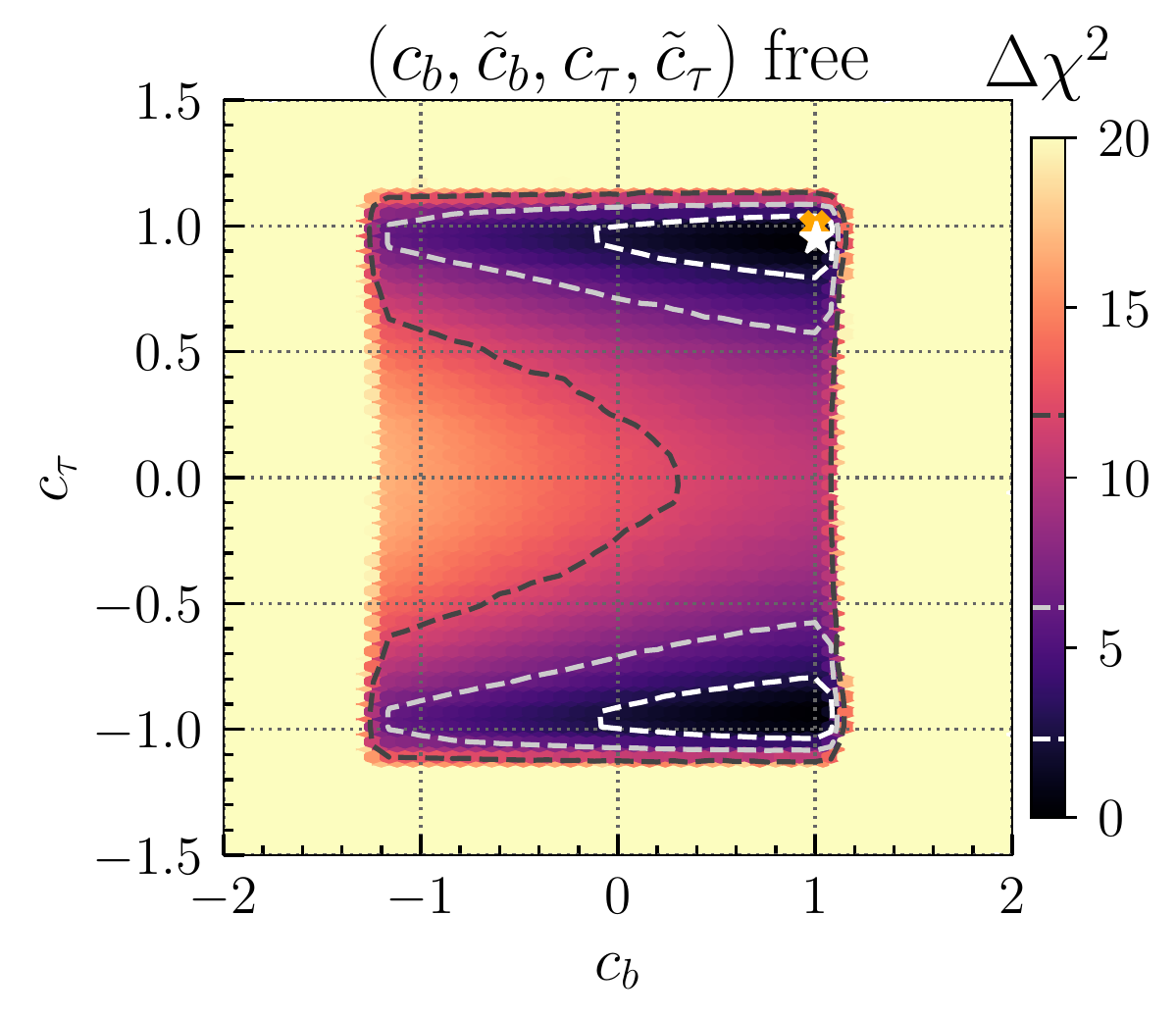}\label{fig:2flavour_2_c}}
    \subfigure[]{\includegraphics[width=.481\linewidth]{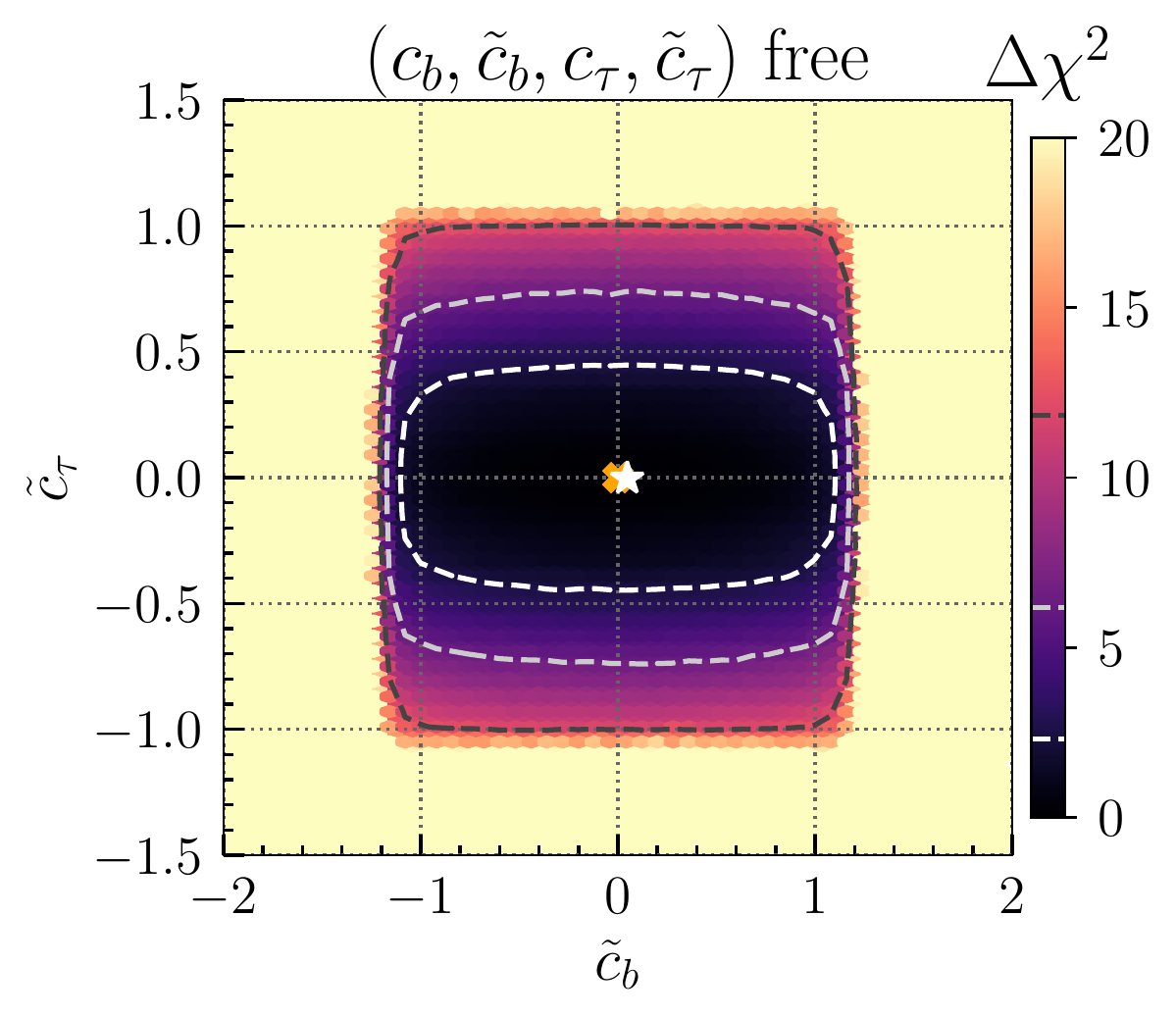}\label{fig:2flavour_2_d}}
    \caption{Results of fits to the LHC measurements in the (a) (\ct, \ctau), (b) (\ctt, \ctaut), (c) (\cb, \ctau), and (d) (\cbt, \ctaut)  parameter planes. 
    In each case, the \cp-even and \cp-odd parts of the couplings shown in the plot are free-floating while all the other parameters are set to their SM values.
    The legend corresponds to the one in \cref{fig:cms_analysis}.}
    \label{fig:2flavour_2}
\end{figure}

Next, we turn to scenarios in which coupling modifiers of two flavors are free-floating simultaneously. The associated results for top and bottom quarks, top quark and tau lepton, and bottom quark and tau lepton are presented in \cref{fig:2flavour_1,fig:2flavour_2}. 

\paragraph{$\boldsymbol{t+b}$ Yukawa couplings}
The constraints on the Higgs--top-quark and Higgs--bottom-quark Yukawa coupling modifiers, see \cref{fig:2flavour_1}, are partially correlated because $ggH$ measurements have an impact on both of them. This opens the possibility of partial cancellations in the BSM contributions, which can lead to a SM-like $ggH$ production cross section even though each coupling deviates significantly from the SM. In particular, low \cb values can be compensated by a slightly reduced \ct parameter (with respect to the SM case), see \cref{eq:kg}. A similar cancellation effect happens in the case of the \cp-odd coupling modifiers, but is suppressed due to the smaller corresponding interference term, as a consequence of the more stringent constraints on \ctt. The best-fit point corresponds to $\chimin = 89.28$.

\paragraph{$\boldsymbol{\tau+t,~\tau+b}$ Yukawa couplings}
For models with free top quark and tau Yukawa couplings, see \cref{fig:2flavour_2_a,fig:2flavour_2_b}, and bottom quark and tau Yukawa couplings, see \cref{fig:2flavour_2_c,fig:2flavour_2_d}, the only source of potential correlation originates from $H\to\gamma\gamma$ decays and is very limited due to the small contribution of the Higgs--tau-lepton Yukawa coupling to this process. In \cref{fig:2flavour_2_c}, the constraints on \cb at $\ctau = 1$ differ from the ones previously shown in \cref{fig:1flavour_b} and \cref{fig:2flavour_1_a}. This is a direct effect of the CMS $H \to \tau^+\tau^-$ analysis, where -- as noted in the discussion of \cref{fig:cms_analysis_b} -- the \chimin value of the best-fit point is slightly lower than the one of the SM point, $\chi^2_{\mathrm{SM}}=89.36$. The best-fit point for the free top-quark (bottom-quark) and tau-lepton couplings corresponds to 87.53 (87.54).
This difference of $\Delta\chi^2 \approx 1.8$ between the best-fit point and the SM point gives rise to a corresponding increase of the $\Delta\chi^2$ value compared to scenarios in which $\ctau = 1$ and $\ctaut = 0$ (as in \cref{fig:1flavour_b,fig:2flavour_1_a}). It has also a small impact on the \ct constraint in \cref{fig:2flavour_2_a}, though the effect is barely visible.

\begin{figure}
    \centering
    \subfigure[]{\includegraphics[width=.479\linewidth]{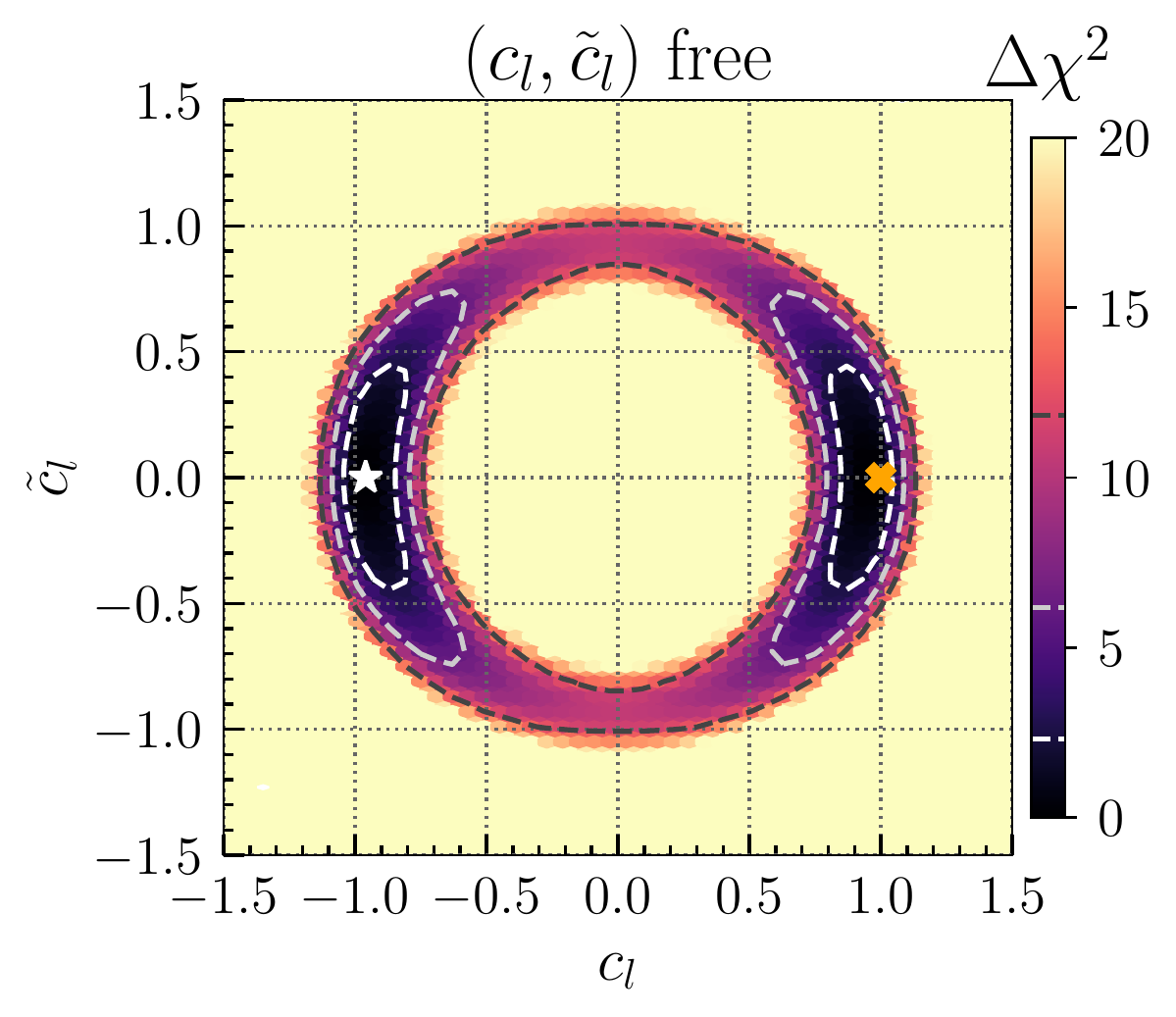}\label{fig:shared_phases_l}}
    \subfigure[]{\includegraphics[width=.481\linewidth]{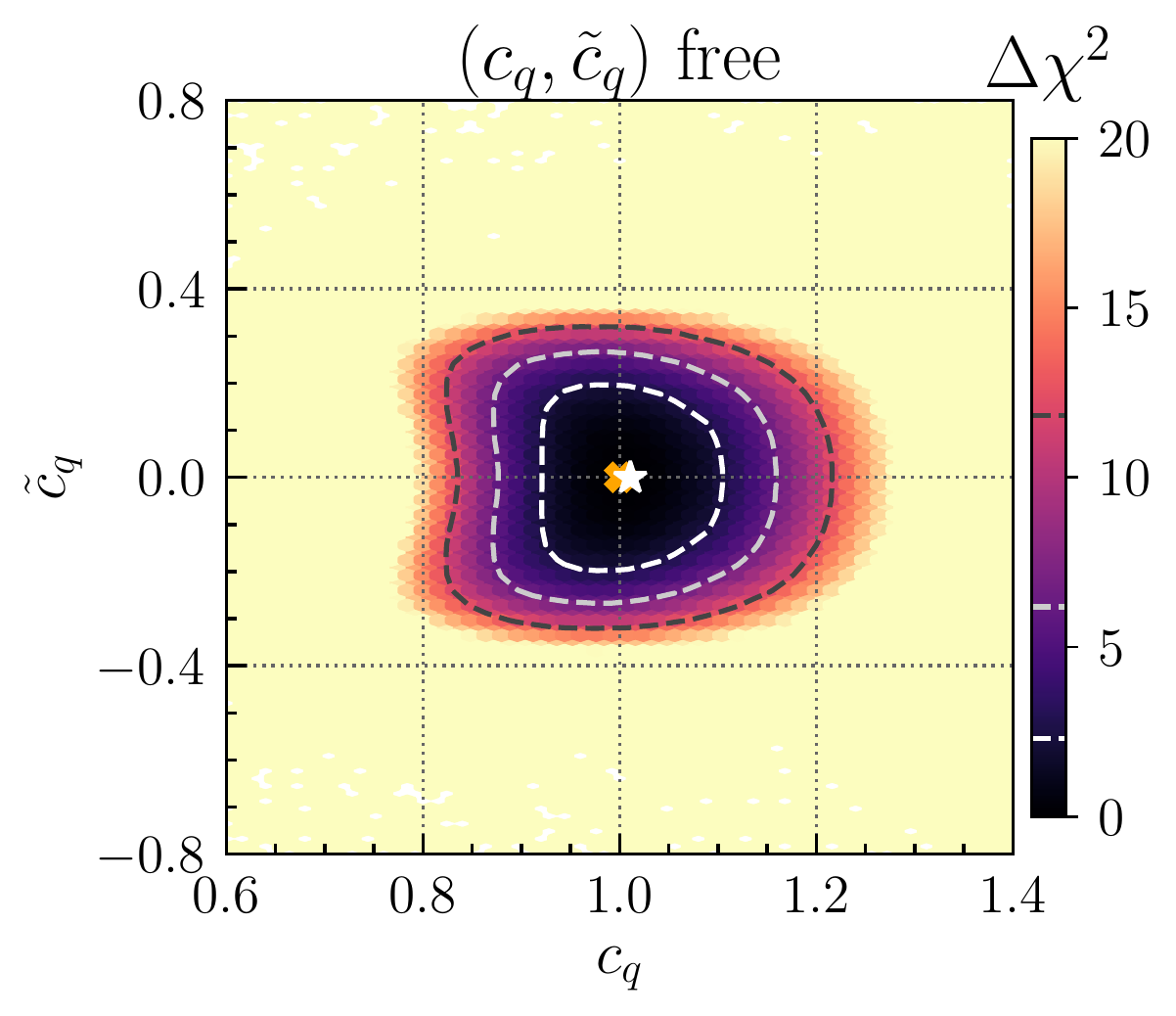}\label{fig:shared_phases_q}}
    \subfigure[]{\includegraphics[width=.479\linewidth]{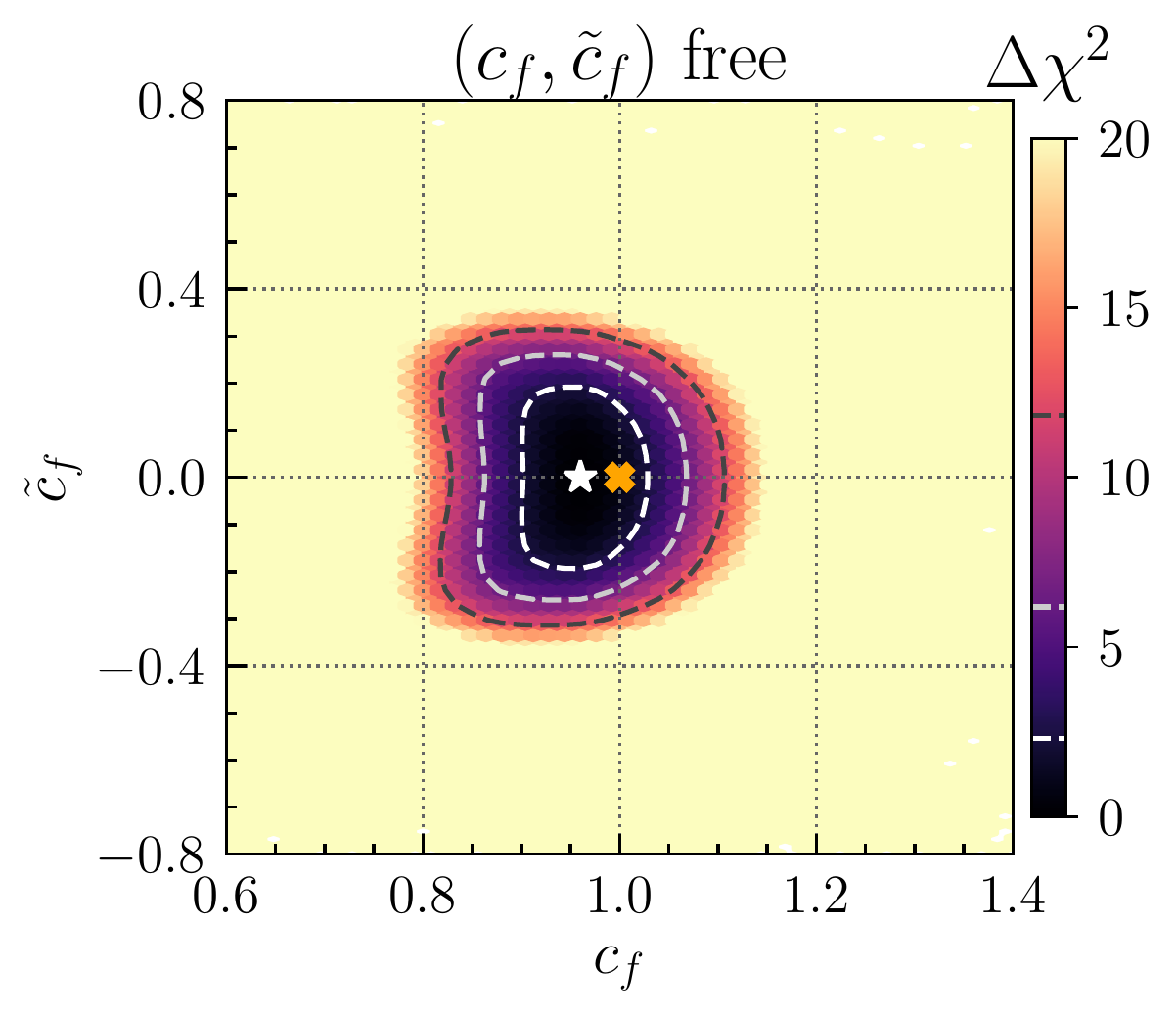}\label{fig:shared_phases_f}}
    \caption{Results of fits to the LHC measurements in the (a) (\cl, \clt), (b) (\cq, \cqt), and (c) (\cf, \cft) parameter planes. 
    In each case, the \cp-even and \cp-odd parts of the couplings shown in the plot are free-floating while all the other parameters are set to their SM values.
    The legend corresponds to the one in \cref{fig:cms_analysis}.}
    \label{fig:shared_phases}
\end{figure}

\subsubsection{Lepton, quark and fermion models}
Next, we investigate models in which the Higgs couplings to leptons, quarks, or all fermions are modified simultaneously with the two coupling modifiers (\cl, \clt), (\cq, \cqt) and (\cf, \cft), respectively. 
The results are shown in \cref{fig:shared_phases}. In \cref{fig:shared_phases_l}, where the lepton couplings \cl and \clt are varied, only the constraints provided by $H\to\tau^+\tau^-$ measurements play a significant role. Consequently, the results are very similar to the $(\ctau,\ctaut)$ results presented in \cref{fig:cms_analysis_b}. When varying the quark couplings \cq and \cqt, see \cref{fig:shared_phases_q}, the constraints are dominated by the limits on the third generation couplings. The form of the constraints qualitatively resembles the one of the 1-flavor top quark Yukawa fit shown in \cref{fig:1flavour_a}. As a consequence of simultaneously varying the bottom Yukawa coupling and the respective effect on the $H\to b\bar b$ decay rate, the constraints are somewhat tighter in comparison to \cref{fig:1flavour_a}. We obtain very similar results if not only the quark couplings are varied simultaneously, but all \hife couplings, see \cref{fig:shared_phases_f}. In comparison to \cref{fig:shared_phases_q}, the additionally relevant constraints originate from $H\to\tau^+\tau^-$ and change the exclusion boundaries only slightly. The best-fit points for the three discussed models are found at \chimin-values of 87.87, 89.20 and 88.26, respectively.

\subsubsection{Universal and fermion+V models}
\begin{figure}
    \centering
    \subfigure[]{\includegraphics[width=.481\linewidth]{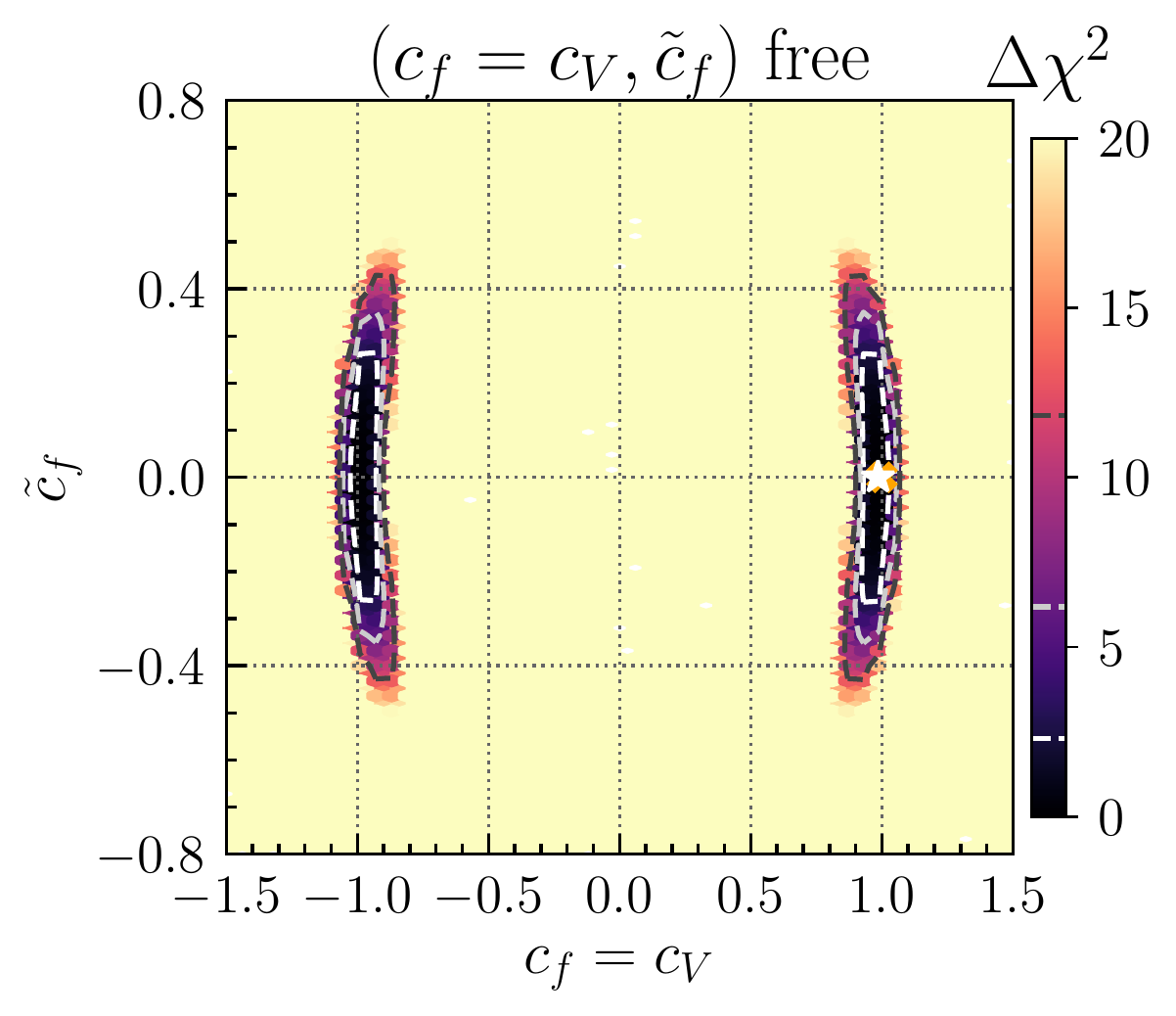}\label{fig:shared_phases_f=V}}
    \subfigure[]{\includegraphics[width=.48\linewidth]{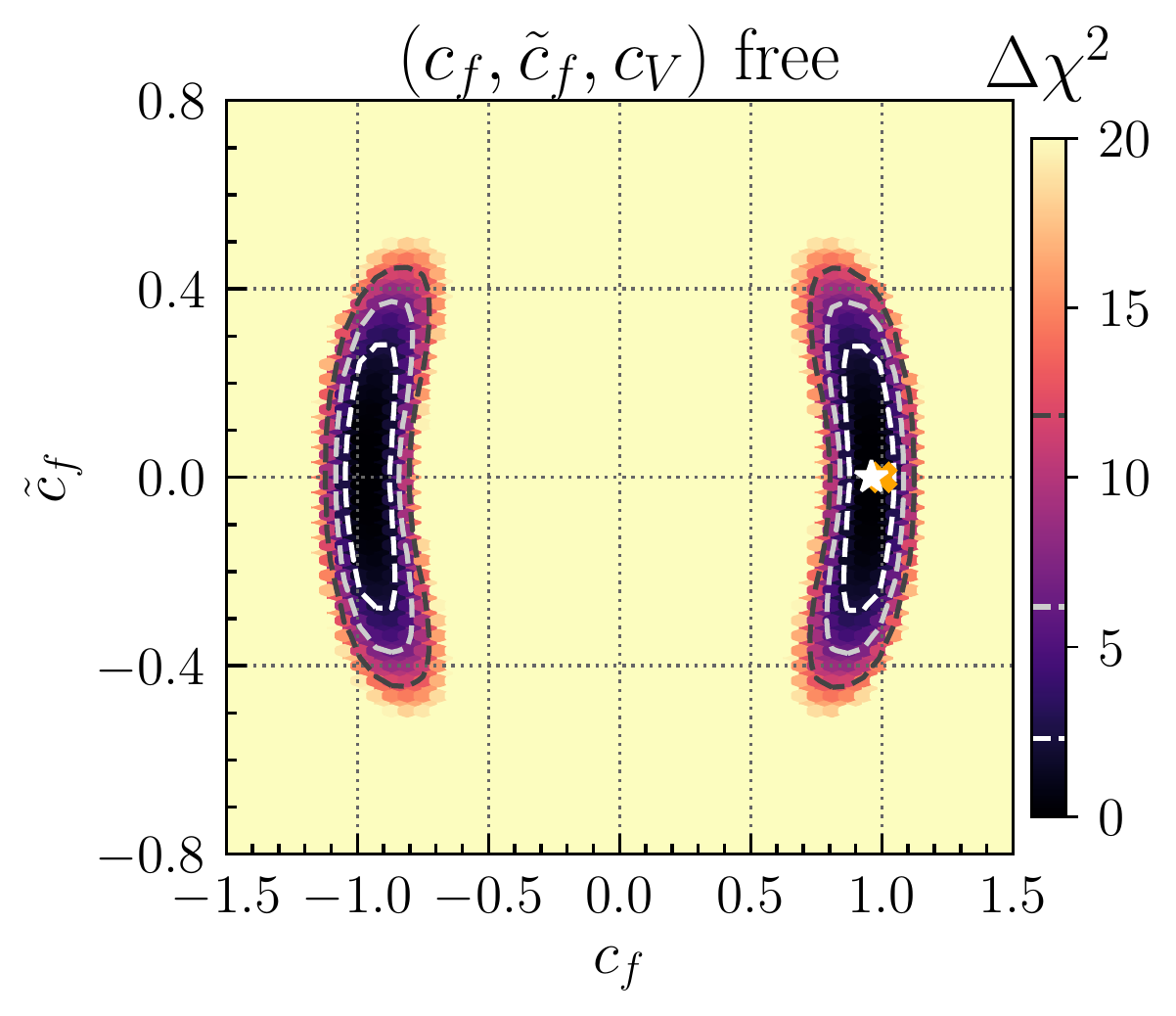}\label{fig:shared_phases_f_V}}
    \caption{Results of fits to the LHC measurements in the (\cf, \cft) parameter plane with (a) $\cf = \cv$ and (b) \cv free-floating.
    In each case, the \cp-even and \cp-odd parts of the couplings shown in the plot are free-floating while all the other parameters are set to their SM values.
    The legend corresponds to the one in \cref{fig:cms_analysis}.}
    \label{fig:shared_phases_V}
\end{figure}

We now consider the case where, as in \cref{fig:shared_phases_f}, 
all \hife couplings are varied simultaneously by the two parameters 
\cf and \cf, but in addition $\cf = \cv$ holds, see 
\cref{fig:shared_phases_f=V}, or \cv is free-floating, see \cref{fig:shared_phases_f_V}. In the scenario with $\cf = \cv$, measurements sensitive to the Higgs coupling to massive vector bosons (e.g.\ of $H\to WW^*, ZZ^*$ or Higgs boson production via vector boson fusion) constrain \cv and therefore also have an impact on \cf. On the other hand, only the $H\to\gamma\gamma$ decay (as well as $tH$ and $tWH$ production, for details see \ccite{Bahl:2020wee}) have a significant dependence on the sign of the \cp-even \hife couplings (i.e.\ on the sign of \ct). This dependence is proportional to $\cv \ct$. As a consequence, the preference for a positive sign of \ct vanishes if \cv is allowed to have negative values. 
Accordingly, in the fit for the case $\cf = \cv$ shown in
\cref{fig:shared_phases_f=V} the preferred region for \cf is found close to $\pm 1$. If instead \cv is floated independently, see \cref{fig:shared_phases_f_V}, the result is qualitatively similar. As a consequence of the free-floating \cv, the constraints are somewhat weaker than in \cref{fig:shared_phases_f=V}. The best-fit point in the two models with $\cf = \cv$ or free-floating \cv has \chimin = 88.40 and \chimin = 87.97, respectively.

\subsubsection{Fermion-vector and up-down-lepton-vector models}
\begin{figure}
    \centering
    \subfigure[]{\includegraphics[width=.479\linewidth]{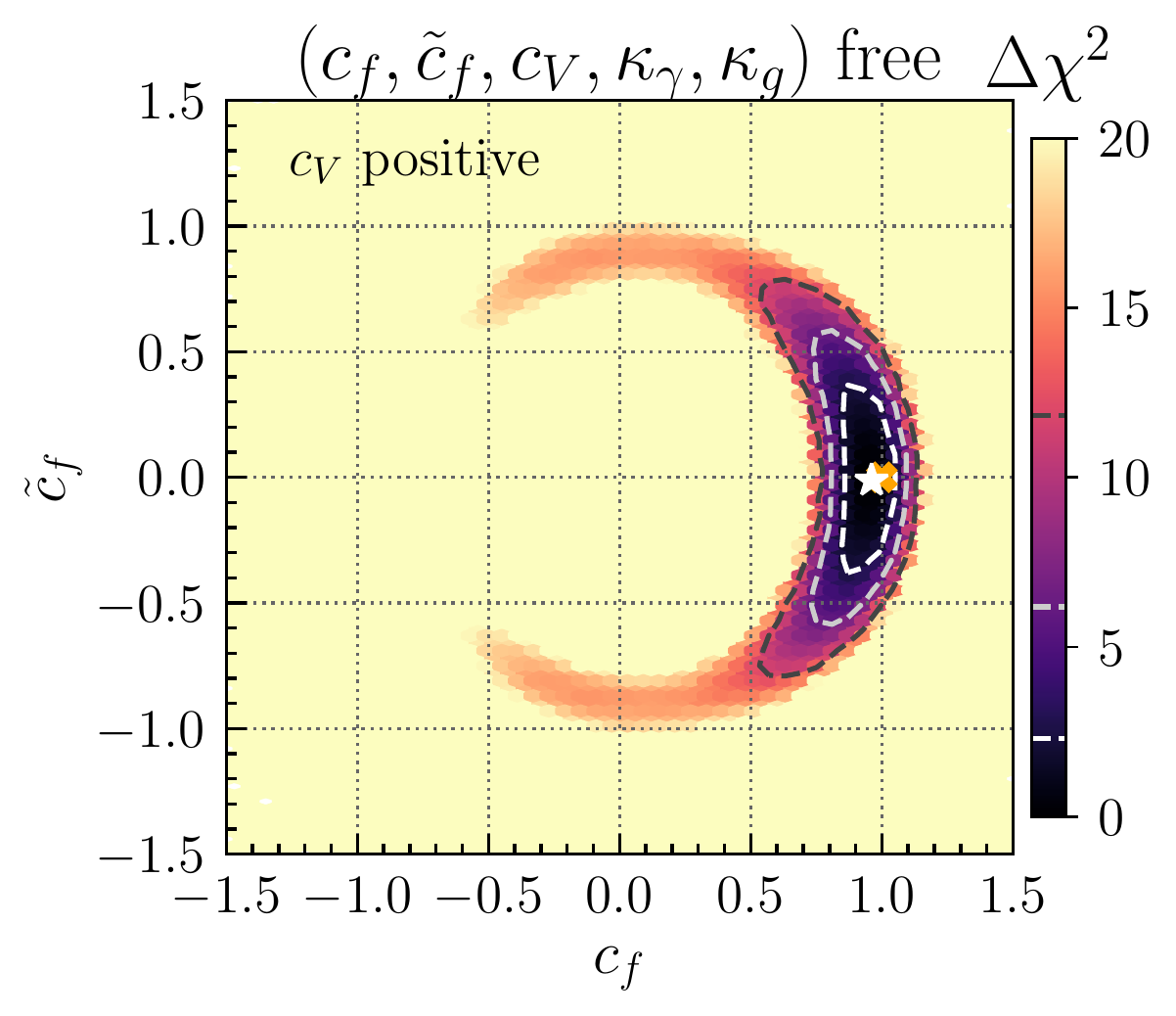}\label{fig:cv_constraints_cV+}}
    \subfigure[]{\includegraphics[width=.481\linewidth]{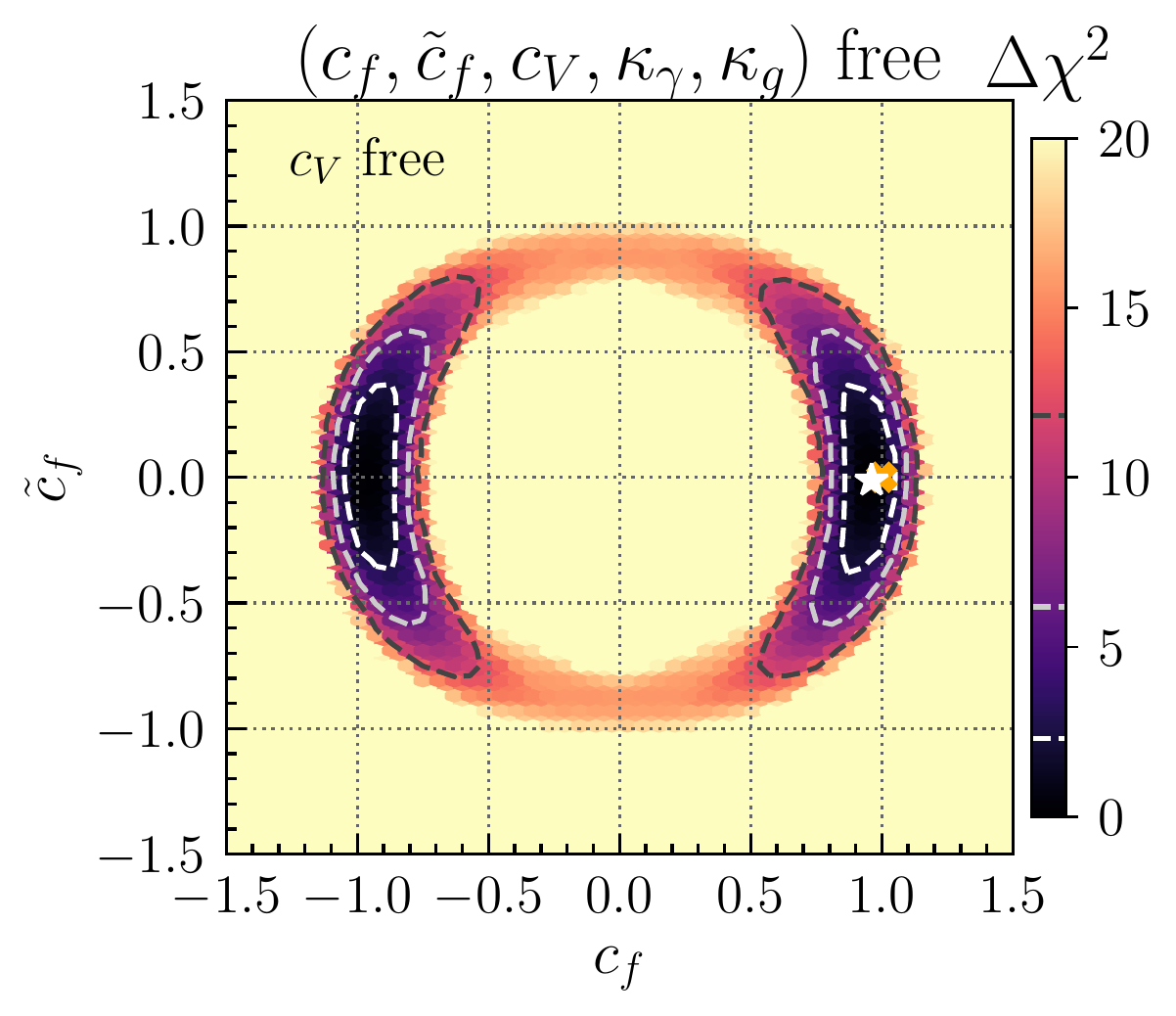}\label{fig:cv_constraints_cV-}}
    \caption{Results of fits to the LHC measurements in the 
    (\cf, \cft) parameter plane. 
    The coupling modifiers \cf, \cft, \kgamma, \kg are free-floating in both plots, while
    (a) \cv is restricted to positive values and (b) \cv is free-floating. 
    In each case, the \cp-even and \cp-odd parts of the couplings shown in the plot are free-floating while all the other parameters are set to their SM values. The legend corresponds to the one in \cref{fig:cms_analysis}.
    }
    \label{fig:cv_constraints}
\end{figure}

So far, we have treated the effective Higgs couplings to gluons and photons (\kg and \kgamma) as dependent parameters using \cref{eq:kg,eq:kgamma}. We can, however, also treat them as free parameters, thus allowing for possible effects of unknown colored or charged BSM particles. This case is studied in \cref{fig:cv_constraints}, in which \cf, \cft, \cv, \kg, and \kgamma are floated freely. If \cv is restricted to be positive, see \cref{fig:cv_constraints_cV+}, we  expect in principle a similar result as obtained in \ccite{Bahl:2020wee}, where \ct, \ctt, \cv, \kg, and \kgamma were floated (assuming \cv to be positive). The CMS $H\to\tau^+\tau^-$ \cp analysis, however, leads to additional constraints limiting $|\cft|\lesssim 0.37$ at the $1\,\sigma$ level, while in \ccite{Bahl:2020wee} a range of $\sim [-1.1,1.1]$ was found for \ctt. If we allow \cv to be negative, see \cref{fig:cv_constraints_cV-}, also negative \cf values are allowed (see the discussion of \cref{fig:shared_phases_V}). The best-fit point corresponds to \chimin = 87.94 in both cases.


\subsection{Impact of EDM and BAU constraints}
\label{sec:results_complementarity}

In this Section, we will investigate the impact of the eEDM constraint by employing the ACME result~\cite{Andreev:2018ayy} (see \cref{sec:EDM_constraints}). Additionally, the amount of the BAU that can be reached based on the (optimistic)  
VIA approach for the displayed parameter regions will be indicated in the plots. 
As discussed in \cref{sec:BAU_constraints}, we treat parameter regions with $\YBratio \geq 1$ as favored by baryogenesis.

The LHC constraints will be applied at the 90\% CL in this section in order to treat them at the same level as the eEDM constraint whose 90\% CL cannot be translated into a 95\% CL bound without further information. Regions in the parameter space that are within the limits from the LHC and ACME measurements at the $90\%$ CL and for which $\YBratio \geq 1$ holds are colored in green.

\subsubsection{1-flavor models}
We first investigate models in which only the Higgs couplings to one fermion species are allowed to float freely. The same fermions as in Sec.~\ref{subsubsec:1-flavor_LHC} will be considered. Contributions from the individual coupling modifiers on the total predicted values for eEDM and BAU are calculated according to the formulas given in \cref{eq:EDM_fit_formula,eq:BAU_fit_formula}, where the Higgs--electron coupling is assumed to be SM-like ($\ce = 1$, $\cet = 0$). The case where this assumption on the Higgs--electron coupling is relaxed will be discussed in Sec.~\ref{sec:results_complementarity_e}. 

\begin{figure}
    \centering
    \subfigure[]{\includegraphics[width=.485\linewidth]{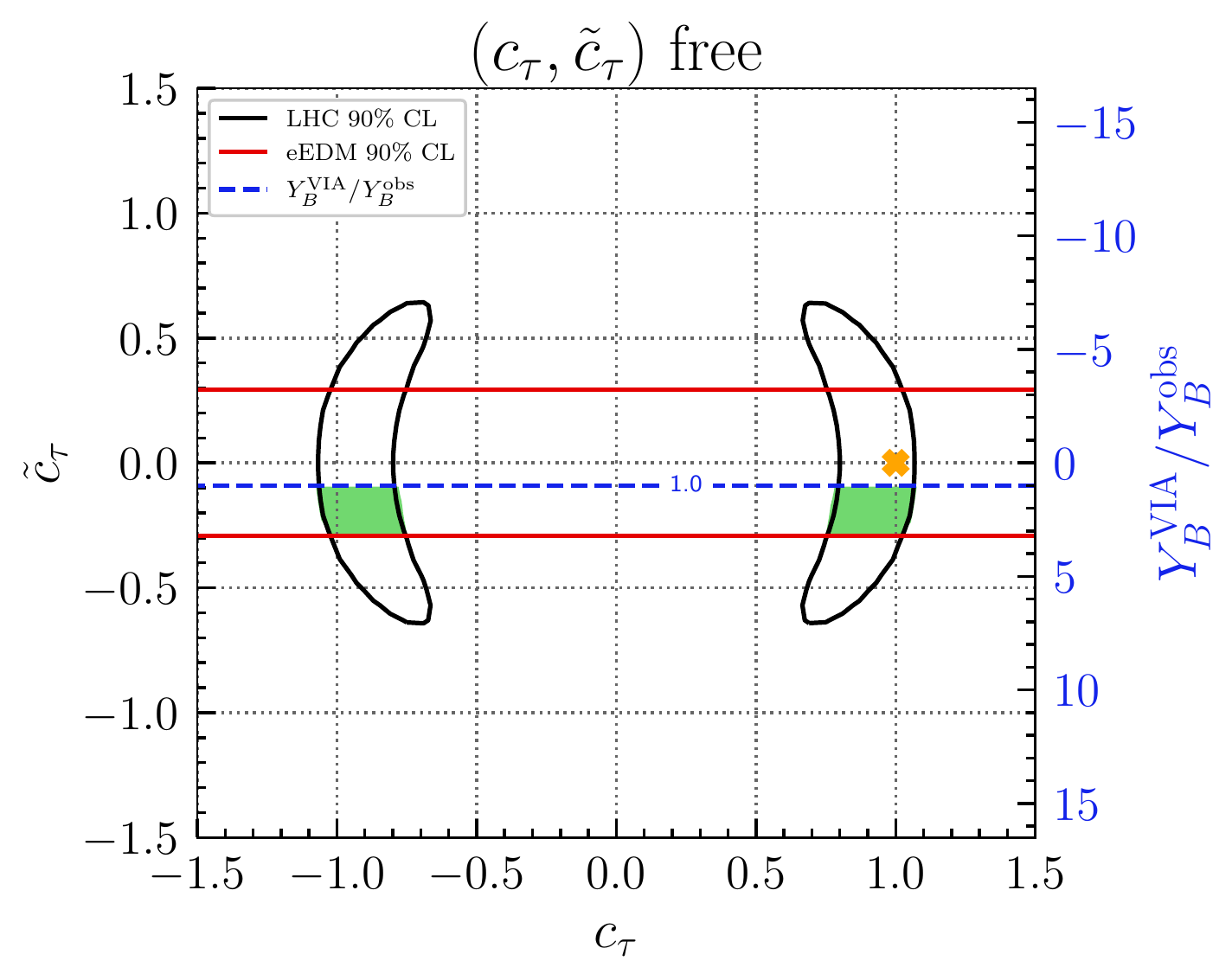}\label{fig:edm_1flavour_tau}}
    \subfigure[]{\includegraphics[width=.475\linewidth]{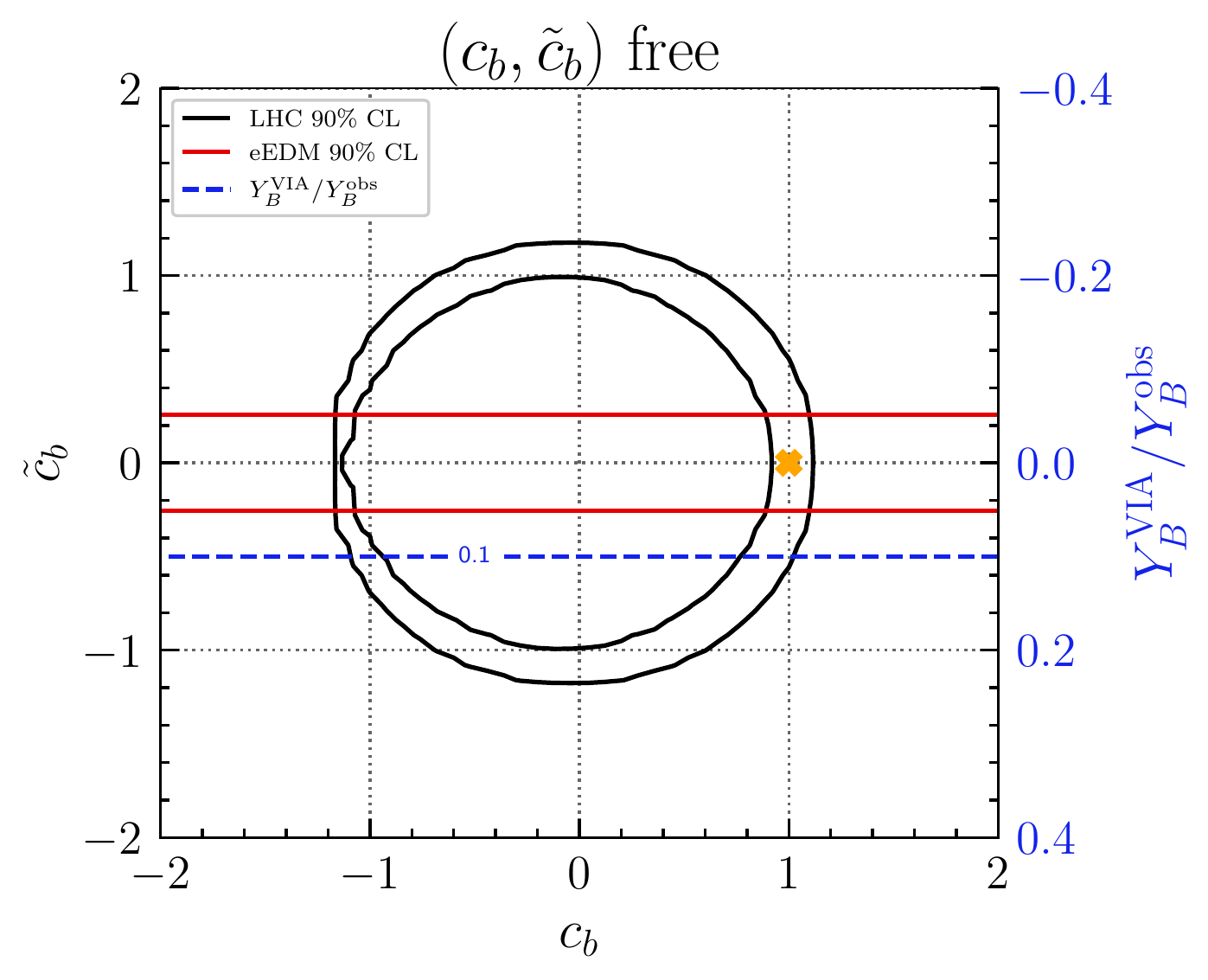}\label{fig:edm_1flavour_bottom}}
    \subfigure[]{\includegraphics[width=.48\linewidth]{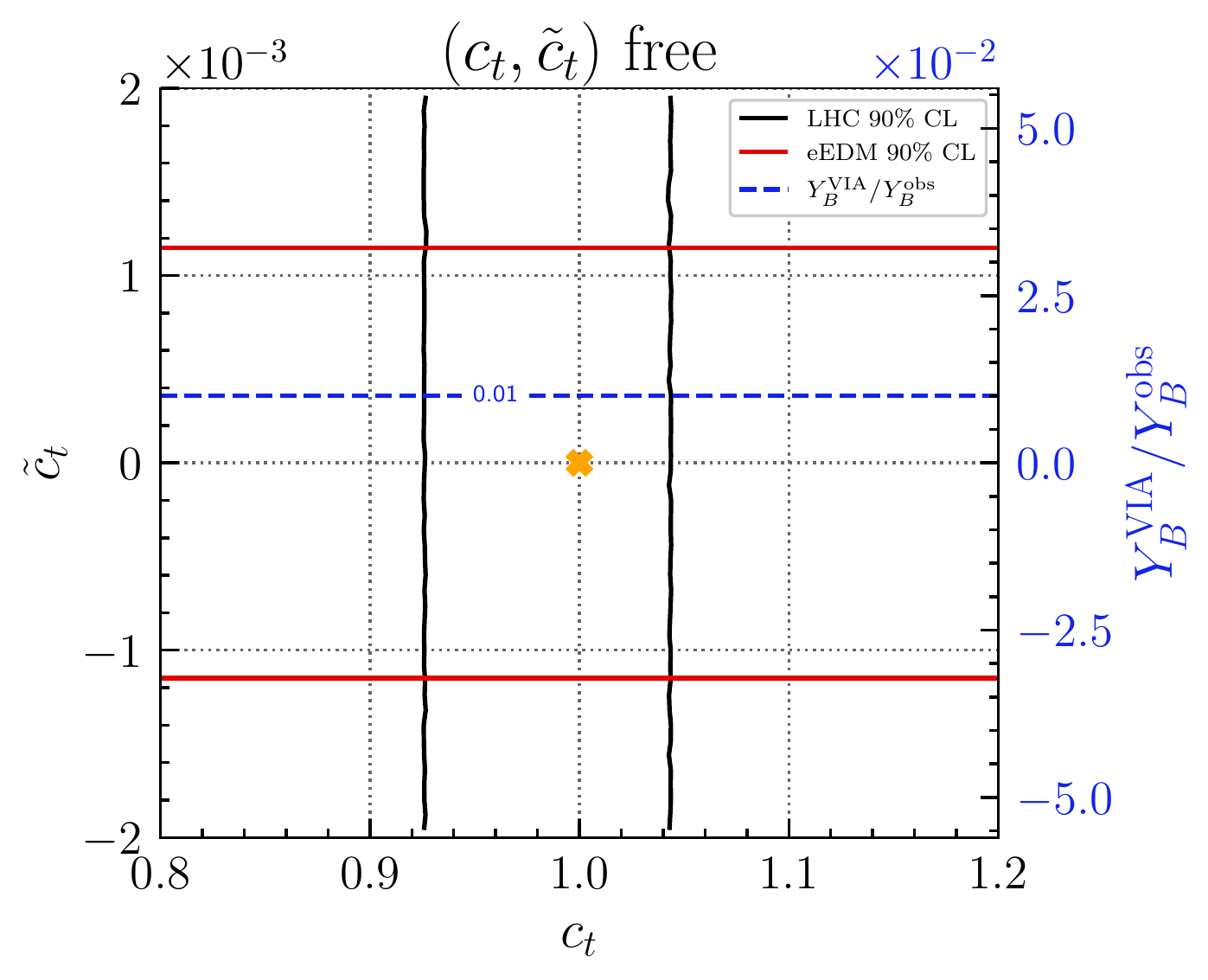}\label{fig:edm_1flavour_top}}
    \caption{Constraints on the \cp-even and \cp-odd modifiers of (a) the tau-Yukawa, (b) the bottom-Yukawa, as well as (c) the top-Yukawa interactions based on LHC measurements (black), eEDM limits (red), and the ratio \YBratio (blue contours and vertical scale on the right). The green colored areas indicate the parameter regions satisfying the LHC and eEDM constraints for which $\YBratio \geq 1$.} 
    \label{fig:edm_1flavour_a}
\end{figure}

\begin{figure}
    \centering
    \subfigure[]{\includegraphics[width=.48\linewidth]{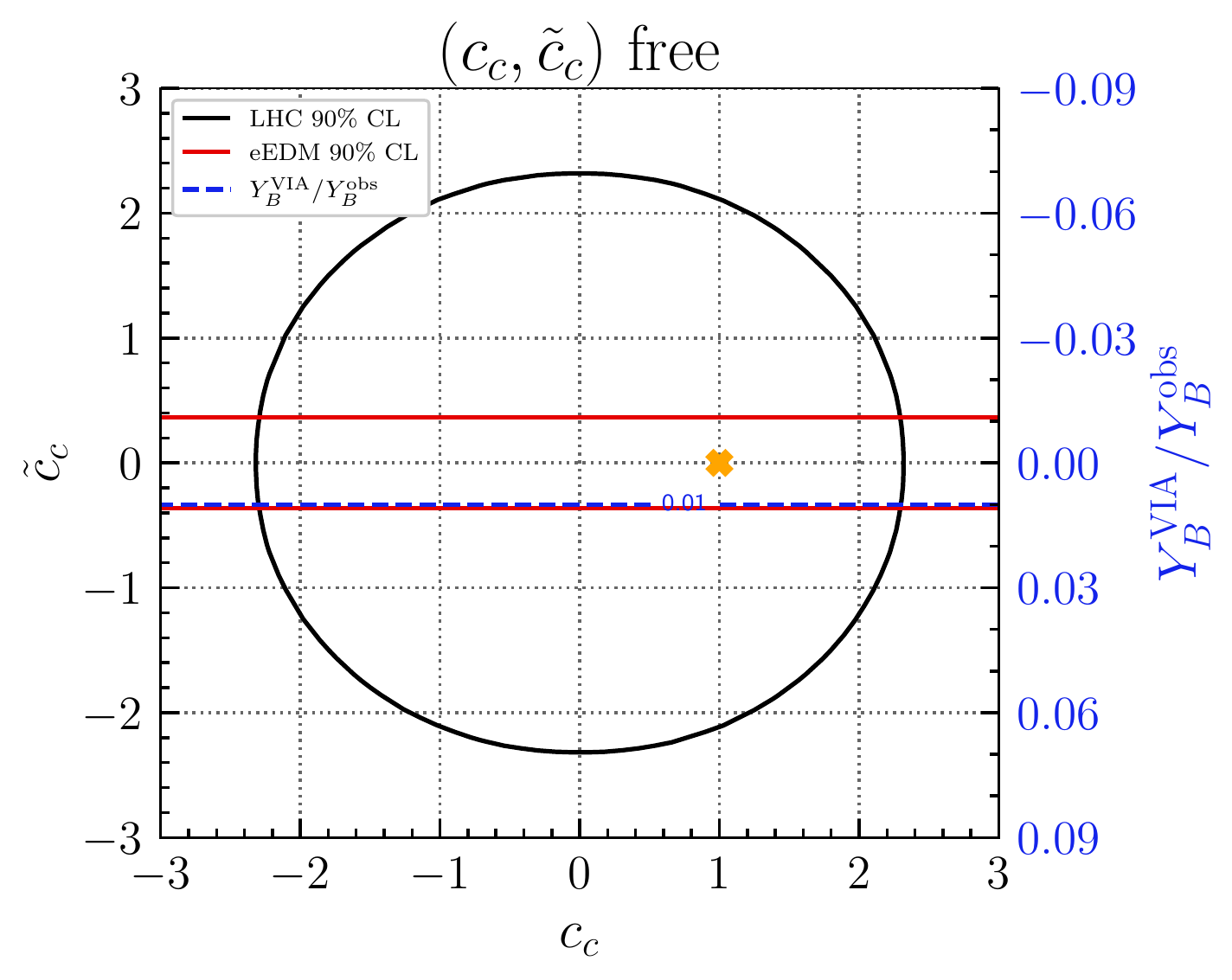}\label{fig:edm_1flavour_charm}}
    \subfigure[]{\includegraphics[width=.48\linewidth]{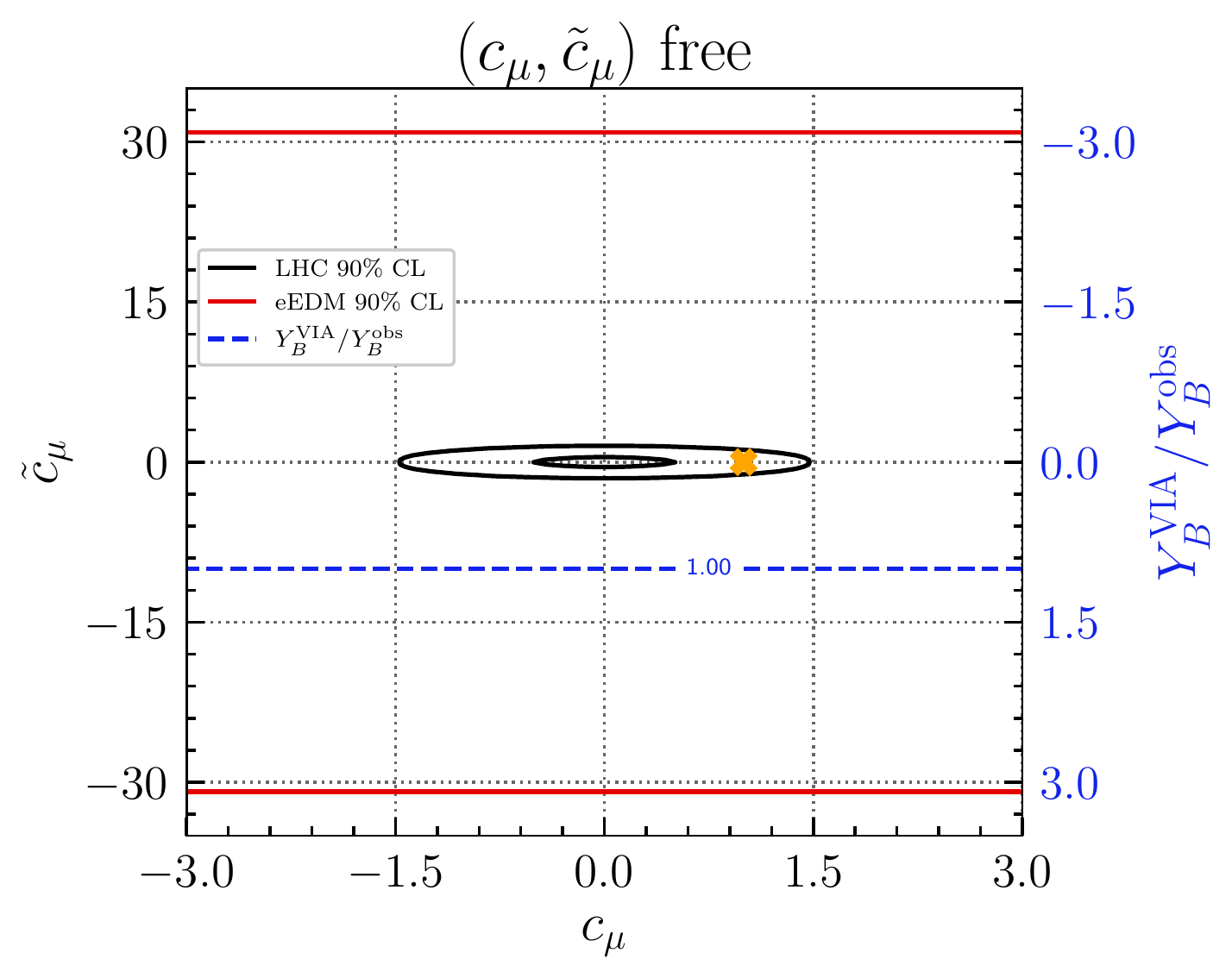}\label{fig:edm_1flavour_muon}}
    \caption{Constraints on the \cp-even and \cp-odd modifiers of (a) the charm-Yukawa as well as (b) the muon-Yukawa interactions. The legend corresponds to the one in \protect\cref{fig:edm_1flavour_a}.}
    \label{fig:edm_1flavour_b}
\end{figure}

Our results are presented in \cref{fig:edm_1flavour_a,fig:edm_1flavour_b}. The eEDM contour lines (red) show where the measured upper limit of ACME is reached. The amount of BAU that can be reached, see \cref{eq:BAU_fit_formula}, is indicated in the plots as a second axis on the right with an additional blue dashed horizontal line at specific values of \YBratio for illustration. Like the eEDM, it only depends on the respective \cp-odd coupling modifiers. Negative \YBratio values imply that more anti-baryons than baryons would have been produced in the early universe and are therefore strongly disfavored.

\paragraph{\boldmath{$\tau$} Yukawa coupling}
In the model with free-floating tau Yukawa couplings as in Sec.~\ref{sec:results_LHC}, see \cref{fig:edm_1flavour_tau}, constraints on \ctau arise mainly due to the CMS $H \to \tau^+\tau^-$ \cp analysis. The strongest constraint on \ctaut, on the other hand, is the eEDM measurement, limiting $|\ctaut| \lesssim 0.29$ at the 90\% CL. This directly translates into $\YBratio \lesssim 3.2$, meaning that \cp violation in the Higgs--tau coupling alone would be sufficient to explain the observed BAU, based on the (optimistic) VIA approach. 

\paragraph{\boldmath{$b$} Yukawa coupling}
Similarly, \cbt is also predominantly constrained by the eEDM, see  \cref{fig:edm_1flavour_bottom}. As a consequence of the smaller contribution of \cbt to the baryon asymmetry, however, $\YBratio$ is limited to be $\lesssim 0.05$ within the parameter region that is allowed by the eEDM and LHC constraints. For illustration, we have indicated the line with $\YBratio = 0.1$, which lies outside of the region that is allowed by the eEDM constraint.

\paragraph{\boldmath{$t$} Yukawa coupling}
The eEDM measurement has an even stronger impact on \ctt, as can be seen in \cref{fig:edm_1flavour_top}. We scale the vertical axis by a factor of $10^{-3}$ in this panel in order to make the eEDM constraint visible. Note, however, that the eEDM constraint strongly depends on the electron Yukawa coupling --- as will be investigated below ---, which here is assumed to be SM-like. As a consequence of the rescaled vertical axis, constraints from the LHC appear as straight lines. The realizable amount of BAU in a scenario where just the Higgs--top coupling deviates from the SM is therefore very small. Within the region that is allowed by the eEDM constraint we find $\YBratio \lesssim 0.033$.

\paragraph{\boldmath{$c$} Yukawa coupling}
Floating the charm quark Yukawa coupling modifiers (see \cref{fig:edm_1flavour_charm}), we observe that the eEDM measurement imposes the dominant constraint on the respective \cp-odd coupling, as it was the case for the third-generation fermion couplings. Within the parameter region that is allowed by the eEDM constraint, only $\YBratio \lesssim 0.01$ can be reached.

\paragraph{\boldmath{$\mu$} Yukawa coupling}
For the case where the Yukawa coupling of the muon is allowed to float, see \cref{fig:edm_1flavour_muon}, we find qualitatively different results. Due to the small muon Yukawa coupling, the eEDM constraint on $\cmut$ is weak, allowing for $\cmut<31$ which corresponds to $\YBratio \lesssim 3.1$. However, the measurement of the $H\to \mu^+\mu^-$ decay at the LHC outperforms the eEDM by constraining the imaginary part of the muon Yukawa coupling to maximally $\cmut < 1.6$ (for $\cmu=0$), corresponding to $\YBratio \lesssim 0.16$. Hence, the sensitivity to this rare decay already provides the dominant information on the \cp-odd part of the muon Yukawa coupling, in agreement with the findings of Ref.~\cite{Fuchs:2019ore}.

\subsubsection{2-flavor models}
\begin{figure}
    \centering
    \subfigure[]{\includegraphics[width=.484\linewidth]{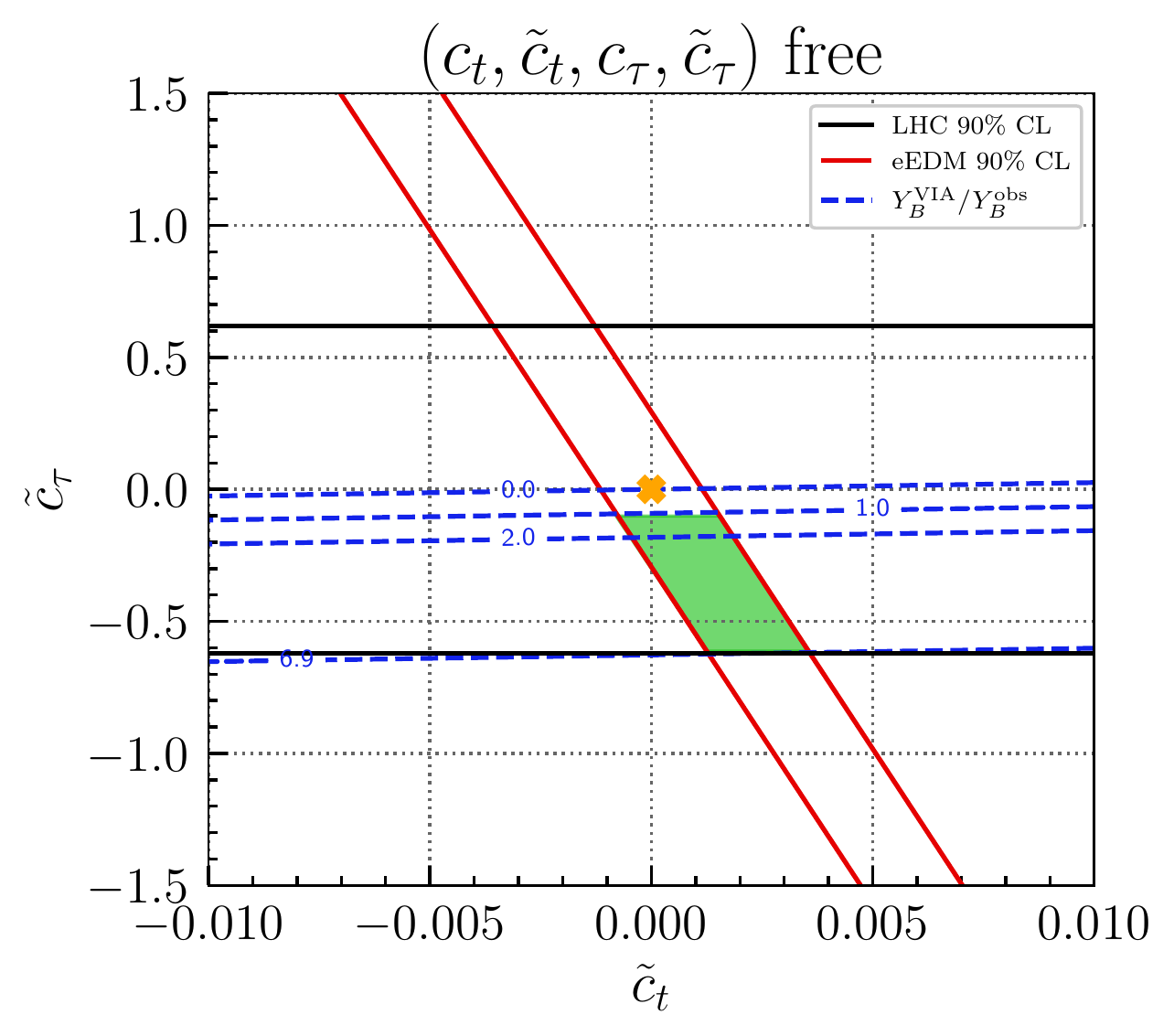}\label{fig:edm_2flavour_toptau}}
    \subfigure[]{\includegraphics[width=.476\linewidth]{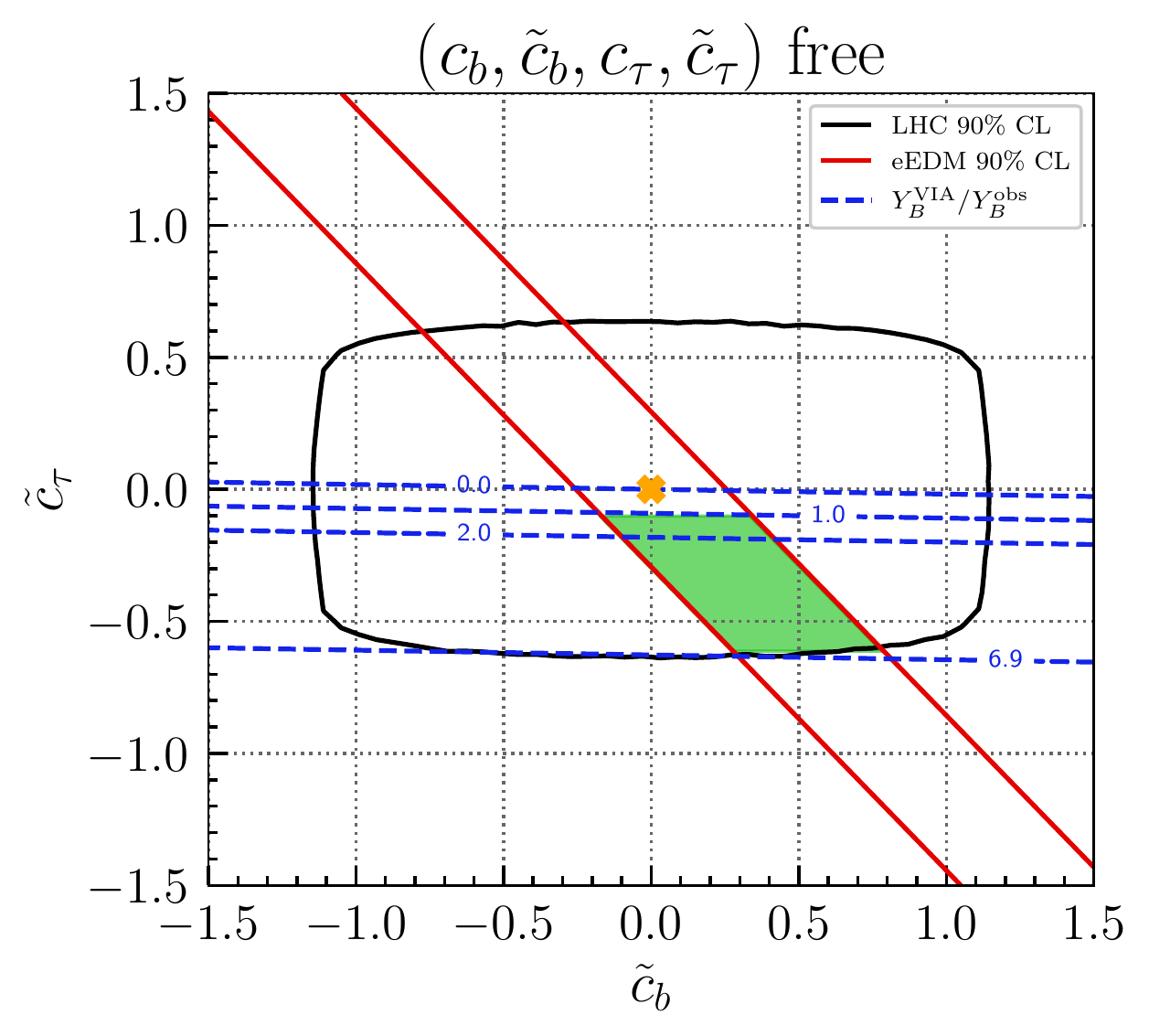}\label{fig:edm_2flavour_bottomtau}}
    \subfigure[]{\includegraphics[width=.48\linewidth]{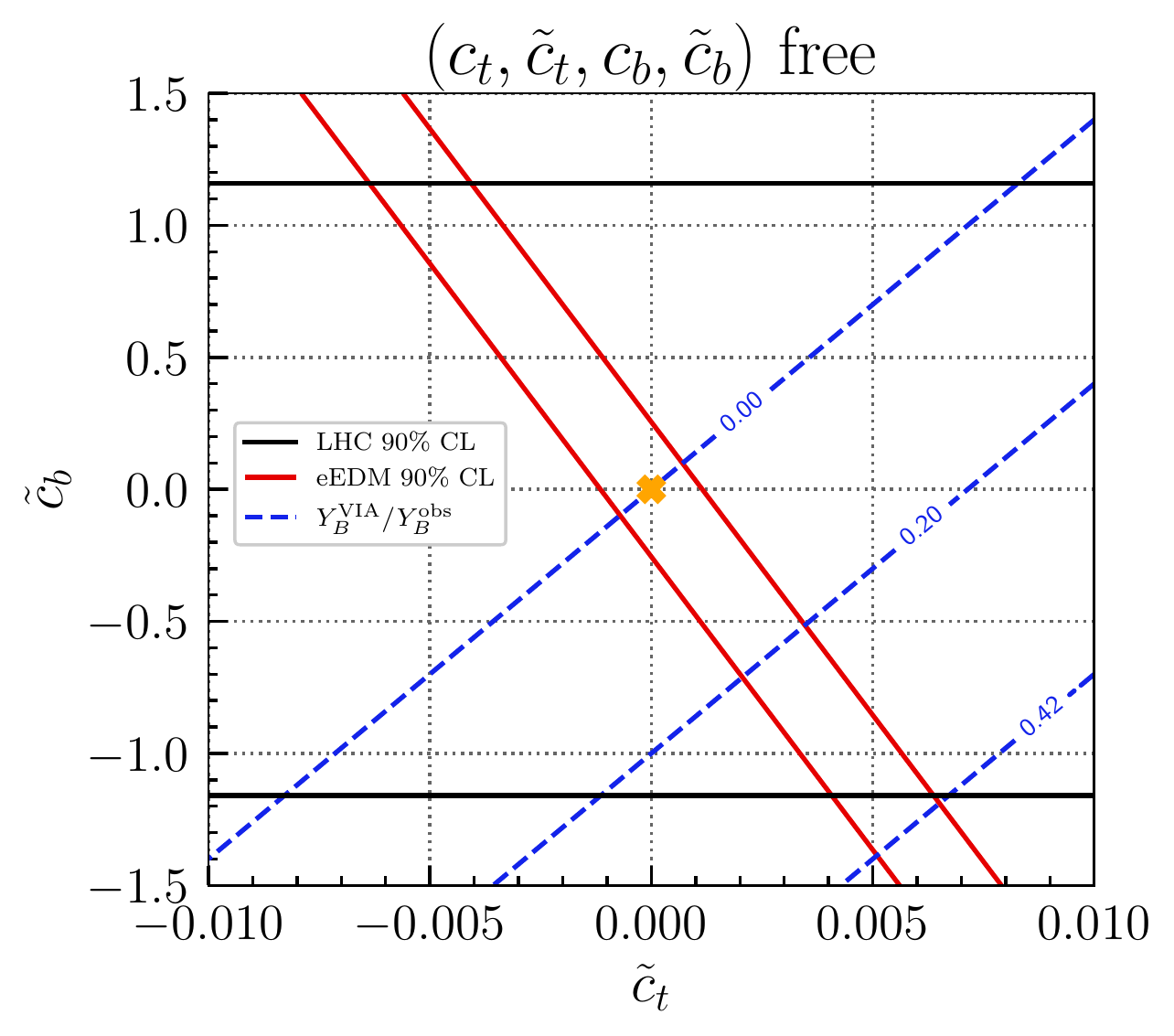}\label{fig:edm_2flavour_topbottom}}
    \caption{Constraints on the \cp-odd modifiers of (a) the top- and tau-, (b) the bottom- and tau-, as well as (c) the top- and bottom-Yukawa interactions based on LHC measurements (black), eEDM limits (red), and the ratio \YBratio (blue). For the LHC constraints, the corresponding \cp-even modifiers are profiled. The legend corresponds to the one in \cref{fig:edm_1flavour_a}.} 
    \label{fig:edm_2flavour_a}
\end{figure}

\begin{figure}
    \centering
    \subfigure[]{\includegraphics[width=.485\linewidth]{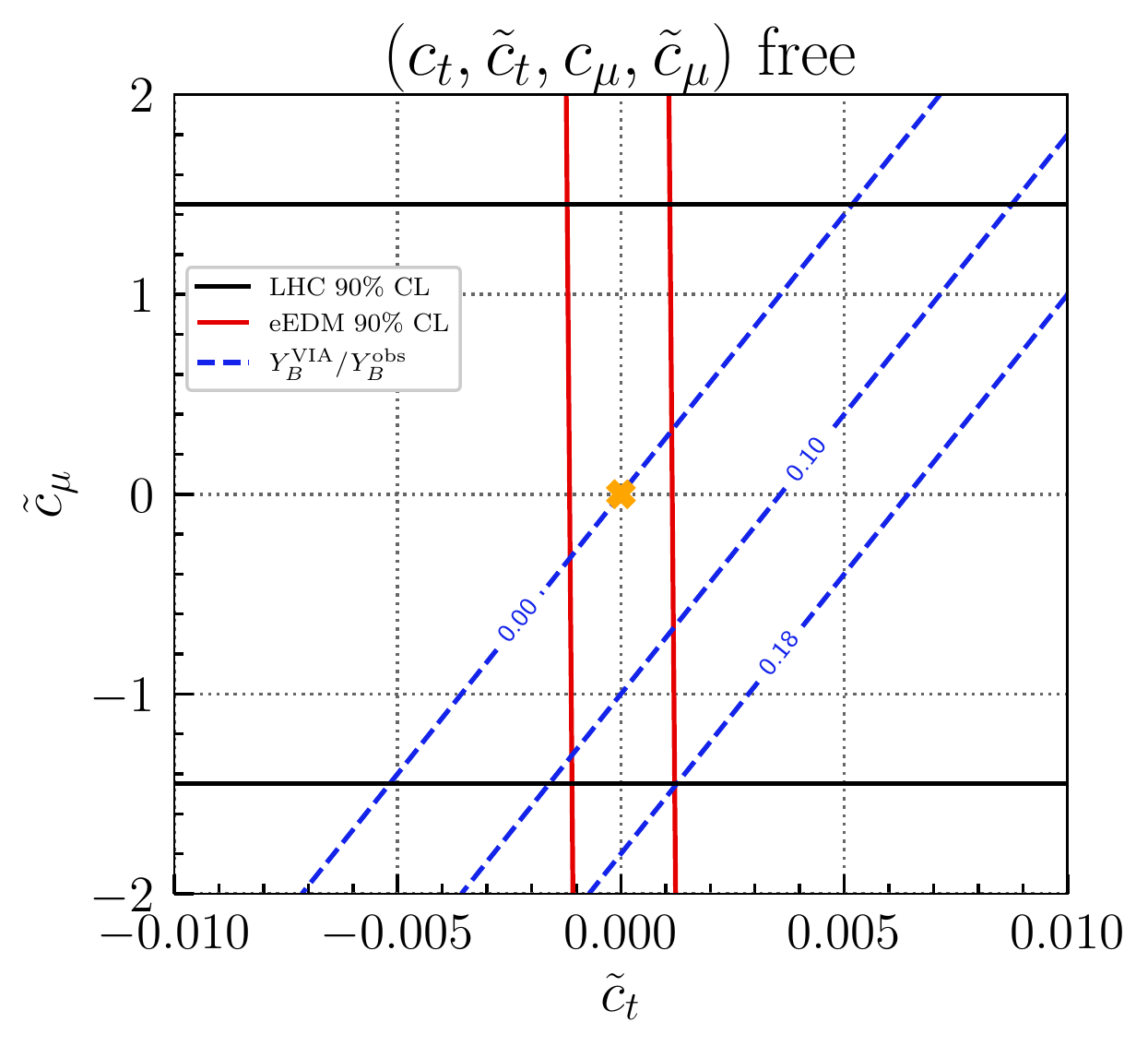}\label{fig:edm_2flavour_topmu}}
    \subfigure[]{\includegraphics[width=.475\linewidth]{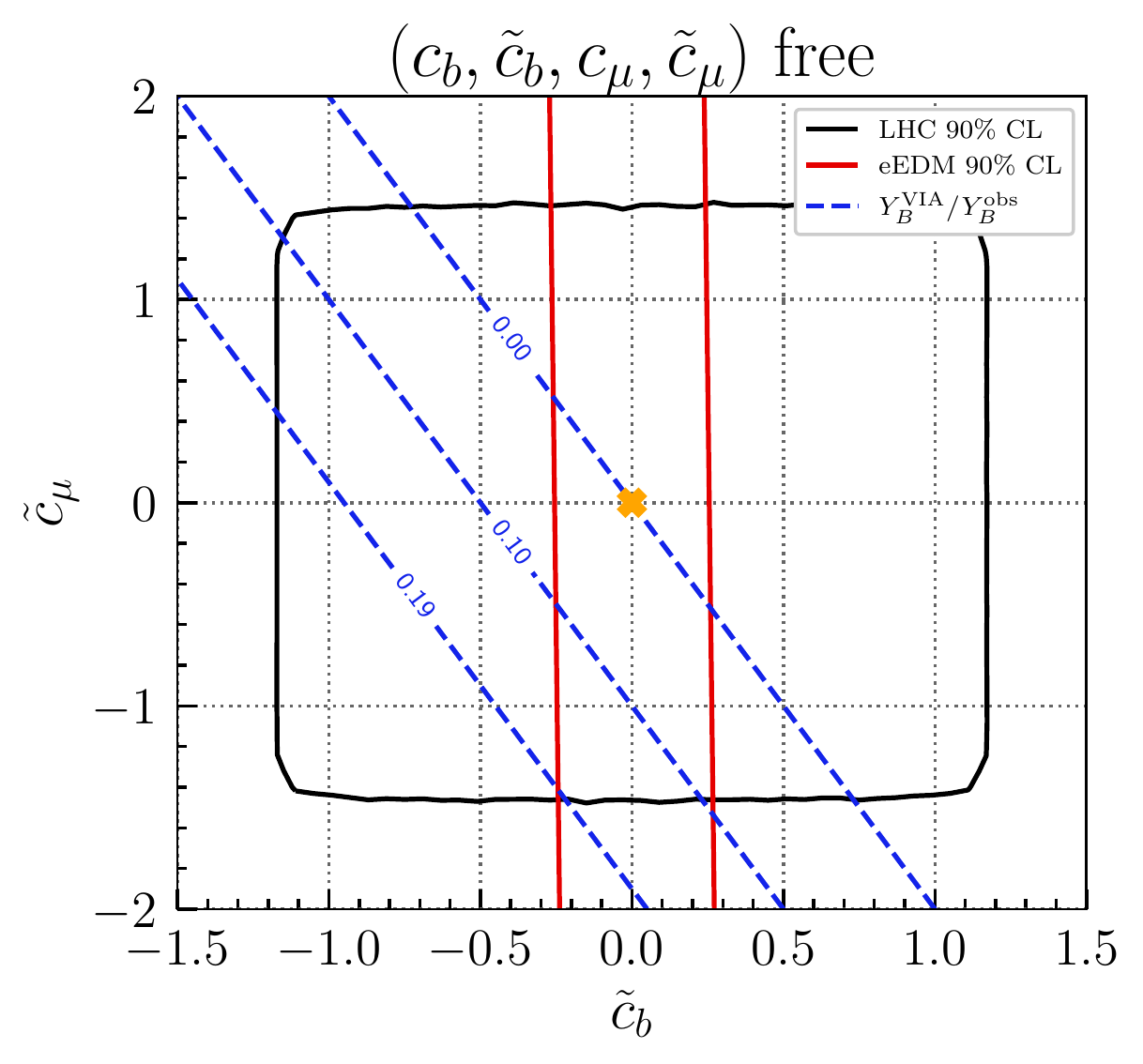}\label{fig:edm_2flavour_bottommu}}
    \caption{Constraints on the \cp-odd modifiers of (a) the top- and muon-, as well as (b) the bottom- and muon-Yukawa interactions. The legend corresponds to the one in \cref{fig:edm_1flavour_a}.}
    \label{fig:edm_2flavour_b}
\end{figure}

In \cref{fig:edm_2flavour_a,fig:edm_2flavour_b} we consider modifications of the Higgs interactions with two different flavors. 

\paragraph{$\boldsymbol{t+\tau}$ Yukawa couplings}
In \cref{fig:edm_2flavour_toptau}, we investigate the possibility of \cp violation in the top quark and tau Yukawa couplings. Since a sufficient amount of \cp violation to explain the BAU (in the VIA approach) can already be generated from the tau Yukawa coupling alone,
see \cref{fig:edm_1flavour_tau}, it can also be achieved when combining a free tau Yukawa coupling with an additional source of \cp violation. The effects of complex tau and top quark Yukawa couplings can cancel each other in the prediction for the eEDM resulting in the diagonal red eEDM contours. Hence, larger values of $\YBratio \lesssim 6.9$ are accessible as compared to the case where only the couplings of one fermion flavor are allowed to float.

\paragraph{$\boldsymbol{b+\tau}$ Yukawa couplings}
A very similar behavior is observed when allowing for \cp violation in the bottom quark and tau Yukawa couplings, with $\YBratio$ again reaching maximally 6.9, as shown in \cref{fig:edm_2flavour_bottomtau}. Since the overall contribution of \cbt to the eEDM is smaller, larger values of $|\cbt|$ are possible in comparison to the allowed $|\ctt|$ values in \cref{fig:edm_2flavour_toptau}. 

\paragraph{$\boldsymbol{t+b}$ Yukawa couplings}
The possibility of \cp violation in the bottom quark and top quark Yukawa interactions is investigated in \cref{fig:edm_2flavour_topbottom}. While in the 1-flavor case \cbt (\ctt) can only reach a small fraction of the observed BAU, $\YBratio = 0.05~(0.03)$, their combination can amount for up to $\YBratio = 0.42$ within the LHC and eEDM limits due to large cancellations in the eEDM prediction. Although this remains short of the full baryon asymmetry, the contribution arising from the combination of 
the couplings is several times larger than the sum of the individual contributions.
This is a qualitatively different result compared to Fig.~5 of \ccite{Fuchs:2020uoc}, where the maximal value of \YBratio was found to be 0.12. The larger \YBratio value determined here is a consequence of profiling over \cb and \ct instead of fixing $\ct = \cb = 1$ as done in \ccite{Fuchs:2020uoc}. 

\paragraph{$\boldsymbol{t+\mu},~\boldsymbol{b+\mu}$ Yukawa couplings}
When allowing for \cp violation in the top quark and muon Yukawa interactions, see \cref{fig:edm_2flavour_topmu}, or in the bottom quark and muon Yukawa interactions, see \cref{fig:edm_2flavour_bottommu}, no sufficient amount of \cp violation can be generated while satisfying the LHC and eEDM constraints ---  the maximal reachable \YBratio\ values are $\sim 0.18$ and  $\sim 0.19$, respectively, i.e.\ just a small increase compared to the contribution of the muon alone.

\subsubsection{Fermion and fermion+V models}
\begin{figure}
    \centering
    \subfigure[]{\includegraphics[width=.482\linewidth]{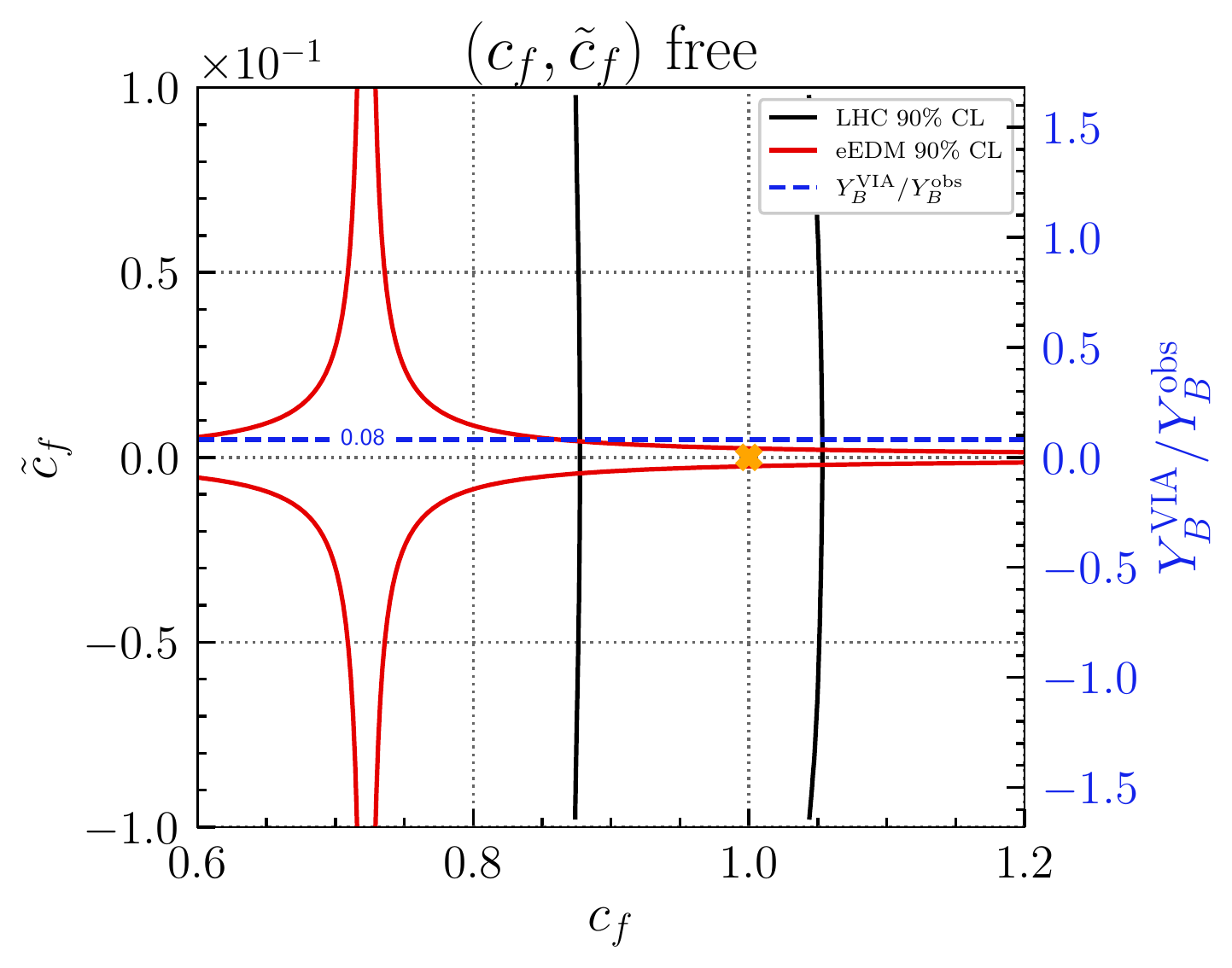}\label{fig:edm_fermions}}
    \subfigure[]{\includegraphics[width=.478\linewidth]{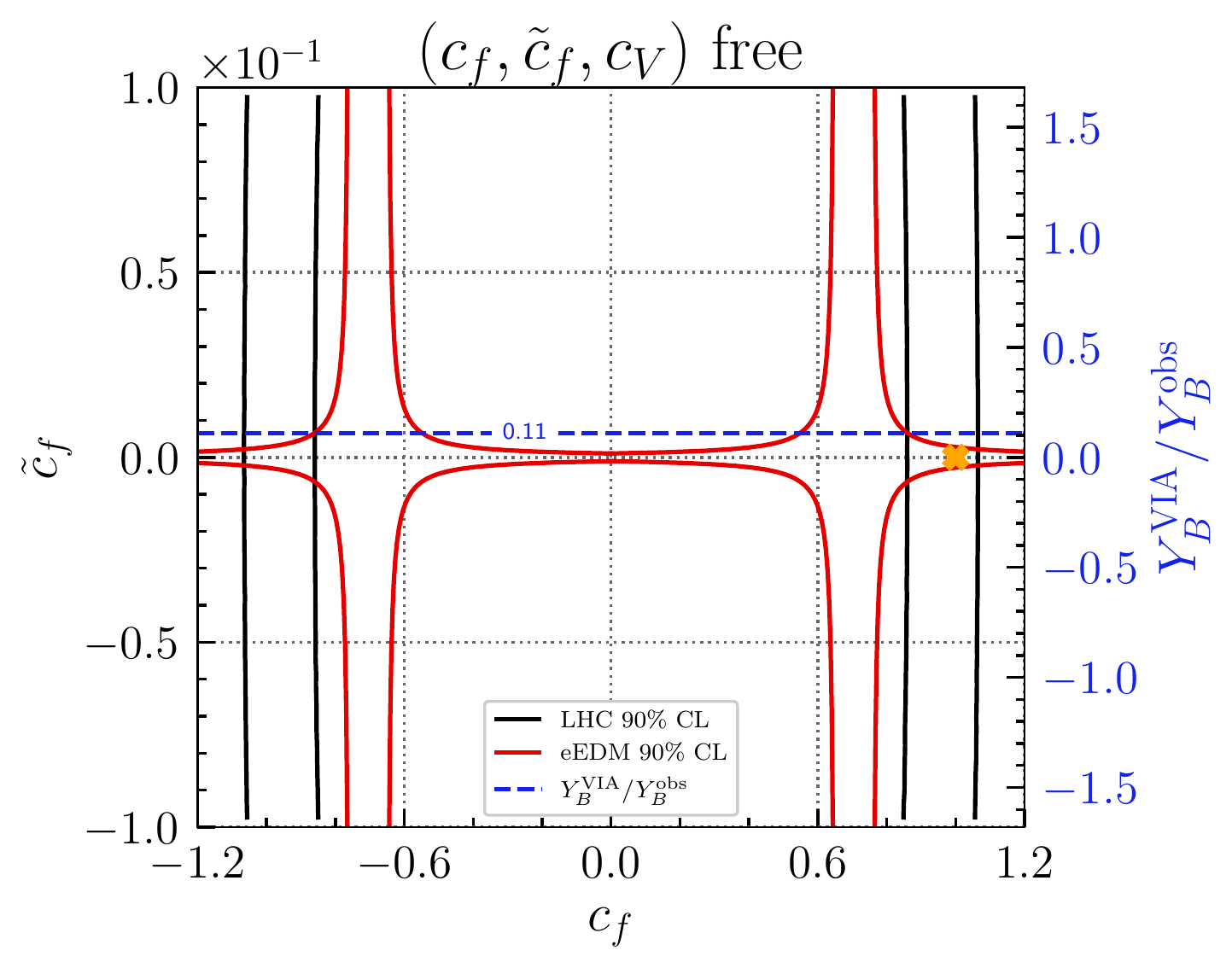}\label{fig:edm_global_cV}}
    \caption{Constraints on the global \hife coupling modifiers \cf and \cft where \cv is (a) set to its SM value or (b) free-floating in the fit. The legend corresponds to the one in \cref{fig:edm_1flavour_a}.}
    \label{fig:edm_global}
\end{figure}

Instead of varying the coupling modifiers of one or two \hife interactions, we now consider the case where all \hife coupling modifiers are varied in the same way by floating the global modifiers \cf and \cft, see \cref{fig:edm_fermions}. In this scenario, the coupling modifiers are varied simultaneously for the top quark Yukawa and the electron Yukawa coupling, which can potentially give rise to cancellations of the different contributions in the eEDM calculation. Indeed, the star-like shape of the contour of the eEDM constraint arises from a cancellation for $\cf \approx 0.73$, for which sizable contributions of $|\cft| \gtrsim 0.1$ are allowed. The collider bounds in \cref{fig:edm_fermions} correspond to the ones shown in \cref{fig:shared_phases_f}. Since the region with significant cancellations in the eEDM prediction has only a small overlap with the region that is allowed by the LHC constraint, the reachable \YBratio values are only slightly increased to $0.08$ as compared to a single flavor modification of the top quark or bottom quark Yukawa coupling.

In addition to varying \cf and \cft, also \cv is free-floated in \cref{fig:edm_global_cV}. For the eEDM calculation, we vary \cv within its 90\% CL LHC limits and then derive the minimal possible $|d_e|$ value. Similarly to \cref{fig:edm_fermions}, the eEDM constraint gives rise to a star-like shaped contour. Since the relative size of \cf and \cv determines the position of this contour, see \cref{eq:EDM_fit_formula}, floating \cv results in a second star-like allowed eEDM region for negative values of \cf. Furthermore, floating \cv gives rise to a smearing of the star shape in the \cf direction. In comparison to \cref{fig:edm_fermions}, the reachable \YBratio values are slightly increased to 0.11.

\subsubsection{Top and electron Yukawa couplings}
\label{sec:results_complementarity_e}
\begin{figure}
    \centering
    \includegraphics[width=.75\linewidth]{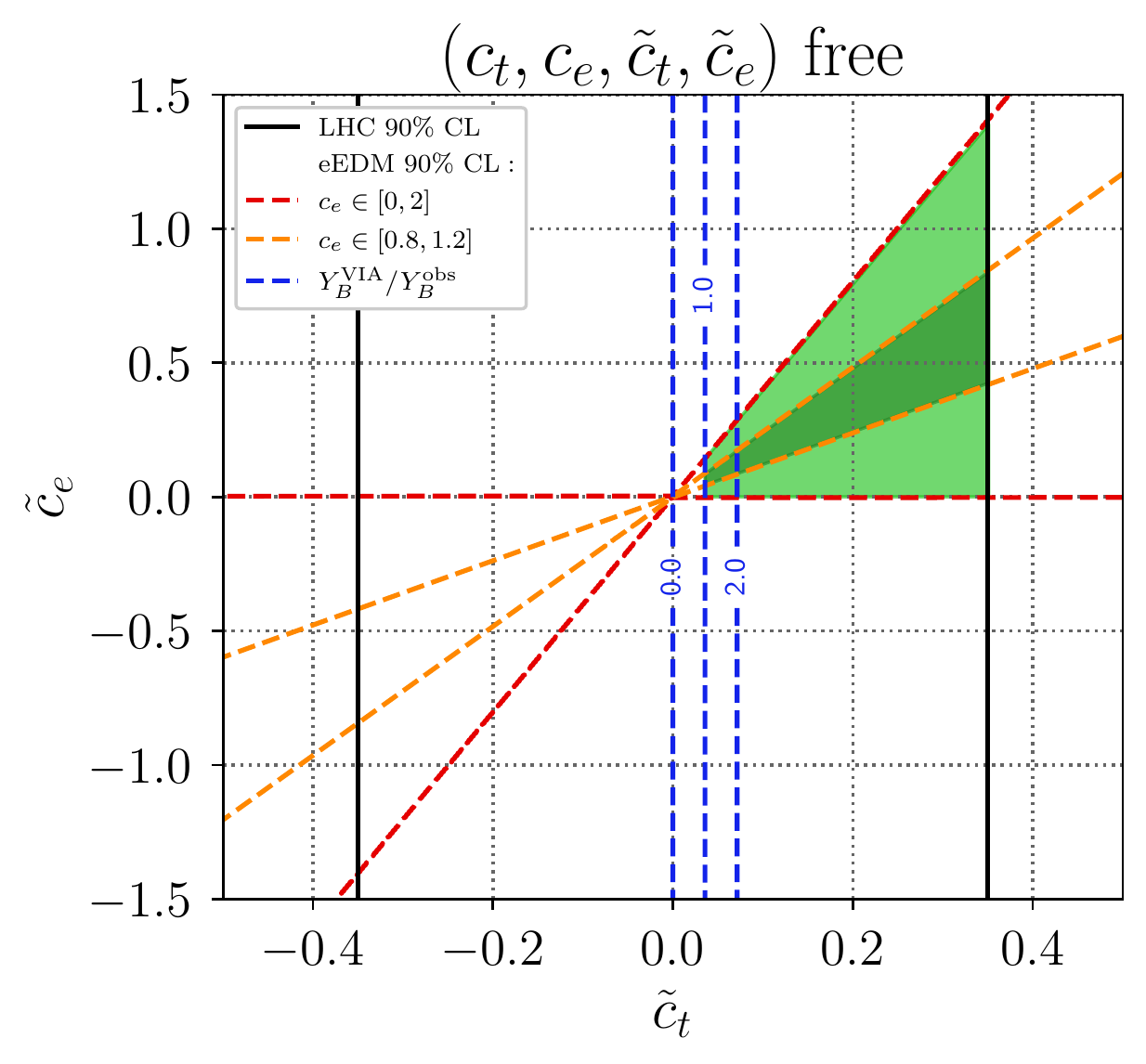}
    \caption{Constraints on the \cp-odd modifiers of the top- and electron-Yukawa interactions with the legend similar to \cref{fig:edm_1flavour_a}. For evaluating the eEDM, \ct is varied within its $90\%$ CL collider limits, and \ce is varied in the interval $[0,2]$ (red contours) and $[0.8,1.2]$ (orange contours).}
    \label{fig:edm_topelectron}
\end{figure}

As discussed above, the eEDM sets strong constraints on \cp violation in the Higgs sector, in particular for the top quark Yukawa interaction, if the electron Yukawa coupling is close to its SM value. 
However, these constraints may vary a lot depending on the value of the electron-Yukawa coupling, which is only very weakly constrained by LHC measurements. 
In the extreme case of a zero electron Yukawa coupling (i.e.\ $\ce = \cet = 0$), the  prediction for the eEDM would be strongly suppressed regardless of the amount of \cp violation in other Higgs couplings. But even in the case of only a small deviation in the electron-Yukawa coupling from the SM value, large cancellations can occur in the eEDM prediction between the contributions involving the components of the top quark and the electron Yukawa couplings.\footnote{The measurement of other EDMs like the neutron EDM can in principle constrain \cp violation in the Higgs sector without relying on assumptions on the electron Yukawa coupling. These EDM measurements, however, either rely on the knowledge of the up quark and down quark Yukawa couplings or are significantly less sensitive than the eEDM (as is the case for the Weinberg operator contribution to the neutron EDM~\cite{Brod:2013cka}).}

This is illustrated in \cref{fig:edm_topelectron}, showing the LHC and eEDM constraints on a model in which the coupling modifiers of the electron and top quark Yukawa interactions are floated freely. The results are depicted in the $(\ctt,\cet)$ parameter plane. Since the LHC limits on the electron Yukawa coupling are not relevant in the displayed parameter space, the LHC constraints appear as vertical lines limiting \ctt to be within $\sim [-0.35,0.35]$ at the 90\% CL
(for the LHC constraints we profile over \ct). Varying the electron Yukawa coupling modifiers, however, strongly affects the eEDM prediction. At every point in the $(\ctt,\cet)$ parameter plane, we vary \ct within its $90\%$ CL LHC limits and \ce in the interval $[0,2]$ (red contours) or in the interval $[0.8,1.2]$ (orange contours); then, we derive the minimal possible $d_e$ value. Since the contribution of a \cp-violating electron Yukawa coupling enters with a coefficient of
similar size as the contribution of a \cp-violating top quark Yukawa coupling, see \cref{eq:EDM_fit_formula}, large cancellations can occur for appropriate values of \ce, \cet, \ct and \ctt. As a consequence, much larger \ctt values become accessible than in \cref{fig:edm_1flavour_top}. As indicated by the different sizes of the regions enclosed by the red and orange contours, varying \ce in a larger region would make an even larger portion of the shown parameter space compatible with the eEDM constraint. As long as \cet floats freely, a sufficient amount of \cp violation to explain the BAU can be generated (see the green colored regions), which corresponds to $\YBratio \lesssim 9.8$. Note, however, that at sizeable values of \ctt or \cet a large amount of fine-tuning is necessary for those cancellations to occur.

\subsubsection{Comparison of the maximal contributions to the BAU}

We summarize some of the results of this section in \cref{tab:YBmax}, where we list the maximal values for \YBratio that are obtained within the regions allowed by the LHC and eEDM constraints at 90\% CL for all combinations of two out of the five considered fermion flavors, where either the modifiers for one or two couplings are varied. The modifiers for the electron Yukawa coupling in this table are fixed to $\ce = 1$, $\cet = 0$. The largest amount of \YBratio with a single \cp-violating \hife coupling can be reached by allowing for \cp violation in the tau Yukawa coupling. This feature of the tau Yukawa coupling is due to a combination of several reasons. First, leptons are not affected by the strong sphalerons that would wash out the initial asymmetry~\cite{deVries:2018tgs}. Moreover, the diffusion of the asymmetry across the bubble wall is more efficient for leptons~\cite{Joyce:1994zt,Cline:2001rk}. Finally, the smaller Yukawa coupling of the tau lepton compared to the top quark leads to a weaker bound from the eEDM.

Adding \cp violation in a second \hife coupling increases the reachable \YBratio by a factor of $\sim 2.2$.\footnote{Regardless whether \cp violation in the tau Yukawa coupling is combined with \cp violation in the top quark, the bottom quark, or the charm Yukawa couplings, we obtain a maximal value of $\YBratio \sim 6.9$. For each of these cases, the amount of baryon asymmetry is almost completely due to \cp violation in the tau Yukawa coupling. The presence of an additional source of \cp violation, however, can reduce the impact of the eEDM constraint. The maximal value for \YBratio is then determined by the collider constraints on the tau Yukawa coupling (see also \cref{fig:edm_2flavour_a}).} Allowing for \cp violation only in up to two \hife couplings excluding the tau and the electron Yukawa couplings, a sufficient amount of \cp violation to explain the baryon asymmetry of the universe cannot be reached -- not even in the optimistic VIA framework. The highest reachable value of \YBratio in this case is $\sim 0.42$, obtained by allowing for \cp violation in the top quark and bottom quark Yukawa couplings. For the scenario with global Higgs-fermion coupling modifiers, we obtained a maximally reachable value of $\YBratio \sim 0.08$. If in addition also \cv is varied, a slightly higher value of $\YBratio \sim 0.11$ can be reached.

\begin{table}[t!]
\renewcommand{\arraystretch}{1.4}
    \centering
    \begin{tabular}{|c||c|c|c|c|c|}
    \hline
        &$t$    &$b$    &$c$    &$\tau$ &$\mu$ \\
    \hline \hline
    $t$ &0.03       &       &       &       &       \\
    $b$ &0.42       &0.05       &       &       &       \\
    $c$ &0.37       &0.19       &0.01       &       &       \\
    $\tau$ &6.9       &6.9       &6.9       &3.2       &       \\
    $\mu$ &0.18       &0.19       &0.16       &3.2       &0.16       \\ \hline
    \end{tabular}
\caption{Maximal values of \YBratio within the regions that are allowed by the LHC and eEDM constraints at 90\% CL for all combinations of the five considered fermion flavors, where the modifiers of up to two couplings are varied. The electron-Yukawa coupling is fixed at
$\ce = 1$, $\cet = 0$. The diagonal entries of the Table represent the cases where only one Yukawa coupling is modified.}
\label{tab:YBmax}
\end{table}


\section{Conclusions}
\label{sec:conclusions}

\cp violation in \hife interactions is an intriguing possibility, since it could play an important role in explaining the observed asymmetry between matter and antimatter in the universe. In this work, we have explored this option by taking into account inclusive and differential experimental results from LHC measurements of Higgs production and decay processes and the eEDM limit in combination with theoretical predictions in an effective model description. After determining the parameter space that is favored by these constraints, we have assessed to which extent \cp violation in the \hife couplings can contribute to the observed baryon asymmetry.

The first part of this work focused on the LHC constraints on \cp-violating \hife couplings. The Higgs characterization model has been used, allowing not only a variation of the various \hife couplings but also of the Higgs coupling to massive vector bosons as well as of the effective Higgs couplings to gluons and photons. Besides including total and differential rate measurements, we also included into our global fit the recent dedicated $H\to\tau^+\tau^-$ \cp analysis performed by the CMS collaboration~\cite{CMS:2021sdq}. This yields $|\ctaut| < 0.75$ at the 95\% CL considering the scenario where only the components of the tau Yukawa coupling are allowed to float freely (1-flavor model), whereas $|\ctaut| < 1.1$ would have been allowed if only the \cp-conserving rate measurements had been taken into account.

We studied the LHC constraints by investigating models with increased complexity. The simplest cases are the 1-flavor models, in which we allowed for deviations from the SM in the interaction of the Higgs boson with only one fermion species. Here, we found the strongest constraints for \cp-violating tau and top quark Yukawa couplings, where the latter is mainly constrained by Higgs production and the decay into photons (for the constraints on the tau Yukawa coupling see the discussion above). On the other hand, the LHC constraints on the \cp-violating bottom quark
and muon Yukawa couplings are currently driven by the \cp conserving observables from the Higgs decay into these fermions. In contrast, the complex charm quark Yukawa coupling is most stringently constrained by the precise measurement of the $H\to\gamma\gamma$ decay rate via the modification of the total Higgs width.
These constraints from Higgs decays give rise to rings of allowed parameters in the plane of the modifiers of the real and imaginary parts of the couplings. The first generation Yukawa couplings are still almost unconstrained. 

When allowing for \cp violation in the Higgs interactions with two fermion species, we found the LHC constraints on the different species to be only weakly correlated, so that the constraints from the 1-flavor modification fits were largely recovered. As an alternative approach, we studied models in which the couplings of a specific group of fermions are modified universally (e.g.\ of all quarks or all leptons). As expected, we found the constraints on the third generation couplings to always dominate the fit results with the top quark Yukawa constraints being the most important among these.
We extended these fit results by increasing the number of coupling modifiers that are allowed to float independently of each other to up to nine (see the discussion in~\cref{sec:appendix_additional_fits}). Generally, our fit confirms the expectation that the allowed parameter region is enlarged in models with additional freedom.

As complementary constraints, we then studied the impact of the eEDM bound and assessed to which extent the BAU can be explained within the parameter regions that are in agreement with the LHC and eEDM constraints. Our approach in this context accounts for the fact that the BAU predictions are affected by large theoretical uncertainties that are in particular related to the choice made in the calculation framework regarding the use of the VIA or the WKB approach~\cite{Cline:2021dkf}. We therefore did not require in our analysis that the BAU prediction using the VIA has to match the observed value, but we have treated all values of $\YBratio \geq 1$ as theoretically allowed. Since the results for the BAU obtained via the WKB approach are significantly smaller than the ones based on the VIA, even for $\YBratio \ge 1$ further sources of \cp violation besides the couplings of the observed Higgs signal at $125\gev$ might be needed. On the other hand, we regard parameter regions with 
$\YBratio < 1$ as disfavored by the observed BAU because the bubble wall parameters used in the VIA calculation are near the values that maximize the predicted BAU.

Using this approach, we found that the amount of \cp violation in the tau Yukawa coupling that is allowed by the latest LHC and eEDM constraints would suffice to explain the BAU (if calculated in the VIA framework) even if it occurs as the only source of \cp violation in addition to the CKM phase. While similar conclusions were drawn previously, see \ccite{Guo:2016ixx,DeVries:2018aul,Fuchs:2020uoc,Shapira:2021mmy}, we reevaluated this statement based on a non-trivial global fit taking into account the very significantly improved constraints on the imaginary part of the tau Yukawa coupling that arise in particular from the inclusion of the recent angular analysis performed by the CMS Collaboration.
Still, the eEDM remains the strongest bound on \ctaut, yielding \ctaut<0.3 at the 90\% CL in the 1-flavor case, i.e.\ about a factor of 2 stronger than the angular CMS analysis.
Moreover, we have confirmed that the feature that \cp violation in the tau Yukawa coupling could account for the whole BAU is unique to this coupling. Our results show that this is not possible for \cp violation in any other single Yukawa coupling and it also cannot be realized for the case where \cp violation in two other third- or second-generation Yukawa couplings (i.e.\ excluding the tau Yukawa coupling) is allowed. 

Regarding the eEDM, it should be noted that the impact of this constraint crucially depends on the chosen input value for the electron Yukawa coupling, which is still almost completely unconstrained by LHC measurements\footnote{For a discussion of the technical challenges involved in obtaining limits on $g_e$ at possible future lepton colliders, see \ccite{Barger:1995hr,dEnterria:2021xij}.}. Treating this unknown quantity as a free parameter reduces the contributions to the eEDM for the case where $|\ce|$ is below the SM value and gives rise to possible cancellations between the different contributions to the eEDM for the case of \cp violation in the electron Yukawa coupling. Accordingly, in those cases substantial parts of the considered parameter space are phenomenologically viable even in view of the latest improvement of the eEDM limit.

Our analysis has demonstrated that the analysis of possible \cp violation in the Higgs sector is of particular interest, since \cp-violating Yukawa couplings can potentially explain the BAU while satisfying all relevant experimental and theoretical constraints. The further exploration of this issue will greatly profit from the complementarity between the information obtainable at colliders, from the EDMs of the electron and of other systems, as well as from improvements in the predictions for the BAU.


\section*{Acknowledgements}
\sloppy{
We thank P.~Bechtle for collaboration in the early phase of this work; T.~Stefaniak and J.~Wittbrodt for help with \texttt{HiggsSignals}; J.~Brod, G.~Panico, M.~Riembau and E.~Stamou for clarifications regarding their 
EDM calculations; as well as A.~Cardini and D.~Winterbottom for useful discussions regarding the CMS $H\to\tau\tau$ \cp analysis. J.~K., K.~P., and G.~W.\ acknowledge support by the Deutsche Forschungsgemeinschaft (DFG, German Research Foundation) under Germany's Excellence Strategy --- EXC 2121 ``Quantum Universe'' --- 390833306. E.~F.\ and M.~M.\ acknowledge support by the Deutsche Forschungsgemeinschaft (DFG) under Germany's Excellence Strategy --– EXC-2123 ``QuantumFrontiers'' --- 390837967. H.~B.\ acknowledges support by the Alexander von Humboldt foundation. The work of S.~H.\ is supported in part by the grant PID2019-110058GB-C21 funded by MCIN/AEI/10.13039/501100011033 and by "ERDF A way of making Europe", and in part by the grant CEX2020-001007-S funded by MCIN/AEI/10.13039/501100011033.
}


\appendix


\section{Additional fit results}
\label{sec:appendix_additional_fits}

In this Appendix, we collect our results of additional 
fits to LHC data which are supplementary to the results presented in \cref{sec:results}.

\begin{figure}
    \centering
    \subfigure[]{\includegraphics[width=.48\linewidth]{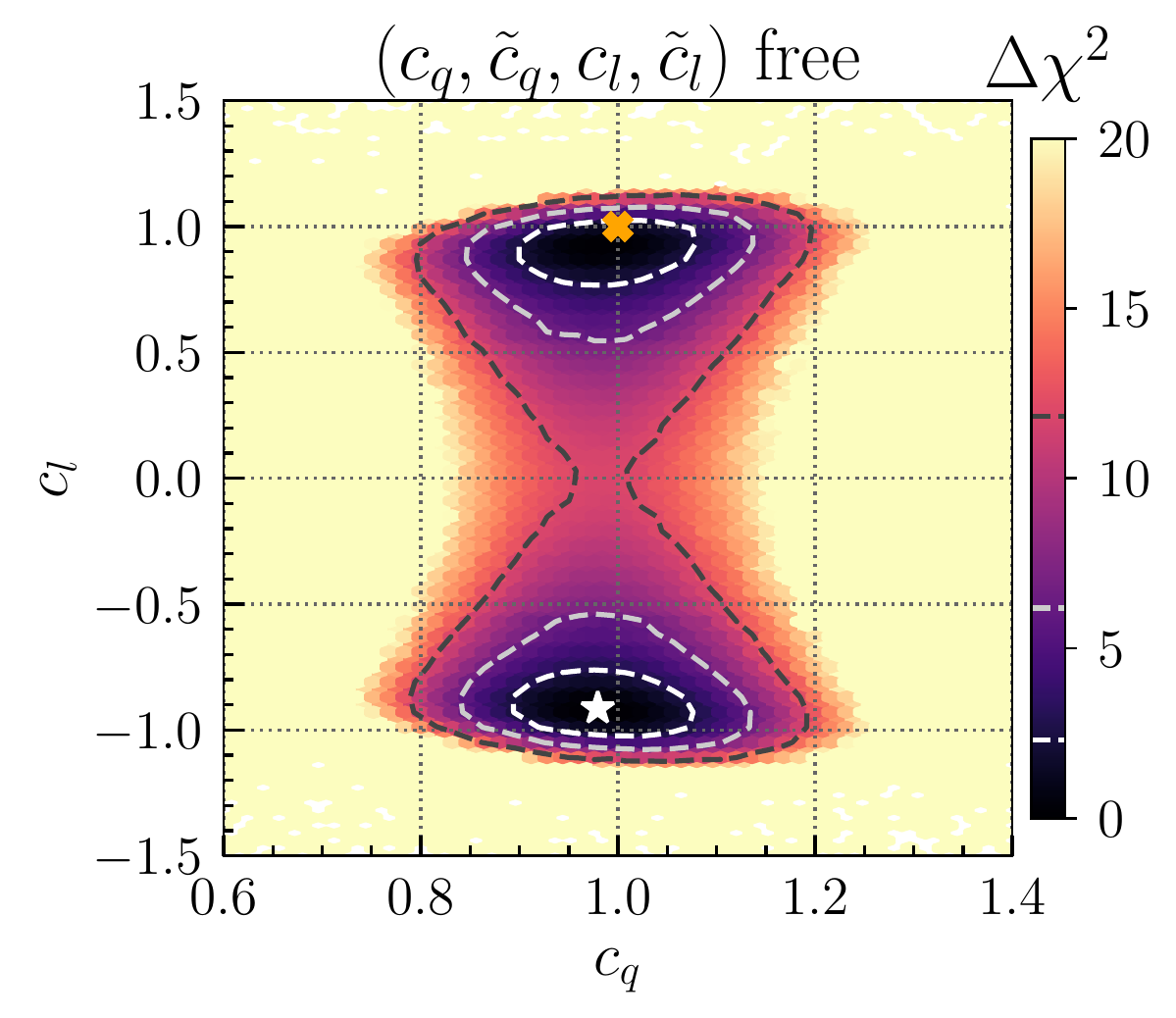}\label{fig:quark_lepton_even}}
    \subfigure[]{\includegraphics[width=.48\linewidth]{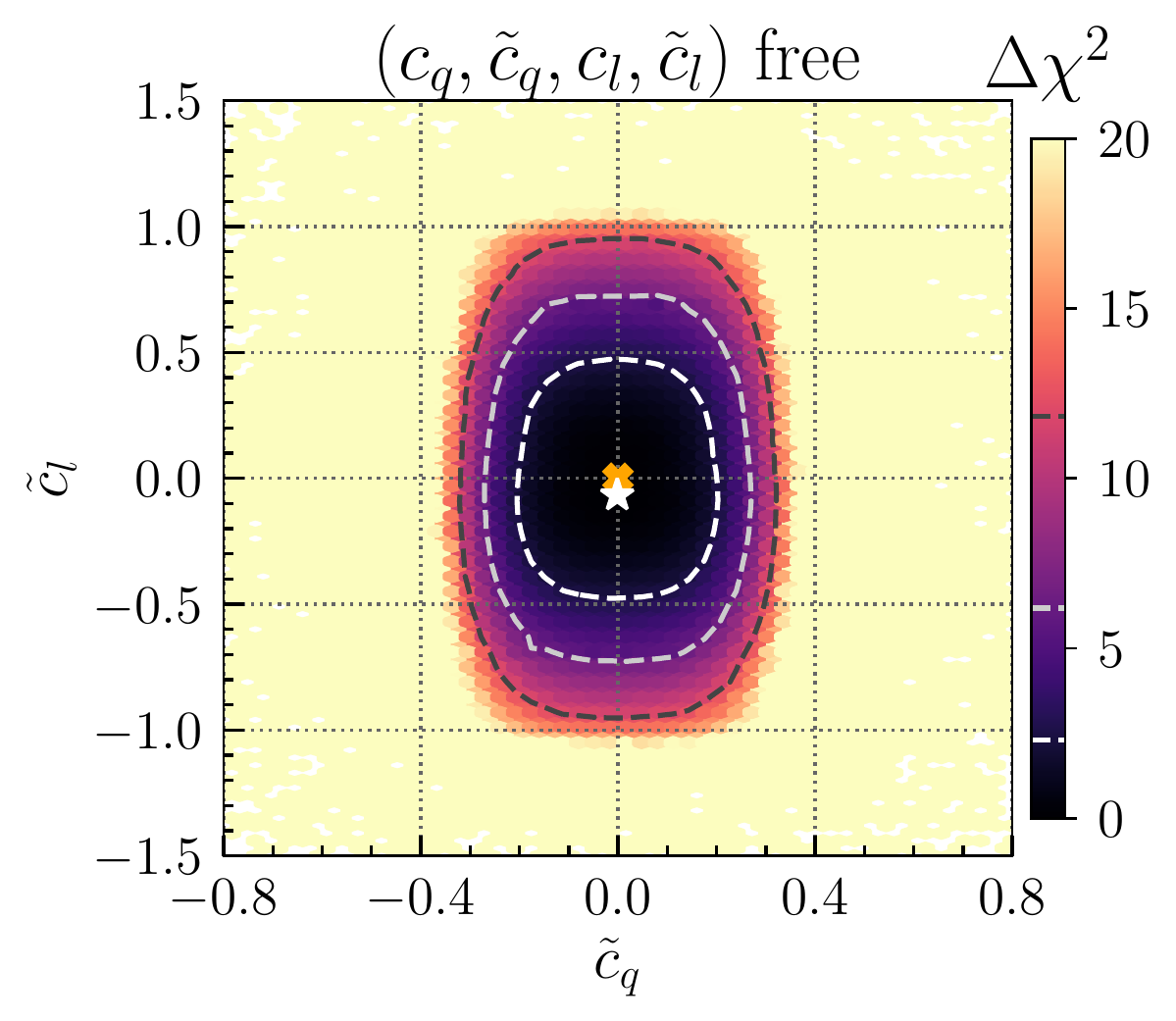}\label{fig:quark_lepton_odd}}
    \caption{
    Results of fits to the LHC measurements in the (a) (\cq, \cl) and (b) (\cqt, \clt)  parameter plane. 
    In both cases, the four parameters shown in the plot are free-floating while all the other parameters are set to their SM values. The legend corresponds to the one  in \cref{fig:cms_analysis}. 
   }
    \label{fig:quark_lepton}
\end{figure}

\paragraph{Quark--lepton model} One possibility that was not explored in \cref{sec:results} is to allow for separate modifications in the quark and lepton sector (varying \cq, \cqt, \cl, and \clt). The corresponding fit results are shown in \cref{fig:quark_lepton}. The results are similar to the two-flavor fit in which \ct, \ctt, \ctau, and \ctaut are varied (see \cref{fig:2flavour_2_a,fig:2flavour_2_b}), with \chimin = 87.84. The quark and lepton sectors are again only weakly correlated via the $H\to\gamma\gamma$ decay process. Setting $\cb = \ct \equiv \cq$ and $\cbt = \ctt \equiv \cqt$ slightly tightens the constraints in comparison to \cref{fig:2flavour_2_a,fig:2flavour_2_b}.

\begin{figure}
    \centering
    \subfigure[]{\includegraphics[width=.49\linewidth]{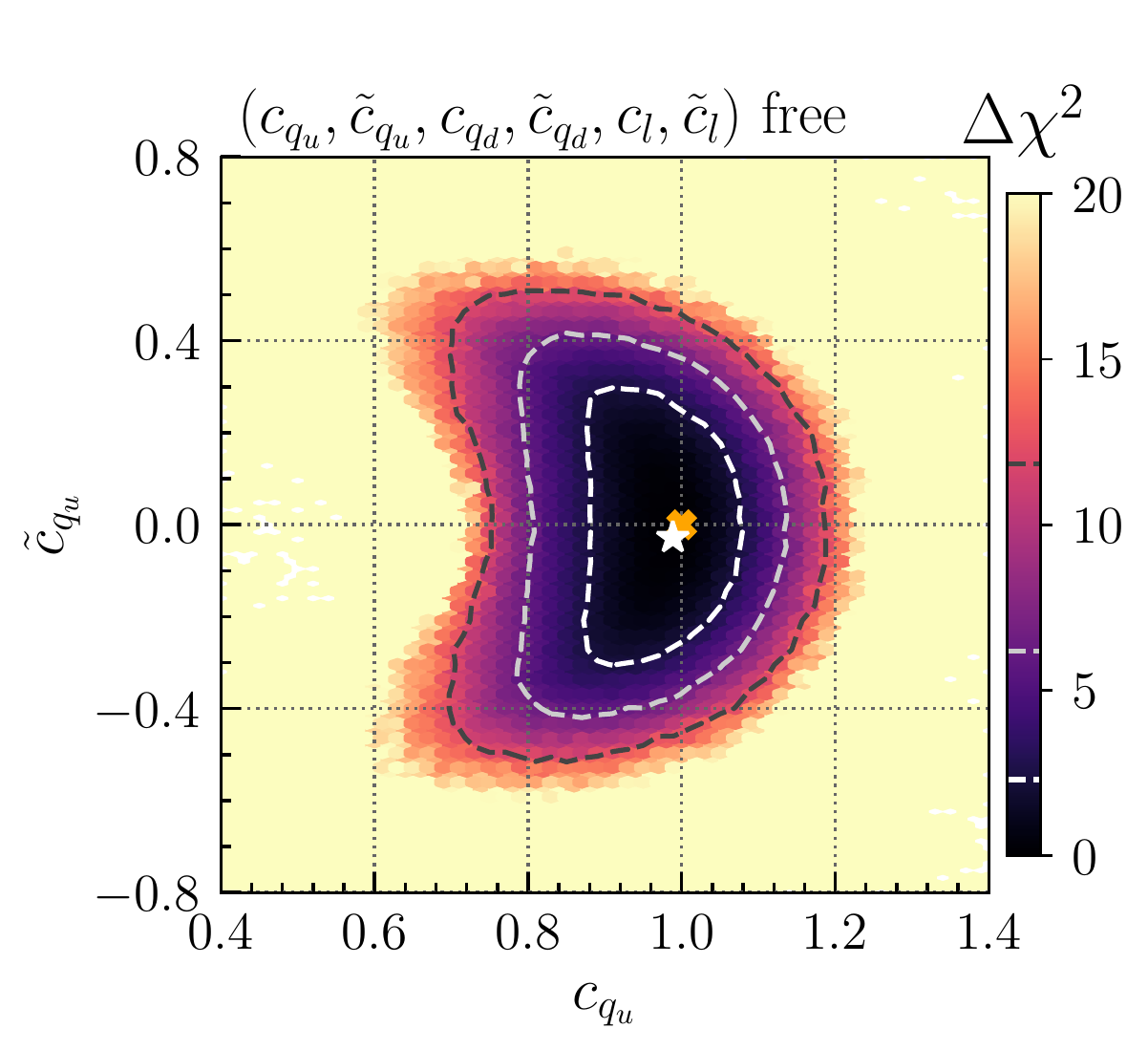}\label{fig:up_down_lepton_qu}}
    \subfigure[]{\includegraphics[width=.48\linewidth]{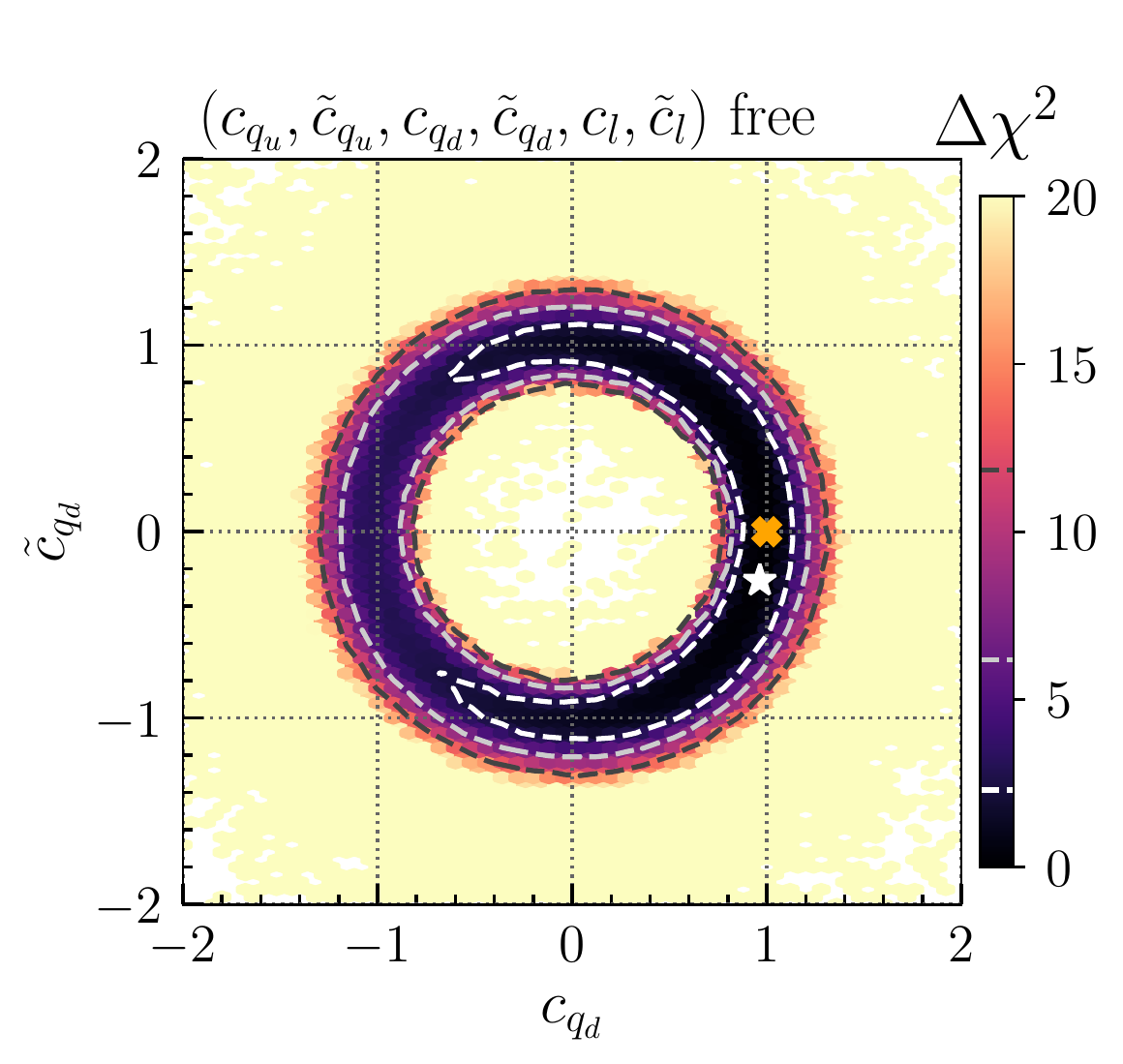}\label{fig:up_down_lepton_qd}}
    \subfigure[]{\includegraphics[width=.48\linewidth]{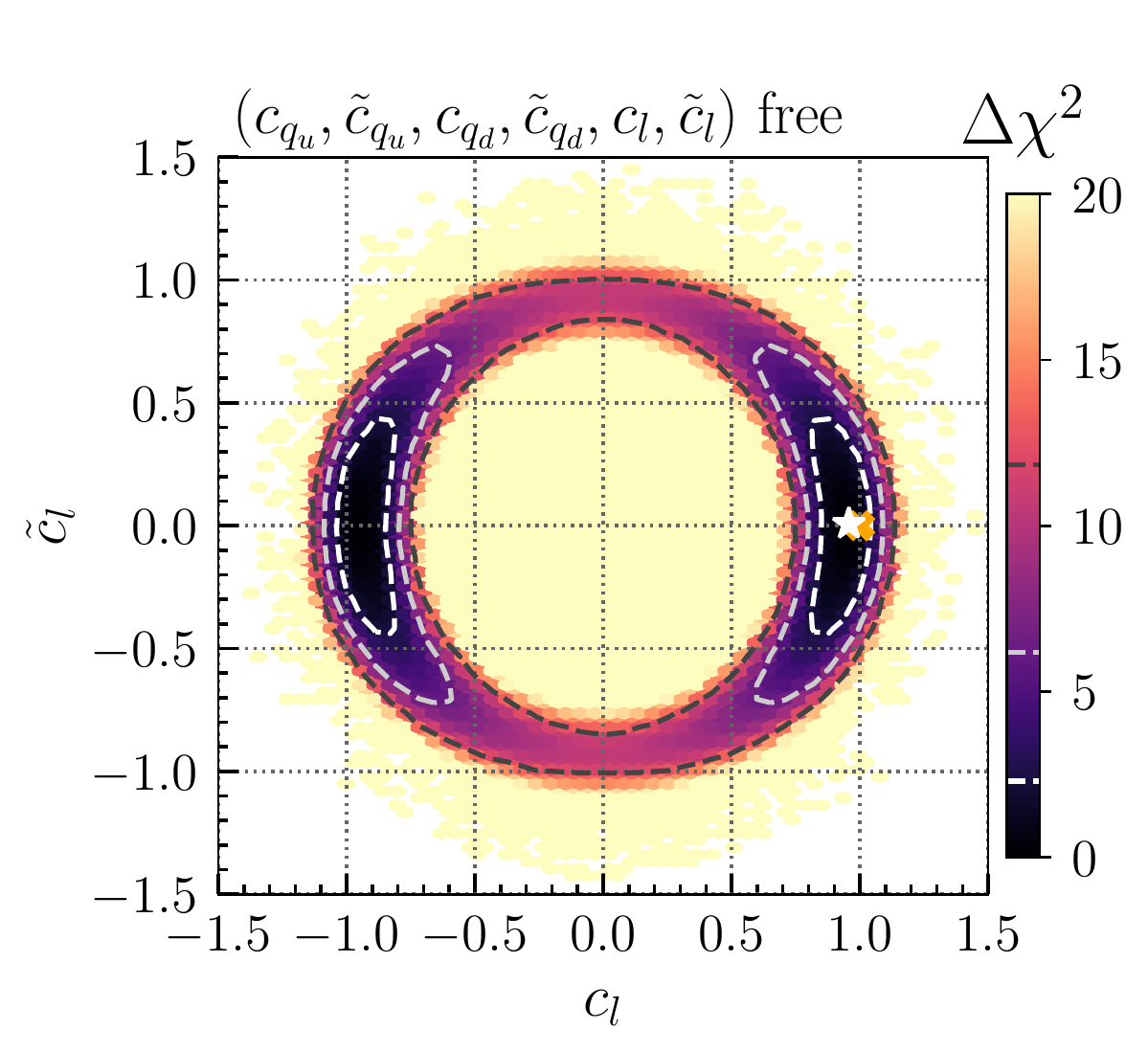}\label{fig:up_down_lepton_l}}
    \caption{
    Results of fits to the LHC measurements in the (a) (\cqu, \cqut), (b) (\cqd, \cqdt), and (c) (\cl, \clt)  parameter plane. 
    For each plot the indicated coupling modifiers are free-floating while all the other parameters are set to their SM values.
    The legend corresponds to the one in \cref{fig:cms_analysis}.
    }
    \label{fig:up_down_lepton}
\end{figure}

\paragraph{Up--down--lepton model} The results can be further generalized by treating the up- and down-type quark sector separately. The fit results of varying \cqu, \cqut, \cqd, \cqdt, \cl, and \clt are shown in \cref{fig:up_down_lepton}. The constraints on the different sectors are again dominated by the constraints on the third generation couplings and only weakly correlated: the constraints on \cqu and \cqut, see \cref{fig:up_down_lepton_qu}, resemble the $(\ct,\ctt)$ fit shown in \cref{fig:1flavour_a} with the bounds being slightly looser due to the additional variation of the bottom-Yukawa coupling; the constraints on \cqd and \cqdt, see \cref{fig:up_down_lepton_qd}, resemble the $(\cb,\cbt)$ fit shown in \cref{fig:1flavour_b} with the bounds being slightly looser due to the additional variation of the top-Yukawa coupling; the constraints on \cl and \clt, see \cref{fig:up_down_lepton_l}, resemble the $(\cl,\clt)$ fit shown in \cref{fig:shared_phases_l}. The best-fit point is found at \chimin = 87.80.

\begin{figure}
    \centering
    \subfigure[]{\includegraphics[width=.49\linewidth]{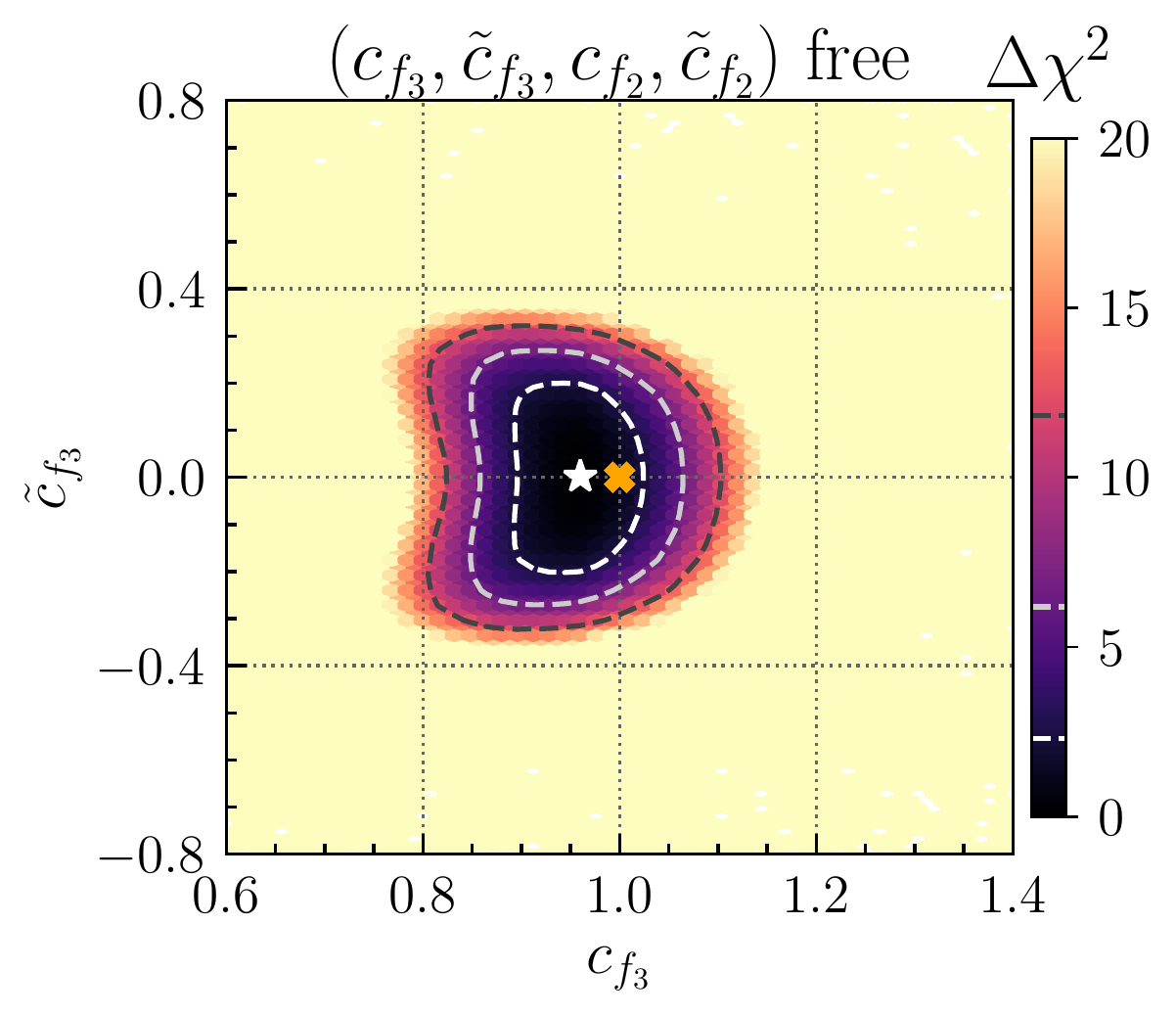}\label{fig:generations_3}}
    \subfigure[]{\includegraphics[width=.48\linewidth]{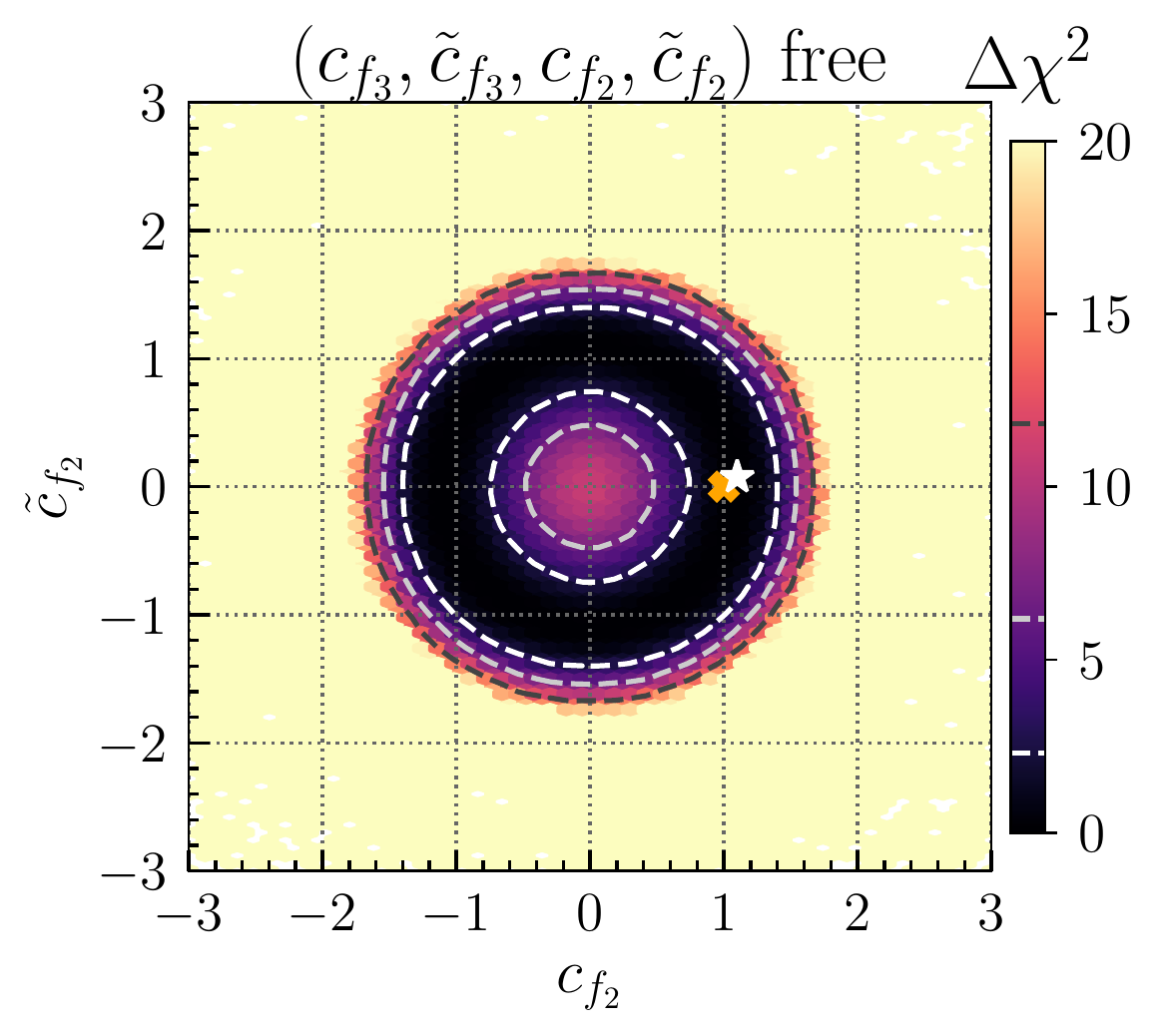}\label{fig:generations_2}}
    \caption{
    Results of fits to the LHC measurements in the (a) (\cfff, \cffft) and (b) (\cff, \cfft)  parameter plane.
    For each plot the indicated coupling modifiers are free-floating while all the other parameters are set to their SM values.
    The legend corresponds to the one in \cref{fig:cms_analysis}.
    }
    \label{fig:generations}
\end{figure}

\paragraph{2$^\text{nd}$/3$^\text{rd}$ generation model} Another possibility to generalize the fit results is to differentiate the second and third generation (we do not refer to the first generation here, since the collider limits on the first generation obtained so far are very weak). The corresponding fit results floating \cfff, \cffft, \cff, and \cfft are shown in \cref{fig:generations}. There is hardly any correlation between the constraints on the second and third generation. Therefore, the constraints on the third generation, see \cref{fig:generations_3}, are very similar to the $(\cf,\cft)$ fit presented in \cref{fig:shared_phases_f}. The constraints on the second generation, shown in \cref{fig:generations_2}, are dominated by the constraints on the muon-Yukawa coupling (see \cref{fig:1flavour_d}). The best-fit point has \chimin = 87.68.

\begin{figure}
    \centering
    \subfigure[]{\includegraphics[width=.48\linewidth]{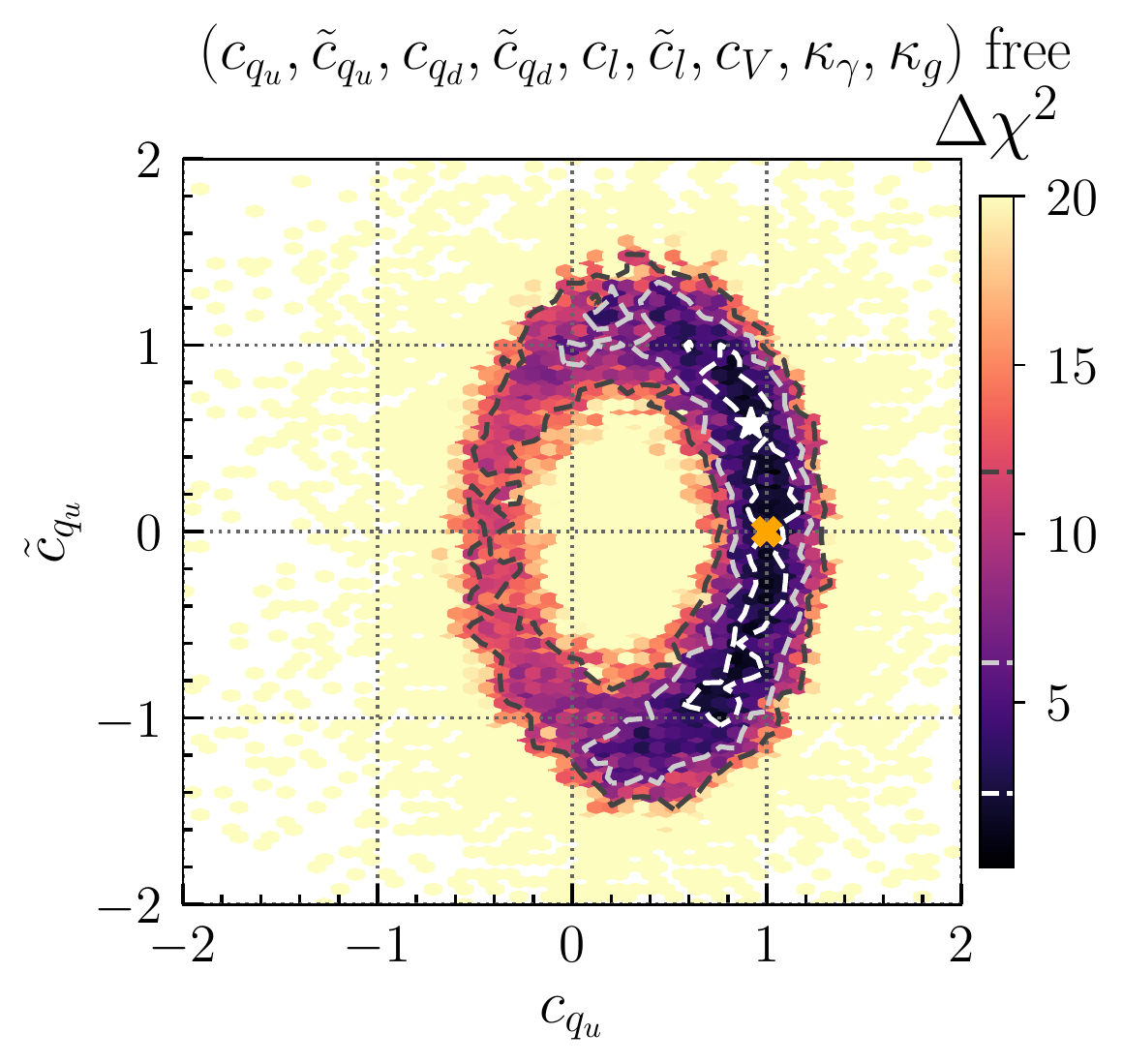}\label{fig:up_down_lepton_full_qu}}
    \subfigure[]{\includegraphics[width=.48\linewidth]{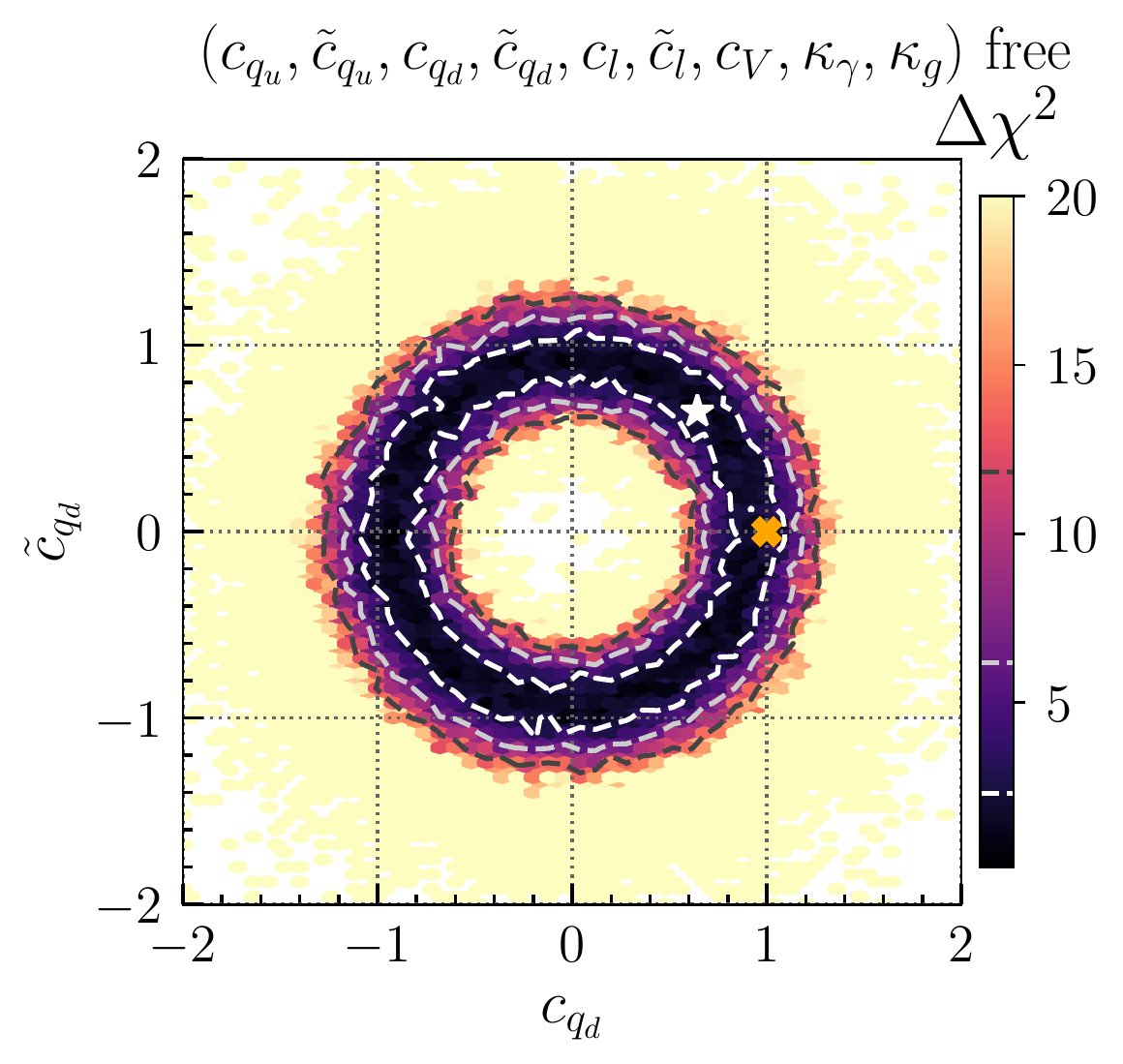}\label{fig:up_down_lepton_full_qd}}
    \subfigure[]{\includegraphics[width=.48\linewidth]{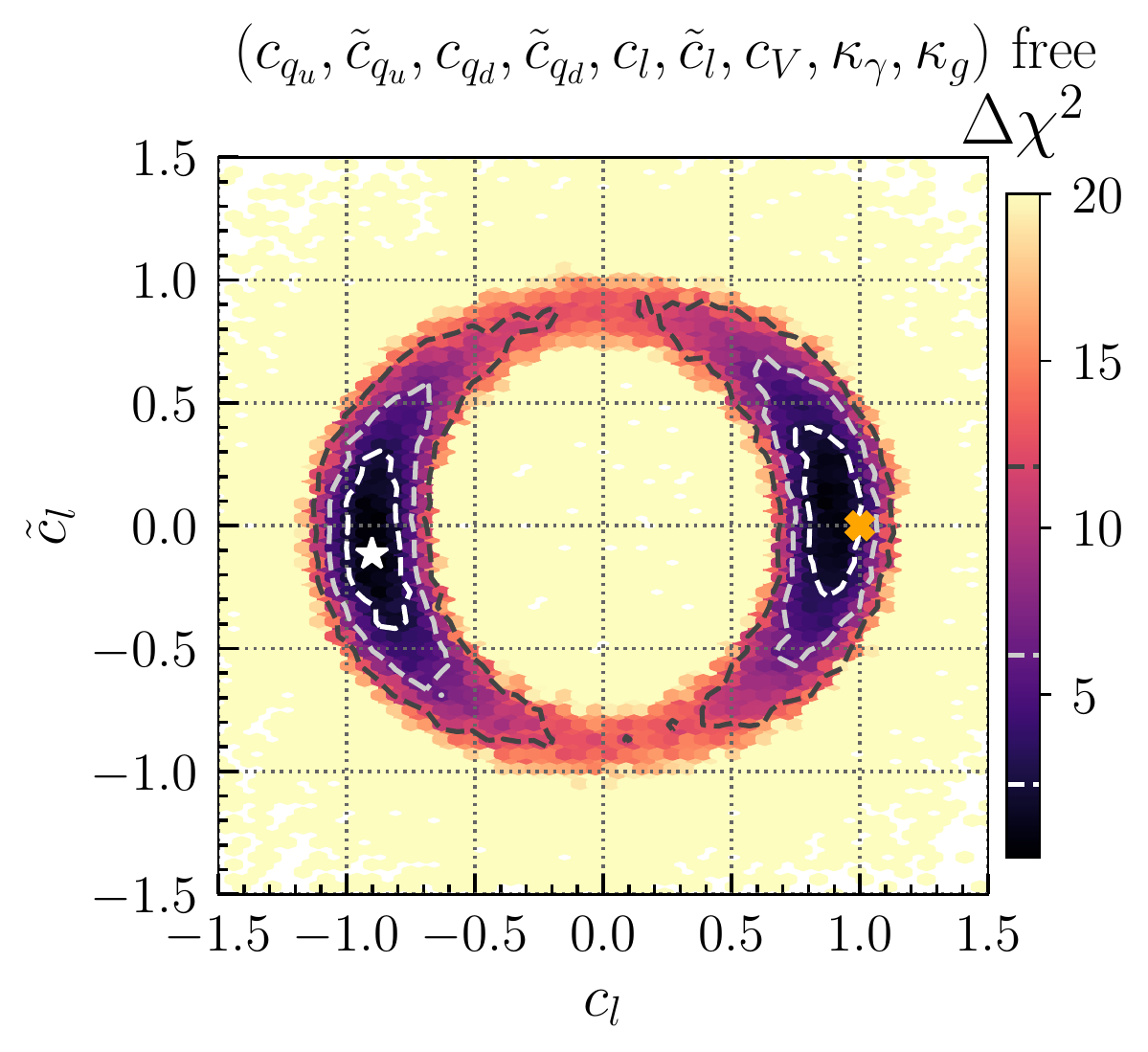}\label{fig:up_down_lepton_full_l}}
    \caption{
    Results of fits to the LHC measurements in the (a) (\cqu, \cqut), (b) (\cqd, \cqdt), and (c) (\cl, \clt) parameter plane. 
    All parameters listed above the plot panels are floated freely. 
    The legend corresponds to the one in \cref{fig:cms_analysis}.
    }
    \label{fig:up_down_lepton_full}
\end{figure}

\paragraph{Up--down--lepton--vector model} As the most general model considered in this work, we do not only vary the up-, down-, and lepton-Yukawa couplings separately, but in addition we also freely float \cv, \kg, and \kgamma. The resulting constraints in the $(\cqu, \cqut)$, $(\cqd, \cqdt)$, and $(\cl, \clt)$ parameter planes are shown in \cref{fig:up_down_lepton_full}, with \chimin = 87.80.\footnote{The rather fuzzy boundaries shown are due to the high dimensionality of the analyzed parameter space, which makes the numerical evaluation very costly. The slightly higher value of \chimin compared to other models 
is likely to originate from the larger step size in the sampling.} 
In \cref{fig:up_down_lepton_full_qu}, showing the constraints for the up-type Yukawa couplings, we observe a significantly enlarged allowed region in comparison to \cref{fig:up_down_lepton_qu}. This is mainly a consequence of freely floating \kg and \kgamma (for a more detailed discussion see \ccite{Bahl:2020wee}). The constraints on the down-type Yukawa couplings, see \cref{fig:up_down_lepton_full_qd}, are only slightly weaker in comparison to \cref{fig:up_down_lepton_qd}. As a consequence of freely floating \kg, $\cqd \simeq -1$ is allowed at the one-sigma level. For the constraints on the lepton-Yukawa couplings, see \cref{fig:up_down_lepton_full_l}, the boundaries of the $3\,\sigma$ region are slightly tighter than in \cref{fig:up_down_lepton_l}. This is most likely an artefact of the coarse sampling, which could be avoided with an increased sample size.



\clearpage
\printbibliography

\end{document}